\documentclass[onecolumn,journal,noadjust]{IEEEtran}
\usepackage{cite}
\ifCLASSINFOpdf
\else
\fi
\usepackage{amssymb}
\usepackage{dsfont}
\usepackage{float}
\usepackage{graphicx}
\newtheorem{lemma}{Lemma}

\newtheorem{proposition}{Proposition}
\newtheorem{theorem}{Theorem}
\newtheorem{remark}{Remark}
\newtheorem{corollary}{Corollary}
\usepackage{xcolor}
\usepackage{multirow}
\usepackage{makecell}

\usepackage [autostyle, english = american]{csquotes}
\MakeOuterQuote{"}

\usepackage{amsmath}

\begin{document}
%\title{$d$-Semifaithful Coding with Universal Distortion}
\title{Minimax Rate-Distortion}

\author{Adeel Mahmood and Aaron B. Wagner\\
School of Electrical and Computer Engineering, Cornell University}

\maketitle

\begin{abstract}
We show the existence of variable-rate rate-distortion codes
that meet the disortion constraint almost surely and are
minimax, i.e., strongly, universal with respect to an unknown
source distribution and a distortion measure that is revealed
only to the encoder and only at runtime. If we only require
minimax universality with respect to the source distribution
and not the distortion measure, then we provide an 
achievable $\tilde{O}(1/\sqrt{n})$ redundancy rate, which
we show is optimal. This is in contrast to prior work on
universal lossy compression, which provides $O(\log n/n)$ 
redundancy guarantees for weakly universal codes under various 
regularity conditions. We show that either eliminating
the regularity conditions or upgrading to strong universality while keeping these regularity conditions entails an inevitable increase in the 
redundancy to $\tilde{O}(1/\sqrt{n})$. Our construction
involves random coding with non-i.i.d.\ codewords 
and a zero-rate uncoded transmission scheme. The proof
uses exact asymptotics from large deviations, acceptance-rejection sampling, 
and the VC dimension of distortion measures.
\end{abstract}


% \fi
\IEEEpeerreviewmaketitle

\section{Introduction}
Consider the problem of lossy compression of a memoryless source on a finite alphabet. Let $X^n$ be an independent and identically distributed (i.i.d.) source taking values on a finite source alphabet $A$ with cardinality $J$. Let $B$ be a finite reconstruction alphabet with cardinality $K$. The fidelity criterion we consider is a single-letter distortion measure $\rho$ between source and reconstruction alphabets. We fix a distortion level $d > 0$ and consider variable-rate codes that meet the distortion constraint almost surely; such codes
are sometimes called \emph{$d$-semifaithful}~\cite{Ornstein:Semifaithful}, \cite{yu1}.
It is well-known that 
%The traditional lossy source coding theorem \cite{berger1} asserts that the 
%best distortion performance achievable asymptotically by block coding at a fixed rate $R$, when optimized for a particular source distribution and (single-letter) distortion measure, is given by the distortion-rate function $d(p,R,\rho)$, where $p$ is the source distribution and $\rho$ is the distortion measure. Likewise, the 
the minimum expected rate achievable asymptotically by a \textit{prefix} code optimized for a particular source distribution $p$ and distortion measure $\rho$ is given by the rate-distortion function $R(p,d,\rho)$. In source coding theory, both lossless and lossy compression, joint descriptions are more efficient than individual descriptions \cite{thomas_cov}; hence, the source sequence $X^n$ is compressed as an $n$-length block and past works have analyzed the convergence of the average\footnote{Average expected rate means the expected rate divided by the blocklength $n$.} expected rate to the rate-distortion function as a function of $n$. The resulting performance metric, i.e., the difference between the average expected rate and the rate-distortion function, is known as the \textit{rate redundancy}. When both the source $p$ and the distortion measure $\rho$ are known ahead of time, \cite[Theorem 5]{yang1} has established an achievable rate redundancy of $\ln n / n  + o(\ln n / n)$ under some regularity conditions while \cite[Theorem 4]{yang1} has given a converse result of $1/2\ln n / n + o(\ln n / n)$. These results stand in contrast to a rate redundancy of $O(1/n)$ \cite[Thm.~5.4.2]{thomas_cov} for prefix lossless codes when the source $p$ is known, where the Shannon entropy $H(p)$ replaces the rate-distortion function in the definition of the rate redundancy.       

In practice, the source distribution is rarely known, and thus one seeks \emph{universal}
codes that do not require knowledge of the source distribution and achieve
the same asymptotic performance of those that do. For an unknown i.i.d. source $p$, let $R(C_n,p)$ denote the expected rate of a prefix lossless code $C_n$ and let $R(\tilde{C}_n,p, d, \rho)$  denote the expected rate of a prefix, $d$-semifaithful lossy code $\tilde{C}_n$. Within the class of 
universal codes, a distinction is made between weakly universal and strongly universal codes \cite{1055092}. A weakly universal code is one with a rate that is guaranteed to converge to the minimum asymptotic limit for each source distribution $p$,
with no guarantee that this convergence is uniform over $p$.
A strongly universal code is one whose rate converges to the minimum asymptotic limit uniformly over all source distributions. This distinction is analogous to the pointwise versus uniform convergence of functions if we consider the expected rate, $R(C_n, p)$ or $R(\tilde{C}_n, p, d, \rho)$, as a function of $p$, where $R(C_n, p)$ converges to $H(p)$ and $R(\tilde{C}_n, p, d, \rho)$ converges to $R(p,d,\rho)$. For lossless compression, the
existence of strongly universal codes is well known \cite{rissanen1, beirami1, kosut1}. In fact,
practical codes are known that approach the entropy limit uniformly over 
the unknown source distribution, and the optimal rate of convergence has been characterized with precision~\cite{rissanen1, beirami1, kosut1}: 
\begin{align}
    \inf_{C_n} \, \sup_{p} \left[ R(C_n, p) - H(p) \right ] = \frac{J - 1}{2} \frac{\ln n}{n} + O\left(\frac{1}{n} \right). 
\end{align}
Less is known, on the other hand, about universal lossy codes, especially the minimax rate of convergence for 
\begin{align}
    \inf_{\tilde{C}_n} \, \sup_{p} \left[ R(\tilde{C}_n, p, d, \rho) - R(p,d,\rho) \right ]. \label{trad_univ}
\end{align}
The existence
of weakly universal, prefix $d$-semifaithful codes that achieve the rate-distortion function
for any source distribution under certain constraints is known, and their speed of convergence to
the rate-distortion function has been 
bounded. Under various regularity conditions,  \cite[Theorem 2]{yu1}
gives an achievable weakly universal convergence rate of 
\begin{align}
    \inf_{\tilde{C}_n} \left[ R(\tilde{C}_n, p, d, \rho) - R(p,d,\rho) \right ] \leq (JK + J + 4) \frac{\ln n}{n} + O\left(\frac{1}{n} \right). \label{yukomaruuu} 
\end{align}
The pre-log factor in $(\ref{yukomaruuu})$ has been improved in an unpublished paper by Yang and Zhang \cite{yang3} in which an achievable weakly universal convergence rate of 
\begin{align}
    \inf_{\tilde{C}_n} \left[ R(\tilde{C}_n, p, d, \rho) - R(p,d,\rho) \right ] \leq \left (\frac{K + 2}{2}\right) \frac{\ln n}{n} + o\left(\frac{\ln n}{n} \right) \label{chodogdskj} 
\end{align}
is shown under some regularity conditions. Furthermore, a converse result in the same paper provides a lower bound of 
\begin{align}
    \inf_{\tilde{C}_n} \left[ R(\tilde{C}_n, p, d, \rho) - R(p,d,\rho) \right ] \geq \left (\frac{K}{2}\right) \frac{\ln n}{n} + o\left(\frac{\ln n}{n} \right) \label{chodogdskj2} 
\end{align}
for most sources $p$ (but see Appendix~\ref{newresult447}). Universal lossy coding has also been considered with a fixed rate constraint instead of a fixed distortion constraint. In this framework, the performance metric used is called the distortion redundancy which is defined as the difference between the expected distortion and the distortion-rate function; see \cite{linder1} and \cite{yang2} which give a weakly universal convergence rate of  $O(\ln n / n)$ for fixed-rate codes which is the same order of decay as the weakly universal convergence rate known for optimal $d$-semifaithful codes. In addition to the convergence of  expected rate, one can also analyze rates of almost-sure convergence. Kontoyiannis~\cite{kontoyiannis1} and Kontoyiannis and Zhang~\cite{kontoyiannis2}
give bounds for almost-sure convergence 
to the rate-distortion function instead of the convergence of expected rate.

None of the aforementioned results for lossy compression are minimax, however; that is, the convergence to the rate-distortion 
function is pointwise for each source distribution as opposed to being uniform over
the set of all possible source distributions. With the exception of Kontoyiannis~\cite{kontoyiannis1},
the above achievability results also apply only to source distributions satisfying 
certain technical conditions. Some universal results of a minimax nature are available \cite[Problem 9.2]{korner1}, \cite{Gray:Ornstein:AP,Garcia-Munoz:Strong,Neuhoff:Shields:Markov,silva1,Neuhoff:Gray:Universal,MacKenthun:Universal,Ziv:Unknown:II}, although 
none provides an explicit bound on the minimax rate of convergence to the 
rate-distortion function for $d$-semifaithful codes. Some works have succeeded in obtaining minimax convergence rates for operational rate redundancy \cite{mahmood2021lossy} which is defined as the difference between the average expected rate and the minimum expected rate of an optimal $n$th order $d$-semifaithful code. Let $R^*(n,p,d,\rho)$ denote the optimal rate for a given $n$, $p$, $d$ and $\rho$:
\begin{align}
     R^*(n,p,d,\rho) = \inf_{\tilde{C}_n} R(\tilde{C}_n,p, d, \rho), \label{hellooperational}
\end{align}
where the infimum is over all prefix codes that are $d$-semifaithful
under $\rho$. Note that $R^*(n,p,d,\rho) \geq R(p,d,\rho)$ for all $n$.  The operational nature of $(\ref{hellooperational})$ makes it an easier target in some ways than the rate-distortion function. For instance, it is easy to show (e.g., \cite[Lemma 5]{silva1}) that \begin{align}
    \inf_{\tilde{C}_n} \sup_{p} \left [R(\tilde{C}_n, p, d, \rho) - R^*(n,p,d,\rho) \right ] \leq (J - 1)\frac{\ln n}{n} + O\left(\frac{1}{n} \right). 
    \label{qppprice}
\end{align}
With respect to (w.r.t.) this operational rate redundancy, minimax results in more advanced settings have been shown. Silva and Piantanida \cite{silva1} have given convergence rates for 
\begin{align*}
    \inf_{\tilde{C}_n} \sup_{p \in \mathcal{P}^\infty} \left [R(\tilde{C}_n, p, d, \rho) - R^*(n,p,d,\rho) \right ],
\end{align*}
where the supremum is over memoryless sources over countably infinite alphabets whose probability mass functions are dominated by summable envelope functions and where the exact rate of convergence depends on the envelope function.

In a different setting called the generalized universal distortion framework, \cite{mahmood2021lossy} has given the following minimax rate of convergence, 
\begin{align}
     \inf_{\tilde{C}_n} \sup_{p, \rho, d} \left [R(\tilde{C}_n, p, d, \rho) - R^*(n,p,d,\rho) \right ] = \left( J^2 K^2 + J -2\right) \frac{\ln n}{n} + O\left(\frac{1}{n} \right), \label{generalizeduniv}  
\end{align}
where the supremum is over memoryless sources over a finite alphabet, all (unbounded)
distortion measures $\rho$ and 
all distortion levels $d > 0$. The universal distortion framework was comprehensively introduced in \cite{mahmood2021lossy} and is a more general setting in which the distortion measure $\rho$ is not available at design time and is available only at runtime and available only to the encoder as an input. This introduces another dimension of universality of the prefix $d$-semifaithful code, namely one over the space of distortion measures, on top of its universality w.r.t. $p$. The practical applications of a universal distortion code are described in detail in \cite{mahmood2021lossy}; briefly stated, it allows for a flexible compression system which can meet the discordant notions of distortions for different users and it also has use in nonlinear transform coding. In a recent paper, under certain technical assumptions, Merhav \cite{https://doi.org/10.48550/arxiv.2203.03305} proved the existence of a universal distortion, prefix, $d$-semifaithful code for i.i.d. sources whose average rate for each source sequence and input distortion measure converges in a pointwise sense to the empirical rate-distortion function $R(t,d,\rho)$, where $t$ is the empirical distribution or the type of the source sequence. Furthermore, under some regularity conditions, \cite[Theorem 3]{mahmood2021lossy} proved the existence of a universal distortion, prefix $d$-semifaithful code whose expected rate converges to the rate-distortion function in a pointwise sense: 
\begin{align}
    \inf_{\tilde{C}_n} \left[ R(\tilde{C}_n, p, d, \rho) - R(p,d,\rho) \right ] \leq \left (\frac{K + 2}{2}\right ) \frac{\ln n}{n} + o\left(\frac{\ln n}{n} \right), \label{yukomaruuu33} 
\end{align}
where pointwise means for every source $p$ and input distortion measure $\rho$. Note that $(\ref{yukomaruuu33})$ is a strengthening of the traditional weakly universal result in $(\ref{chodogdskj})$ in the sense that it includes universality over distortion measures; both are weakly universal results, however. 

In this paper, we obtain strongly universal (or minimax) $d$-semifaithful codes in the universal distortion setting whose expected rate converges uniformly to the rate-distortion function, i.e., \begin{align}
    \lim_{n \to \infty}\, \inf_{\tilde{C}_n} \, \sup_{p, \rho} \left[ R(\tilde{C}_n, p, d, \rho) - R(p,d,\rho) \right ] = 0, \label{non_trad_univ}
\end{align}
where the infimum is over prefix, $d$-semifaithful codes in the universal distortion setting. We consider strong universality in the absence of any regularity conditions on the source $p$ or distortion measure $\rho$, except the assumption that the distortion measures are uniformly bounded by some constant. Note that the guarantee in $(\ref{non_trad_univ})$ is stronger than that obtained by showing that the redundancy in $(\ref{trad_univ})$ tends to zero. For the quantity in $(\ref{trad_univ})$, we give an achievability result (Corollary \ref{corr:prefix2}) with an explicit decay rate of $O(\ln^{3/2} (n) / \sqrt{n})$. We also establish a converse result (Corollary \ref{corollary:redundancylower}) which says that the worst-case redundancy of the best $d$-semifaithful code, even in the non-universal setting, cannot be better than $\Omega(1/\sqrt{n})$. 
%This converse result 
%should be compared with \cite[Theorem 5]{yang1}, which gave an achievable convergence rate of $O(\ln n / n)$ for non-universal $d$-semifaithful codes under certain regularity conditions\footnote{Specifically, the regularity conditions included full-support $p$, $R(p,d,\rho) > 0$ and the assumption of a unique, full-support optimal output distribution of the rate-distortion function.}. 
%In contrast, our results do not rely on any regularity conditions or constraints on the source $p$ or distortion measure $\rho$, except that the distortion measures are uniformly bounded by some constant.   

The rate redundancy in $(\ref{non_trad_univ})$ is evidently upper bounded by the sum of two limits,
\begin{align}
\begin{split}
    &\limsup_{n \rightarrow \infty} \inf_{\tilde{C}_n} \sup_{p, \rho} \Big[ R(\tilde{C}_n,p,d,\rho) - R^*(n,p, d,\rho) \Big] \\
    & + \limsup_{n \rightarrow \infty} \sup_{p, \rho} \left[ R^*(n,p,d, \rho) - R(p,d,\rho)\right],
    \end{split}
    \label{splitintotwo}
\end{align}
both nonnegative, the first of which one might call the price of 
universality \cite{yang2}, \cite{508836aa}. The price of universality is zero and the rate of convergence for the first term is $O(\ln n / n)$, which follows from $(\ref{generalizeduniv})$. Indeed, 
the encoder can communicate the type $t$ of the source sequence and the equivalence class\footnote{Although there is a continuum of distortion measures, for a given distortion level, they can be divided into a polynomial number of equivalence classes so that within an equivalence class, all distortion measures agree on which sequences satisfy the distortion constraint. 
See \cite[Proposition 1]{mahmood2021lossy}.} of the distortion measure to the decoder,
and then employ an optimal $d$-semifaithful code w.r.t. a suitable representative distortion measure from the equivalence class for sources that are uniformly distributed over the 
type class $t$. Thus, if the goal is to establish (\ref{non_trad_univ}), one need only show that
the second term in $(\ref{splitintotwo})$ vanishes, namely that the worst-case redundancy of the optimal prefix, $d$-semifaithful code in a non-universal setup tends to zero. 

%However, $R^*(n,p,d,\rho)$ is a difficult object to characterize and upper bounding the second term in $(\ref{splitintotwo})$ via a standalone non-universal achievability result will not reveal much insight into the universal code design problem. In particular, it will not be clear if conventional random coding approaches are sufficient to achieve strong universality over both $p$ and $\rho$.  

Following precedent~\cite{yu1,yang3}, we shall 
adopt a more convenient decomposition which upper bounds the rate redundancy in $(\ref{non_trad_univ})$ as  
\begin{align}
\begin{split}
    &\limsup_{n \rightarrow \infty} \inf_{\tilde{C}_n} \sup_{p, \rho} \Big[ R(\tilde{C}_n,p, d, \rho) - \mathbb{E}[R(T,d,\rho)]\Big] \\
    & + \limsup_{n \rightarrow \infty} \sup_{p, \rho} \left[ \mathbb{E}[R(T,d,\rho)] - R(p,d,\rho)\right],
    \end{split}
    \label{ultimate_decom}
\end{align}
where $T$ is the $n$-type of the source sequence generated i.i.d. according to $p$. Such a decomposition naturally arises in universal source coding where, in the absence of the knowledge of the underlying source $p$, the type of the source sequence is used as a proxy for $p$ and convergence to the asymptotic limit associated with $T$ is achieved. Indeed, we show that unless the difference between the expected rate of a code and the expected rate-distortion function $\mathbb{E}[R(T,d,\rho)]$ tends to zero \textit{uniformly} over both $p$ and $\rho$ as $n$ tends to infinity, it is not possible to have uniform convergence to the rate-distortion function. Regarding the first term in $(\ref{ultimate_decom})$, we show (Theorems \ref{upper_bound}-\ref{upper_bound4}) that there exists a sequence of codes $\tilde{C}_n$ satisfying
\begin{align}
    \label{eq:intro:redun1}
    \lim_{n \rightarrow \infty} \sup_{p, \rho} \Big[ R(\tilde{C}_n,p, d, \rho) - \mathbb{E}[R(T,d,\rho)]\Big] \cdot n^{\alpha} = 0 \quad \text{if} \  
        \alpha < 5/8. 
\end{align}
Furthermore, it follows from our results (specifically Theorems \ref{upper_bound}-\ref{upper_bound4} and Lemmas \ref{lemmagolden} and \ref{lemmagolden2}) that 
\begin{align}
   \lim_{n \rightarrow \infty} \sup_{p, \rho} \big | R^*(n, p, d, \rho) - \mathbb{E}[R(T,d,\rho)] \big | \cdot n^{\alpha} = 0 \quad \text{if} \ 
        \alpha < 5/8. \label{hellogenius}
\end{align}
Since the worst-case convergence to the rate-distortion function cannot be any faster than $\Omega(1/\sqrt{n})$, as noted above, it follows that, at least retrospectively, using the decomposition in $(\ref{ultimate_decom})$ instead of $(\ref{splitintotwo})$ does not entail any loss in the order of convergence.

The second term in $(\ref{ultimate_decom})$ poses a challenge since the rate-distortion function $R(p,d,\rho)$ is not well-behaved as a function of $p$, e.g., it is not necessarily concave in $p$ or differentiable w.r.t. $p$. This makes the analysis more challenging than in the lossless case, where the entropy function $H(p)$ is concave in $p$, which enables a simple upper bound of $\mathbb{E}[H(T)] \leq H(p)$. This also partially explains why results in universal lossy coding are less well-developed, frequently relying on various regularity conditions to obtain pointwise convergence\footnote{See \cite[Lemma 5]{mahmood2021lossy} which extracts from \cite{yang1} a pointwise $o(\ln n / n)$ convergence of $\mathbb{E}[R(T,d,\rho)]$ to $R(p,d,\rho)$. } of $\mathbb{E}[R(T,d,\rho)]$ to $R(p,d,\rho)$. Nevertheless, we show (Lemma \ref{modcont}) that  
\begin{align}
    \lim_{n \rightarrow \infty} \sup_{p,\rho} \Big |\,\mathbb{E}[R(T,d,\rho)] - R(p,d,\rho)\,\Big | = 0,
\end{align}
where the above result relies on a type concentration result (Lemma \ref{lemmatypes}) and uniform continuity of the rate-distortion function w.r.t. $p$ and $\rho$ (Lemma \ref{lemmaunifcont}). We
thus conclude (Corollary \ref{corr:prefix}) that codes that approach the rate-distortion
function uniformly with respect to both the source and the 
distortion measure exist, i.e., the result in $(\ref{non_trad_univ})$. 

Note that this result does not provide an explicit bound on the speed of convergence. 
However, using a result of Palaiyanur and Sahai~\cite[Lemma 2]{hari_unifcont}, we show (Lemma \ref{lemmataylor}) that 
\begin{align}
    \lim_{n \rightarrow \infty} \sup_p \Big | \mathbb{E}[R(T,d,\rho)] - R(p,d,\rho) \Big | \cdot n^{\alpha} = 0 \quad \text{if} \ \alpha < 1/2. \label{beathimtolife}
\end{align}
Thus we have proven (Corollary \ref{corr:prefix2}) the existence of strongly universal, prefix, $d$-semifaithful codes in the traditional universal setting with
minimax redundancy at most (essentially) $1/\sqrt{n}$:
\begin{align}
    \lim_{n \rightarrow \infty} \inf_{\tilde{C}_n} \sup_{p} \left [  R(\tilde{C}_n,p, d, \rho) - R(p,d,\rho) \right] \cdot n^{\alpha} =
    0 \quad \text{if} \ \alpha < 1/2. \label{beathimtodeath}
\end{align}
where the rate is controlled by the speed of convergence of the code-independent
quantity $\mathbb{E}[R(T,d,\rho)]$ to the rate-distortion function, because the convergence of the expected rate to $\mathbb{E}[R(T,d,\rho)]$ from $(\ref{eq:intro:redun1})$ is faster. Lastly, we show (Lemma \ref{lemma:expectedlower} and Corollary \ref{corollary:redundancylower}) that the $1/\sqrt{n}$
bound in both $(\ref{beathimtolife})$ and $(\ref{beathimtodeath})$ is tight. Specifically, $(\ref{beathimtolife})$ and $(\ref{beathimtodeath})$ can be strengthened to, for at least some $\rho$ and $d$,
\begin{align}
    \lim_{n \rightarrow \infty} \sup_p \Big | \mathbb{E}[R(T,d,\rho)] - R(p,d,\rho) \Big | \cdot n^{\alpha} = 0 \quad \iff \ \alpha < 1/2 \label{beathimtolife2}
\end{align}
and 
\begin{align}
    \lim_{n \rightarrow \infty} \inf_{\tilde{C}_n} \sup_{p} \left [  R(\tilde{C}_n,p, d, \rho) - R(p,d,\rho) \right] \cdot n^{\alpha} =
    0 \quad \iff \ \alpha < 1/2, \label{beathimtodeath2}
\end{align}
respectively. 

The optimal rate of convergence of $\tilde{O}(1/\sqrt{n})$ stands 
in stark contrast to the $O(\ln n / n)$ optimal convergence rate 
in prior work on universal compression noted above. The 
$\tilde{O}(1/\sqrt{n})$ rate is controlled by the 
worst-case convergence rate of $\mathbb{E}[R(T,d,\rho)]$ 
to $R(p,d,\rho)$ in $(\ref{beathimtolife2})$.
Indeed, $R(T,d,\rho)$ has a $1/\sqrt{n}$ spread around 
$R(p,d,\rho)$ from central limit theorem-type arguments.
In typical cases, the positive and negative deviations
tend to cancel, leading to a $O(\log n / n)$ redundancy.
If $R(p,d,\rho)$ is zero or nearly zero, however, then
$R(T,d,\rho)$ has deviations in the positive direction
only, which explains the $1/\sqrt{n}$ redundancy.
Note that this effect does not arise in the lossless
case because when $H(p) = 0$ we have $H(T) = 0$ almost surely.
In the lossy context, 
prior work on $d$-semifaithful coding, both non-universal~\cite{yang1}
and weakly universal~\cite{yu1,yang2,yang3,kontoyiannis2,mahmood2021lossy}, impose regularity conditions
on the source and distortion measure that have the effect
of excluding this phenomenon. We show that in the absence of these 
regularity conditions, the optimal redundancy is
$\tilde{O}(1/\sqrt{n})$, even in the non-universal
case (Corollary~\ref{corollary:redundancylower}). However, the $1/\sqrt{n}$
behavior does not come about solely from relaxing the
regularity conditions from previous works.
In Appendix \ref{newresult447}, we assume the regularity conditions
in \cite{yang1,yang2,yang3}, where \cite{yang3} in particular shows a pointwise rate redundancy of $O(\ln n / n)$ for weakly universal codes. We prove 
that under these conditions, imposing strong universality worsens the redundancy from $O(\ln n / n)$ to $\Omega(1/\sqrt{n})$. 
Thus the $O(\ln n / n)$ redundancy finding is sensitive to
both the regularity conditions and the weak universality assumption.
Table \ref{comparison} summarizes these results. 
\begin{table}[h]
\begin{minipage}{\textwidth} 
\begin{center}
\setcellgapes{3pt}
\makegapedcells
\begin{tabular}{|c|c|c|c|}
\cline{2-4}
    \multicolumn{1}{c|}{} & \multirow{2.5}{*}{Lossless} & \multicolumn{2}{c|}{Lossy ($d$-semifaithful)}\\
    \cline{3-4} 
    \multicolumn{1}{c|}{} & \multicolumn{1}{c|}{} & with regularity conditions & without regularity conditions
    %\footnote{
    %This column compares convergence speeds across different notions of universality under the same regularity conditions; see Appendix \ref{newresult447}.}
    \\
    \hline
Non-universal & $O\left(1/n \right)$ \cite[Thm. 5.4.2]{thomas_cov} & $O( \ln n / n)$ \cite{yang1} &  $\tilde{O}(1/\sqrt{n})$\\
\hline
Weakly universal & $O\left( \ln n/n \right)$ \cite[Thm. 1a]{rissanen1} & $O( \ln n/n)$ \cite{yang3} & $\tilde{O}(1/\sqrt{n})$ \\
\hline 
Strongly universal & $O\left( \ln n/n \right)$ \cite[(61)]{1056355} & $\tilde{O}(1/\sqrt{n})$ & $\tilde{O}(1/\sqrt{n})$\\
\hline
\end{tabular}
\end{center}
\caption{Comparison of the optimal rate redundancy between prefix lossless codes, prefix $d$-semifaithful codes for sources satisfying certain regularity conditions (Appendix~\ref{newresult447}), and prefix $d$-semifaithful codes for arbitrary source distributions, in the classical setting of fixed distortion measure. Results without a citation are from this paper. }
\label{comparison}
\end{minipage}
\end{table}

We prove analogous results to the above for non-prefix codes, following the lossless coding literature. There are different results for prefix and non-prefix universal lossless codes, both in terms of the optimal rate redundancy and the coding scheme used; the prefix constraint leads to a higher optimal rate redundancy, see e.g., \cite[Table 1]{kosut1}. Even though the dominant term in the redundancy bounds we obtain is the same in both prefix and non-prefix codes, the higher-order terms differ in qualitatively the same way as they do with lossless codes.

Table \ref{prevworks} compares the achievability results of this paper with some of the previous work on $d$-semifaithful codes, focusing only on the expected rate analysis.

\begin{table}[h]
\begin{minipage}{\textwidth} 
\begin{center}
\setcellgapes{2pt}
\makegapedcells
\begin{tabular}{|c|c|c|c|c|c|}
\hline
    Paper & Universality w.r.t.\ & Guarantee &
        Redundancy w.r.t.\ & Convergence &  Convergence Rate \\
        \hline
    Yu and Speed \cite{yu1} & $p$ & for most $p$ & $R(p,d,\rho)$ & in expectation
       & $O(\log n/n)$ \\
    Yang, Zhang, and Berger \cite{623145f} & $p$ & for all $p$  & $R(p,d,\rho)$ & almost surely & - \\
    Yang and Zhang \cite{yang3} & $p$ & for most $p$ & $R(p,d,\rho)$ & in expectation & $O(\log n/n)$ \\
    Kontoyiannis \cite{kontoyiannis1} & $p$ & for all $p$ & $R(p,d,\rho)$ & almost surely & - \\
    Kontoyiannis and Zhang \cite{kontoyiannis2} & $p$ & for most $p$ & other & both\footnote{Results on convergence in expectation as well as almost surely are given.} & $O(\log n/n)$ \\
    Silva and Piantanida \cite{silva1}  & $p$ & minimax & $R^*(n,p,d,\rho)$ & in expectation & $O(\log n/n)$ \\
    Mahmood and Wagner \cite{mahmood2021lossy} & $p$, $\rho$, and $d$ & minimax & $R^*(n,p,d,\rho)$ & in expectation & $O(\log n/n)$ \\
    Mahmood and Wagner \cite{mahmood2021lossy} & $p$ and $\rho$ & for most $p$ & $R(p,d,\rho)$ & in expectation & $O(\log n/n)$ \\
    \hline
    This paper & $p$ and $\rho$ & minimax &  $\mathbb{E} \left [ R(T,d,\rho)  \right ]$ & in expectation &
       $O(\frac{\log n}{n^{5/8}})$ \\
    This paper & $p$ & minimax & $R(p,d,\rho)$ & in expectation & 
        $\,\,O(\frac{\log^{3/2} n}{n^{1/2}})$ \\
    This paper & $p$ and $\rho$ & minimax & $R(p,d,\rho)$ & in expectation & - \\
    \hline
    Merhav \cite{https://doi.org/10.48550/arxiv.2203.03305} & $t$ and $\rho$ & for most $t$ &  $ R(t,d,\rho) $ & almost surely & $O(\log n/n)$\\
    \hline
    %TO ADD:
    %  Ziv
    %  Ornstein and Shields: or earlier
    %  E.-h. Yang, “Universal almost sure data compression for abstract alphabets and arbitrary fidelity criterions,” Probl. Contr. Inform. Theory, vol.  20, no. 6, pp. 397–408, 1991.
\end{tabular}
\end{center}
\caption{\textbf{Universal $d$-Semifaithful Achievability Results}
  (particularized to expected rate
      for i.i.d.\ sources with finite alphabets)}
\label{prevworks}
\end{minipage}
\end{table}

Most existing works on universal lossy compression rely on random code constructions
that are analyzed using type-theoretic tools. This analysis can
be quite involved, and it requires various technical conditions.
In contrast, we use a random code construction which relies on acceptance-rejection
sampling and exact asymptotics from large deviations in place of
type-theoretic methods.  The codebook is generated from
a specific mixture distribution called the normalized maximum-likelihood (NML) distribution (given in $(\ref{NMLdist})$). Such mixture distributions have precedent in the context of universal rate-distortion in the work of Kontoyiannis and
Zhang~\cite{kontoyiannis2}. 
Our approach obviates the need for the
technical conditions alluded to earlier. It also has the added advantage that it can readily accommodate universality over the distortion 
measure, albeit with the modification discussed next.    

If one
wishes to achieve universality with respect to the distortion measure,
then the usual random coding approach is insufficient in the following 
sense.  For any given source distribution $p$ and distortion measure
$\rho$, it is well-known
that there exists a distribution over the reconstruction
alphabet, $Q^{p,d,\rho}$, such that if $X^n$ is i.i.d.\ $p$ and $Y^n$
is i.i.d.\ $Q^{p,d,\rho}$, with $X^n$ and $Y^n$ independent, then
\begin{align}
    \label{eq:cuckoo}
    \lim_{n \rightarrow \infty} -\frac{1}{n} \ln
       \mathbb{P}(\rho(X^n,Y^n) \le d) = R(p,d,\rho)
\end{align}
if the rate is measured in nats.
Indeed, several achievability schemes \cite{yang1}, \cite{yang2},  \cite{mahmood2021lossy}, \cite{https://doi.org/10.48550/arxiv.2203.03305} based on random coding rely on lower bounding the probability that a random codeword meets the distortion constraint with a given source sequence. We show (Proposition~\ref{prop:counterexample}) that such an argument cannot provide uniform convergence over all source sequences $x^n$ and all distortion measures $\rho$ because
\begin{align}
    \inf_{x^n,\rho} \mathbb{P}(\rho(x^n,Y^n) \le d) = 0, \label{impossibilityaaron}
\end{align}
where $t$ is the type of $x^n$ and $Y^n$ is i.i.d. $Q^{t,d,\rho}$.

We mitigate this
shortcoming by providing leeway in the allowed distortion  which results in a nonvanishing lower bound (Lemma \ref{lemlowerbnd55} and $(\ref{CODMsucks})$):
\begin{align}
    \inf_{x^n,\rho} \frac{\mathbb{P}\left (\rho(x^n,Y^n) \le d + \frac{2 \rho_{\max}}{n^{5/8}}  \right )}{e^{  -n R(t,d,\rho) } } \geq   \exp \left(- \Omega\left (n^{3/8}\right) \right). \label{post_correction09}
\end{align}
The resulting code is not $d$-semifaithful, however. To make it $d$-semifaithful, we employ a
post-correction scheme that uses uncoded (or uncompressed) transmission from the encoder to the decoder to replace suitable symbols in the reconstruction sequence so that the post-corrected sequence meets the distortion constraint. The word "uncoded" here means that there is no compression and the number of post-correction bits sent is essentially equal to the log of the alphabet size times the number of replacement symbols (see $(\ref{opwxz})$ in the proof of Theorem \ref{upper_bound}).  
This use of uncoded transmission is reminiscent of schemes for 
achieving the rate-distortion function at very low rates \cite{7542580}. 
Uncoded transmission
is also employed in the recent work of the authors, mentioned above, for 
showing that the price of universality over unknown distortion measures
is zero \cite[Theorem 2]{mahmood2021lossy}.  Prior studies considered
uncoded transmission due to its simplicity, not because it outperforms 
other schemes. It has also been considered in the context of joint 
source-channel coding~\cite{1197846}, \cite{1624626}, where it can
outperform other schemes.  Its use in 
achieving universality appears to be unique to this paper 
and~\cite{mahmood2021lossy}.

The remainder of the paper is organized as follows. Section \ref{prelimsss} establishes the notation, definitions and basic properties of various objects related to lossy compression. Section \ref{mainresults} lists and discusses the main results of this paper. Section \ref{lagrangeformulation} states the known results about the Lagrange formulation of the rate-distortion function. Section \ref{mimplemmas} develops the $d$-covering lemmas whose proofs are given in appendices \ref{pandakakuta} and \ref{pandakakuta2}. Sections \ref{thms12}$-$\ref{thm4proof} are devoted to the proofs of the main theorems.    

\section{Preliminaries \label{prelimsss}}

Without loss of generality, we let $A = \{1, 2, ..., J \}$ and $B = \{1,2, ..., K \}$. $\mathcal{P}(A)$ denotes the set of all probability distributions on $A$. $\mathcal{P}(A|B)$ denotes the set of all conditional distributions. In this paper, $\,\ln\,$ represents log to the base $e$, $\,\log\,$ represents log to the base $2$ and $\exp(x)$ is equal to $e$ to the power of $x$. Unless otherwise stated, all information theoretic quantities will be measured in nats. For $p \in \mathcal{P}(A)$, $H(p)$ denotes the Shannon entropy. For $p \in \mathcal{P}(A)$ and $W \in \mathcal{P}(B|A)$, $H(W|p)$ denotes the conditional entropy and $I(p, W) = I(X;Y)$ denotes the mutual information where $(X,Y)$ have the joint distribution given by $p \times W$. For $p_1 \in \mathcal{P}(A)$ and $p_2 \in \mathcal{P}(A)$,
$D(p_1 || p_2)$ denotes the relative entropy between the two probability distributions. For any vector $v \in \mathbb{R}^m$, $||v||_1$ will denote the $l^1$ norm of $v$. For any two $m$-dimensional vectors $u = (u_1, ..., u_m)$ and $v = (v_1, ..., v_m)$, $||v - u||_2$ will denote the Euclidean distance between $u$ and $v$. We use $\Phi(\cdot)$ to denote the standard normal CDF.

For a given sequence $x^n \in A^n$, the $n$-type $t = t(x^n)$ of $x^n$ is defined as
\begin{align*}
    t(j) &= \frac{1}{n} \sum_{i=1}^n \mathds{1}(x_i = j)
\end{align*}
for all $j \in A$, where $\mathds{1}(.)$ is the standard indicator function. $\mathcal{P}_n(A)$ denotes the set of all $n$-types on $A$. For a pair of sequences $x^n \in A^n$ and $y^n \in B^n$, the joint $n$-type $s$ is defined as 
\begin{align*}
    s(j, k) &= \frac{1}{n} \sum_{i=1}^n \mathds{1} \left( x_i = j, y_i = k \right)
\end{align*}
for all $j \in A$ and $k \in B$. $\mathcal{P}_n(A \times B)$ denotes the set of all joint $n$-types on $A \times B$. For two sequences $x^n$ and $y^n$ with $n$-types $t_x = t(x^n)$ and $t_y = t(y^n)$, the joint $n$-type $s $ can also be written as
\begin{align*}
    s(j, k) &= t_x(j) W_y(k|j) = t_y(k) W_x(j|k),
\end{align*}
where $W_y$ is called a conditional type of $y^n$ given $x^n$, and $W_x$ is called a conditional type of $x^n$ given $y^n$. From \cite[Lemma 2.2]{korner1}, we have 
\begin{align}
\begin{split}
    |\mathcal{P}_n(A)| &\leq (n+1)^{J-1}\\
    |\mathcal{P}_n(A \times B)| &\leq (n+1)^{JK - 1}.
\end{split}
\label{ghba}
\end{align}
For a given type $t \in \mathcal{P}_n(A)$, $T_A^n(t)$ is called the type class where
\begin{align}
    T_A^n(t) &= \{x^n \in A^n: t(x^n) = t  \}. \notag 
\end{align}
For any given $P \in \mathcal{P}(A)$ or $P \in \mathcal{P}(B)$, $P^n$ will denote the $n$-fold product distribution induced by $P$. Let $X^n$ be an independent and identically distributed source. Let $p \in \mathcal{P}(A)$ be the generic probability distribution of the source so that $X^n$ is distributed according to $p^n$. The probability that $X^n$ is of type $t$ satisfies \cite[Lemma 2.6]{korner1}
\begin{align}
    \mathbb{P}_p \left( X^n \in T^n_A(t) \right) &= p^n \left( T^n_A(t)\right) \leq \exp \left(  - n D(t||p) \right). \label{opqq}
\end{align}

For a given source distribution $p$, it suffices to focus only on sequence types $t$ satisfying $||t - p||_2 \leq a \sqrt{\ln n/n}$, where $a \geq \sqrt{2 + 2J}$. Source sequence types sufficiently away from source distribution $p$ have negligible probability for large $n$ as quantified by the following lemma (\cite[Lemma 1]{mahmood2021lossy}):
\begin{lemma}
If $a$ satisfies $a \geq \sqrt{2 + 2J}$, then for all $p \in \mathcal{P}(A)$ and all $n \in \mathbb{N}$ , we have 
\begin{align*}
    \sum_{t:||t - p||_2 > a \sqrt{\ln n / n}} p^n(T^n_A(t)) \leq \frac{e^{J-1}}{n^2}.
\end{align*}
\label{lemmatypes}
\end{lemma}

Let $\rho : A \times B \to [0, \rho_{\max} ]$ be a single letter distortion measure and $\rho_n(x^n, y^n)$ be its $n$-fold extension defined as
\begin{align}
    \rho_n(x^n, y^n) = \frac{1}{n} \sum_{i = 1}^n \rho(x_i, y_i),     \label{dist_n_fold}
\end{align}
where $x^n \in A^n$, $y^n \in B^n$. 

Let $\mathcal{D}$ be the space of uniformly bounded distortion measures, i.e., fix some $\rho_{\max} > 0$ and let $\mathcal{D}$ denote those $\rho$ such that $0 \leq \rho(j, k) \leq \rho_{\max}$ for all $j \in A$, $k \in B$. All distortion measures considered in this paper will be in $\mathcal{D}$ and $\rho_{\max}$ will denote the uniform bound on all $\rho \in \mathcal{D}$. Furthermore, we will assume that
\begin{align}
    \max_{j \in A} \min_{k \in B} \rho(j,k) = 0 \,\,\,\,\,\,\,\,\,\,\,\,\,\,\,\,\,\, \text{for all } \rho \in \mathcal{D}. \label{dist_assump}  
\end{align}
When the source distribution and the distortion measure are fixed, (\ref{dist_assump}) is without loss of generality~\cite[p.~26]{berger1}. Here, it is tantamount to having $d$ represent the allowable excess expected distortion above the minimum possible for the given source distribution and distortion measure. For the universal setup, this is preferable to having $d$ represent a constraint on the absolute expected distortion: a given $d$ will be below the minimum achievable expected distortion for some cases, for instance.

For a given $\rho \in \mathcal{D}$, $p \in \mathcal{P}(A)$ and $d > 0$, the rate-distortion function $R(p, d, \rho)$ is defined as \cite[Theorem 10.2.1]{thomas_cov}
\begin{align}
    &R(p,d,\rho) \notag \\
    &= \min_{Q_{B|A} \in \mathcal{Q}_{d,\rho}} I(p, Q_{B|A}) \label{rdfunc} \\
    &= \min_{Q_{B|A} \in \mathcal{Q}_{d,\rho}} \sum_{j, k} p(j) Q_{B|A}(k|j) \ln \left(\frac{Q_{B|A}(k|j)}{Q(k)} \right), \notag \\
    &\text{where } Q(k) = \sum_{j \in A} p(j) Q_{B|A}(k|j) \text{ and }  \label{rdfunc2} \\
    &\mathcal{Q}_{d,\rho} = \left \{ Q_{B|A} : \sum_{j, k} p(j) Q_{B|A}(k|j) \rho(j,k) \leq d \right \}. \label{rdfunc3}
\end{align}
For any given $p$ and $\rho$, $R(p,d,\rho)$ is nonincreasing, convex and differentiable everywhere as a function of $d$ except possibly at $d = \min_{k \in B} \sum_{j \in A} p(j) \rho(j, k)$ \cite[Exercise 8.6]{korner1}, \cite[Lemma 10.4.1]{thomas_cov}. In particular, for $0 < d < \min_{k \in B} \sum_{j \in A} p(j) \rho(j, k)$, $R(p,d,\rho)$ is strictly decreasing in $d$.
The function's dependence on $p$ for given $d$ and $\rho$ is complex \cite{Ahl1}.
In particular, it is not concave in general.

For the given $p, d$ and $\rho$, if $Q^*_{B|A}$ solves $(\ref{rdfunc})$$-$$(\ref{rdfunc3})$, then $Q^{p,d,\rho}$ defined as
\begin{align}
    Q^{p,d,\rho}(k) = \sum_{j \in A} p(j) Q^*_{B|A}(k|j) \label{vqye}
\end{align}
will be called an optimal (output) distribution on $B$ associated with $p, d$ and $\rho$. The optimal transition probability matrix $Q^*_{B|A}$ or the optimal output distribution $Q^{p,d,\rho}$ may not be unique\footnote{Lemma 7 in \cite{yang2} gives sufficient conditions on the distortion measure under which $Q^{p,d,\rho}$ is unique for all full support distributions $p$ and for $K \leq J$. We will not assume these conditions in this paper.} for a given $(p,d,\rho)$. 
\begin{lemma}
For a fixed distortion level $d > 0$, the rate-distortion function $R(p,d,\rho)$ is uniformly continuous on $\mathcal{P}(A) \times \mathcal{D}$. \label{lemmaunifcont}
\end{lemma}
\textit{Proof:} The proof of Lemma \ref{lemmaunifcont} is given in Appendix \ref{facebookdafarhai}. 
\begin{remark}
Throughout this paper, we will adopt the following metric on $\mathcal{P}(A) \times \mathcal{D}$:
\begin{align}
    &||(p_1, \rho_1) - (p_2, \rho_2)|| \notag\\
    &= \sqrt{\sum_{j \in A} (p_1(j) - p_2(j))^2 + \sum_{j \in A, k \in B} (\rho_1(j, k) - \rho_2(j,k))^2}
    \label{newmetric}
\end{align}
for any $(p_1,\rho_1)$ and $(p_2, \rho_2)$. Uniform continuity in Lemma \ref{lemmaunifcont} can be thought of with respect to this given metric.  
\end{remark}

Previous works on lossy coding \cite{linder1}, \cite{yu1}, \cite{yang1}$-$\cite{yang2} have primarily considered two kinds of block codes:
\begin{itemize}
    \item fixed rate codes 
    \item $d$-semifaithful codes
\end{itemize}
As mentioned before, we will focus on the latter. An $n$th order $d$-semifaithful block code is defined by a triplet $C_n = (\phi_n, f_n, g_n)$:
\begin{align}
\begin{split}
    \phi_n & : A^n \to B_{\phi_n} \subset B^n\\
    f_n & : B_{\phi_n} \to \mathcal{B}^*\\
    g_n & : \mathcal{B}^* \to B_{\phi_n},
\end{split} \label{dsemi}
\end{align}
where 
\begin{itemize}
    \item $\mathcal{B}^*$ is a set of binary strings,
    \item $(f_n, g_n)$ is a binary encoder and decoder pair,
    \item $B_{\phi_n}$ is the \text{codebook}, and
    \item $\phi_n$ is a $d$-quantizer, i.e., for all $x^n \in A^n$, we have $\rho_n(x^n, \phi_n(x^n)) \leq d.$
\end{itemize}
We further define the code $C_n$ to be a random code if any one of the functions $\phi_n, f_n$ or $g_n$ is random. 
When considering random codes, we assume that infinite common randomness is
available between the encoder and the decoder.

The performance of a $d$-semifaithful code $C_n$ can be measured by the rate redundancy $\mathcal{R}_n(C_n,p,\rho)$ defined as
\begin{align}
  \mathcal{R}_n(C_n,p,\rho) \triangleq \frac{1}{n} \mathbb{E} \left [l\left( f_n\left(\phi_n(X^n)\right) \right) \ln 2 \right ] - R(p,d,\rho), 
  \label{rate_redunda}
\end{align}
where $\mathbb{E}\left [ l(f_n(\phi_n(X^n))) \right ] $ is the expected length of the binary string $f_n(\phi_n(X^n))$, the expectation being with respect to the product distribution $p^n$ (as well as $C_n$ if the code is itself random) and the factor of $\ln 2$ is because we measure coding rate in nats. 

In the universal distortion framework studied in \cite{mahmood2021lossy}, the modified formulation of a $d$-semifaithful block code $\tilde{C}_n = (\phi_n, f_n, g_n)$ is given by 
\begin{align}
\begin{split}
    \phi_n & : A^n \times \mathcal{D} \to B_{\phi_n} \subset B^n\\
    f_n & : B_{\phi_n}  \to \mathcal{B}^*\\
    g_n & : \mathcal{B}^* \to B_{\phi_n},
\end{split} \label{dsemi_unknown}
\end{align}
where $\mathcal{D}$ is the space of uniformly bounded distortion measures defined earlier. Thus the distortion measure is not known in advance and only revealed to the $d$-quantizer at run-time. Henceforth, we will use $C_n$ to denote a code in the traditional setting as in $(\ref{dsemi})$ and $\Tilde{C}_n$ to denote a code in the universal distortion setting as in $(\ref{dsemi_unknown})$.  The rate redundancy in the universal distortion setting is given by 
\begin{align*}
  \mathcal{R}_n(\tilde{C}_n,p,\rho) \triangleq \frac{1}{n} \mathbb{E} \left [l\left( f_n\left(\phi_n(X^n, \rho)\right) \right) \ln 2 \right ] - R(p,d,\rho). 
\end{align*}

Viewing the codebook $B_{\phi_n} \subset B^n$ as a set of indexed\footnote{Indexed as $1,2,3,\ldots $} codewords available to both the encoder $f_n$ and decoder $g_n$, the encoder $f_n$ can map the integer index of the codeword to a binary string followed by the decoder performing the inverse mapping to recover the codeword. A frequently used integer-to-binary encoding is based on Elias coding \cite{elias1}. If $y^n_i \in B_{\phi_n}$ is a codeword with index $i$, then with Elias coding \cite{elias1}, the length of the binary encoding $f_n(y^n_i)$ satisfies  
\begin{align}
    l(f_n(y^n_i)) &\leq \lfloor \log(i) \rfloor + 2 \lfloor \log \left( \lfloor \log(i) \rfloor + 1 \right) \rfloor + 1. \label{codeen1}
\end{align}
Another integer-to-binary encoding is the fixed-to-variable one given by 
\begin{align}
    f_n : \{y^n_1, y^n_2, y^n_3, \ldots \} \to \{&0, 1, 00, 01, 10, 11, 000, 001, \ldots \}, \label{codeen111}
\end{align}
where the length of the binary encoding $f_n(y^n_i)$ satisfies 
\begin{align}
    l(f_n(y^n_i)) &\leq 1 + \log(i). 
\label{codeen2}
\end{align}
The encoder $f_n$ is said to be a prefix code if for all $i, j \in \mathbb{Z}_{> 0}$, $f_n(y^n_i)$ is not a prefix of $f_n(y^n_j)$ so long as $y_i^n \ne y_j^n$. Otherwise, it is a non-prefix code. Elias encoding in $(\ref{codeen1})$ results in a prefix code while the fixed-to-variable encoding in $(\ref{codeen111})$ and $(\ref{codeen2})$ yields a non-prefix code. Previous works have considered $d$-semifaithful codes with a prefix encoder. In this paper, we give minimax achievability and converse results for both prefix and non-prefix encoders. Imposing the prefix constraint is rarely necessary when considering block coding and actually incurs a loss of optimality. Indeed, applying prefix constraint in universal lossless coding incurs an extra factor of $\ln n/n$ in rate redundancy when compared to non-prefix codes, see, e.g., \cite[Table I]{kosut1}. In universal lossy coding considered in this paper, we observe a similar penalty in the higher-order terms of the rate redundancy although the dominant term is unaffected. 

While the expected rate of a $d$-semifaithful code with a prefix encoder is strictly lower bounded by the rate-distortion function \cite[Secs. 5.4 and 10.4]{thomas_cov}, this is not necessarily true for a $d$-semifaithful code with a non-prefix encoder. However, as we will show later, the rate-distortion function is still an asymptotic lower bound in the non-prefix case. 

Let $\overline{\mathcal{P}}(A^n)$ be a set of probability distributions on $A^n$. Then Shtarkov's sum \cite{shar1} for $\overline{\mathcal{P}}(A^n)$ is defined as 
\begin{align*}
    S_n = \sum_{x^n \in A^n} \sup_{p \in \overline{\mathcal{P}}(A^n)} p(x^n).
\end{align*}
In particular, if $\overline{\mathcal{P}}(A^n)$ is the set of i.i.d. distributions, then we have 
\begin{align}
    S_n &= \sum_{x^n \in A^n} \sup_{p \in \mathcal{P}(A)} p^n(x^n). \label{shariguy}
\end{align}
Shtarkov \cite{shar1} showed the important result that $\log S_n$ is essentially (up to a discrepancy of at most $1/n$) equal to the universal lossless coding redundancy over the set of distributions $\overline{\mathcal{P}}(A^n)$. 
It is known from previous works (\cite{orlitsky1}, \cite{risannen2}, \cite{precise1}, \cite{613240}) that the universal lossless coding redundancy for i.i.d.\ sources taking values in alphabet $A$ of size $J$ is given by   
\begin{align}
    \frac{J - 1}{2} \log (n) - \frac{J - 1}{2} \log (2 \pi) + \log \left( \frac{\Gamma\left( \frac{1}{2}\right)^J}{\Gamma \left(\frac{J}{2} \right)}\right) + o_J(1),
    \label{iidredund}
\end{align}
where $\Gamma(\cdot)$ is the gamma function and $o_J(1) \to 0$ as $n \to \infty$ at the rate determined only by $J$. Combining this with Shtarkov's result and changing base to natural log, we can express $S_n$ from $(\ref{shariguy})$ as
\begin{align}
    &S_n = \sum_{x^n \in A^n} \sup_{p \in \mathcal{P}(A)} p^n(x^n) \notag\\
    &= \exp \left( \frac{J - 1}{2} \ln n + \ln \left(  \frac{\Gamma\left( \frac{1}{2}\right)^J}{(2\pi)^{\frac{J-1}{2}} \, \Gamma \left(\frac{J}{2} \right)} \right) +  o_J(1)\, \ln (2)\right).
    \label{S_n bound }
\end{align}
The above result is used in constructing random codes (Theorems \ref{upper_bound} and \ref{upper_bound2}) which use acceptance-rejection sampling using the normalized maximum-likelihood distribution $Q^{\text{NML}} \in \mathcal{P}(B^n)$, specified by
\begin{align}
   Q^{\text{NML}}(y^n) =  \frac{\sup \limits_{q \in \mathcal{P}(B)} q^n(y^n) }{\sum \limits_{z^n \in B^n} \sup \limits_{p \in \mathcal{P}(B)} p^n(z^n)}, \label{NMLdist}
\end{align}
to generate i.i.d.\ codewords from the optimal distribution $(Q^{t,d,\rho})^n$, where $Q^{t,d,\rho}$ is defined according to $(\ref{vqye})$.

\section{Main Results \label{mainresults}}

In this section, we list the main theorems of the paper. Theorems $\ref{upper_bound}$$-$$ \ref{upper_bound4}$ are minimax achievability results in the universal distortion setting which establish that the difference between the expected rate and  $\mathbb{E}_p \left [ R(T,d,\rho) \right ] $ is upper bounded by a quantity that tends to zero at a rate independent of $p$ and $\rho$. Corollaries $\ref{corr:prefix}$ and $\ref{corr:arb}$ are minimax results which establish uniform convergence of the expected rate to $ R(p,d,\rho)$ over all source distributions $p \in \mathcal{P}(A)$ and distortion measures $\rho \in \mathcal{D}$. Corollaries $\ref{corr:prefix2}$ and $\ref{corr:arb2}$ are minimax results which establish uniform convergence of the expected rate to $ R(p,d,\rho)$ over all source distributions $p \in \mathcal{P}(A)$ with an explicit convergence rate. Finally, Corollary \ref{corollary:redundancylower} establishes that the order of the minimax convergence rate of the previous two corollaries is essentially optimal, ignoring logarithmic factors; see Table \ref{myftable} for a summary of main results. These results encompass both random and deterministic coding schemes as well as both prefix and non-prefix coding schemes. 

\begin{table}[h]
\begin{minipage}{\textwidth}
\begin{center}
\setcellgapes{3pt}
\makegapedcells
\resizebox{\columnwidth}{!}{
\begin{tabular}{ 
   |c | c | c| c | c|c|c| } 
\cline{2-7}
\multicolumn{1}{c|}{} & \multicolumn{2}{c|}{Code Characterization} & \multicolumn{1}{c|}{Universality w.r.t.} &  \multicolumn{1}{c|}{Performance Metric} & \multicolumn{1}{c|}{Bound\footnote{Only the dominant terms omitting the multiplicative constants are specified.  }} & \multicolumn{1}{c|}{ Result Type}   \\ 
\hline
Theorem \ref{upper_bound} & 
 non-prefix & 
    random &  
     $p$ and $\rho$ & $
\sup \limits_{p, \rho}  \big [\, \mathbb{E}\left [\text{Rate} \right] - \mathbb{E}_p\left [R(T,d,\rho) \right] \big ]$ & $\leq  \frac{\ln n}{n^{5/8}}$ &  achievability \\ 
\hline
Theorem \ref{upper_bound2} & prefix & 
    random  &  
     $p$ and $\rho$  & $
\sup \limits_{p, \rho}  \big [\, \mathbb{E}\left [\text{Rate} \right] - \mathbb{E}_p\left [R(T,d,\rho) \right] \big ]$ & $\leq  \frac{\ln n}{n^{5/8}}$  & achievability \\ 
\hline
Theorem \ref{upper_bound3} & non-prefix &  deterministic &  
     $p$ and $\rho$ & $
\sup \limits_{p, \rho}  \big [\, \mathbb{E}\left [\text{Rate} \right] - \mathbb{E}_p\left [R(T,d,\rho) \right] \big ]$ & $\leq  \frac{\ln n}{n^{5/8}}$  & achievability \\ 
\hline
Theorem \ref{upper_bound4} & prefix & deterministic  &  
     $p$ and $\rho$ & $
\sup \limits_{p, \rho}  \big [\, \mathbb{E}\left [\text{Rate} \right] - \mathbb{E}_p\left [R(T,d,\rho) \right] \big ]$ & $\leq  \frac{\ln n}{n^{5/8}}$  & achievability \\ 
\hline
Lemma \ref{lemmagolden} & prefix &  deterministic &  
     non-universal & $
\inf \limits_{p, \rho}  \big [\, \mathbb{E}\left [\text{Rate} \right] - \mathbb{E}_p\left [R(T,d,\rho) \right] \big ]$ & $\geq -  \frac{\ln n}{n}$  & converse \\
\hline
Lemma \ref{lemmagolden2} & non-prefix &  deterministic & non-universal & $
\inf \limits_{p, \rho}  \big [\, \mathbb{E}\left [\text{Rate} \right] - \mathbb{E}_p\left [R(T,d,\rho) \right] \big ]$ & $\geq -  \frac{\ln n}{n}$  & converse \\ 
\hline
Corollary \ref{corr:prefix} & prefix &  deterministic &  $p$ and $\rho$ & $
\sup \limits_{p, \rho}  \big [\, \mathbb{E}\left [\text{Rate} \right] - R(p,d,\rho)  \big ]$ & $\leq o(1)$ & achievability \\ 
\hline
Corollary \ref{corr:arb} & non-prefix &  deterministic &  $p$ and $\rho$ & $
\sup \limits_{p, \rho}  \big [\, \mathbb{E}\left [\text{Rate} \right] - R(p,d,\rho)  \big ]$ & $\leq o(1)$ & achievability \\
\hline
Corollary \ref{corr:prefix2} & prefix &  deterministic &  $p$ & $
\sup \limits_{p}  \big [\, \mathbb{E}\left [\text{Rate} \right] - R(p,d,\rho)  \big ]$ & $\leq   \frac{\ln^{3/2}(n)}{\sqrt{n}}$ & achievability \\
\hline
Corollary \ref{corr:arb2} & non-prefix &  deterministic & $p$ & $
\sup \limits_{p}  \big [\, \mathbb{E}\left [\text{Rate} \right] - R(p,d,\rho)  \big ]$ & $\leq   \frac{\ln^{3/2}(n)}{\sqrt{n}}$ & achievability \\
\hline
Corollary \ref{corollary:redundancylower} & prefix/non-prefix & deterministic & non-universal & $
\sup \limits_{p}  \big [\, \mathbb{E}\left [\text{Rate} \right] - R(p,d,\rho)  \big ]$ & $\geq \frac{1}{\sqrt{n}} $  & converse \\
\hline
\end{tabular}}
\end{center}
\caption{Summary of main results}
\label{myftable}
\end{minipage}
\end{table}

For a given $d > 0$, $p \in \mathcal{P}(A)$ and $\rho \in \mathcal{D}$, let
\begin{align*}
    \mathbb{E}_p \left [ R(T,d,\rho) \right ] &= \sum_{t \in \mathcal{P}_n(A)} p^n(T^n_A(t)) R(t,d,\rho).
\end{align*}
Throughout the rest of the paper, $\mathbb{E}_p$ will denote expectation with respect to the source distribution $p$ as above and $\mathbb{E}_c$ will denote expectation with respect to the random code. 

\begin{theorem}
Fix $d > 0$. Then for sufficiently large $n$, there exists a universal random non-prefix $d$-semifaithful code $\tilde{C}_n = (\phi_n, f_n, g_n)$ for the universal distortion problem such that
\begin{align*}
    &\sup_{p \in \mathcal{P}(A), \rho \in \mathcal{D}} \left [ \frac{1}{n} \mathbb{E}_p \left [ \ln(2) \mathbb{E}_c\left [l (f_n(\phi_n(X^n, \rho))) \right ] \right ] - \mathbb{E}_p \left [R(T,d,\rho) \right ] \right ]   \\
    &  \leq   \frac{2 \rho_{\max} \ln (n)}{d \, n^{5/8}}   + \frac{4 \rho_{\max} \left( \min\left(\ln(K), \ln(J) \right) + \ln(2)\right)}{d \, n^{5/8}}   + \mbox{}\\
    & \,\,\,\,\,\,\,\,\,\,\,\,\,\,\,\,\,\,\,\,\,  \frac{K + 5/4}{2} \frac{\ln n}{n} +   \frac{V_1 + \ln(8)}{n} + \frac{\min\left(\ln(K), \ln(J) \right)}{n},   
\end{align*}
where 
\begin{align*}
    V_1 = \ln \left(  \frac{\Gamma\left( \frac{1}{2}\right)^K}{(2\pi)^{\frac{K-1}{2}} \, \Gamma \left(\frac{K}{2} \right)} \right) + 2\ln(2).
\end{align*}
\label{upper_bound}
\end{theorem}

\begin{theorem}
Fix $d > 0$. Then for sufficiently large $n$, there exists a universal random prefix $d$-semifaithful code $\tilde{C}_n = (\phi_n, f_n, g_n)$ for the universal distortion problem such that
\begin{align*}
    &\sup_{p \in \mathcal{P}(A), \rho \in \mathcal{D}} \left [ \frac{1}{n} \mathbb{E}_p \left [ \ln(2) \mathbb{E}_c\left [l (f_n(\phi_n(X^n, \rho))) \right ] \right ] - \mathbb{E}_p \left [R(T,d,\rho) \right ] \right ]   \\
    &  \leq   \frac{2 \rho_{\max} \ln(n)}{d \, n^{5/8}}   + \frac{4 \rho_{\max} \left(\min\left(\ln(K), \ln(J)\right) + \ln(2)\right)}{d\, n^{5/8}}    + \mbox{} \\
    & \,\,  + \frac{K + 21/4}{2}\frac{ \ln n}{n} +  \mathcal{G} \frac{\ln \ln n}{n} + \frac{\min\left(\ln(K), \ln(J) \right)}{n}  + \frac{\ln(4)}{n} ,
\end{align*}
where $\mathcal{G}$ is a constant depending on $J, K, \rho_{\max}$ and $d$. 
\label{upper_bound2}
\end{theorem}
\noindent

\textit{Proof:} The proofs of Theorems \ref{upper_bound} and \ref{upper_bound2} are given in Section \ref{thms12}.

\textbf{Proof outline:}
A random codebook with codewords drawn according to the normalized maximum-likelihood distribution $Q^{\text{NML}}$ in $(\ref{NMLdist})$ is available to both the encoder and decoder. For any input source sequence $x^n$ with type $t = t(x^n)$ and input distortion measure $\rho$, the encoder uses acceptance-rejection sampling from $Q^{\text{NML}}$ to obtain i.i.d. codewords according to the optimal output distribution $Q^{t,d,\rho}$. The encoder then communicates to the decoder the index of the first accepted codeword which meets the distortion constraint. The proof then primarily relies on lower bounding the probability $\mathbb{P}(\rho_n(x^n, Y^n) \leq d)$ where $Y^n$ is i.i.d. according to $Q^{t,d,\rho}$. For minimax results, such a lower bound must hold uniformly for all source sequences and distortion measures.
As discussed in the Introduction section in $(\ref{impossibilityaaron})$, a nonvanishing lower bound is impossible to obtain as shown by a simple counterexample in Proposition \ref{prop:counterexample}. Thus, as discussed in $(\ref{post_correction09})$, we provide some leeway in distortion and then use post-correction to satisfy the distortion constraint. The lower bound to the probability of meeting the relaxed distortion constraint is developed in Lemmas \ref{rootnode}-\ref{lemlowerbnd55} in Section \ref{mimplemmas}.    

\begin{remark}
The existence of a universal deterministic code does not directly follow from Theorems \ref{upper_bound} and \ref{upper_bound2}. The existence of strongly universal deterministic codes in the universal distortion setting is the subject of the next two theorems. 
\end{remark}

\begin{theorem}
Fix $d > 0$. Then for sufficiently large $n$, there exists a universal deterministic non-prefix $d$-semifaithful code $\tilde{C}_n = (\phi_n, f_n, g_n)$ for the universal distortion problem such that
\begin{align*}
&\sup_{p \in \mathcal{P}(A), \rho \in \mathcal{D}} \left [ \frac{1}{n} \mathbb{E}_p \left [ \ln(2) l (f_n(\phi_n(X^n, \rho)))  \right ] - \mathbb{E}_p \left [ R(T, d, \rho) \right ] \right]\\
&\leq  \frac{2 \rho_{\max} \ln(n)}{d \, n^{5/8}}   + \frac{4 \rho_{\max}\left(  \min \left( \ln(K), \ln(J) \right)  + \ln(2)\right)}{d \, n^{5/8}}    + \mbox{} \notag \\
    & +  \frac{K + 5/4}{2} \frac{\ln n}{n}   + \gamma_n \frac{\ln \ln n}{n} + O\left(\frac{1}{n} \right),
\end{align*}
where $V_1$ is as defined in Theorem \ref{upper_bound}, \begin{align*}
    \gamma_n = 1 + \frac{\ln(J^2 K^2 + J - 1)}{\ln \ln n},
\end{align*}
and the $O(1/n)$ term depends only on $J$ and $K$. 
\label{upper_bound3}
\end{theorem}
\textit{Proof:} The proof of Theorem \ref{upper_bound3} is given in Section \ref{pfthms3}. 

\begin{theorem}
Fix $d > 0$. Then for sufficiently large $n$, there exists a universal deterministic prefix $d$-semifaithful code $\tilde{C}_n = ( \phi_n, f_n, g_n)$ for the universal distortion problem such that
\begin{align*}
    &\sup_{p \in \mathcal{P}(A), \rho \in \mathcal{D}} \left [ \frac{1}{n} \mathbb{E}_p \left [ \ln(2) l (f_n(\phi_n(X^n, \rho)))  \right ] - \mathbb{E}_p \left [ R(T, d, \rho) \right ] \right] \\
    &\leq  \frac{2 \rho_{\max} \ln(n)}{d \, n^{5/8}}  +  \frac{4 \rho_{\max}\left(  \min \left( \ln(K), \ln(J) \right) + \ln(2) \right)}{d \, n^{5/8}}  + \mbox{} \notag \\
    &  \frac{K + 21/4}{2} \frac{\ln n}{n} +  \mathcal{G} \frac{\ln \ln n}{n} + O\left(\frac{1}{n} \right),
\end{align*}
where $\mathcal{G}$ is a constant depending on $J, K, \rho_{\max}$ and $d$, and the $O(1/n)$ term depends only on $J$ and $K$.  
\label{upper_bound4}
\end{theorem}
\textit{Proof:} The proof of Theorem \ref{upper_bound4} is given in Section \ref{thm4proof}. 

\textbf{Proof Outline:} The proofs of Theorems \ref{upper_bound3} and \ref{upper_bound4} again rely on a random coding argument as in the proofs of Theorems \ref{upper_bound} and \ref{upper_bound2}. While Theorems \ref{upper_bound} and \ref{upper_bound2} showed that the random code performs uniformly well in expectation, we must now show that the random code performs uniformly well with high probability. This is the key to derandomization, i.e., inferring the existence of a deterministic code from a random one. To achieve this objective, we used a uniform concentration result for the random rate used to encode a sequence from a given type class w.r.t. a given input distortion measure followed by a union bound over all types and equivalence classes of distortion measures.

Theorems $\ref{upper_bound}$$-$$\ref{upper_bound4}$ establish an $O(n^{-5/8} \ln n)$ achievable rate for uniform convergence of the difference between expected rate and $\mathbb{E}_p \left [ R(T,d,\rho) \right ] $ to zero. Concavity of the rate-distortion function in the source distribution $p$ would enable application of Jensen's inequality and thus, establish convergence to the rate-distortion function. However, the rate-distortion function $R(p,d,\rho)$ is not necessarily concave or even quasiconcave in $p$ \cite{Ahl1}. 

Our ultimate goal is to establish uniform convergence to the rate-distortion function. It may seem that $\mathbb{E}_p \left [ R(T,d,\rho) \right ] $ appearing as an intermediate quantity might be an artifact of our analysis. However, the following lemmas based on \cite[Lemma 5]{mahmood2021lossy}, when combined with Theorems~\ref{upper_bound}-\ref{upper_bound4}, establish the fundamental nature of $\mathbb{E}_p \left [ R(T,d,\rho) \right ] $ in analyzing convergence of the expected rate for any code.
\begin{lemma}
For all $n \in \mathbb{N}$, any prefix $d$-semifaithful code $C_n = (\phi_n, f_n, g_n)$ satisfies
\begin{align*}
    &\frac{1}{n} \mathbb{E}_p \left [ l(f_n(\phi_n(X^n))) \ln 2 \right ]\\
    &\geq \mathbb{E}_p \left [ R(T,d,\rho) \right] - (J K  + J - 2) \frac{\ln n}{n} - \frac{JK + J - 2}{n}
\end{align*}
for all $p \in \mathcal{P}(A)$ and $\rho \in \mathcal{D}$. 
\label{lemmagolden}
\end{lemma}

The next lemma shows that a uniform lower bound involving $\mathbb{E}_p \left [ R(T,d,\rho) \right ] $ holds for non-prefix $d$-semifaithful codes as well.   
\begin{lemma}
Any non-prefix $d$-semifaithful code $C_n = (\phi_n, {f}_n, {g}_n)$  satisfies
\begin{align*}
    &\frac{1}{n} \mathbb{E}_p \left [ l(f_n(\phi_n(X^n) )) \ln 2 \right ]\\   &\geq \mathbb{E}_p \left [ R(T,d,\rho) \right] - (J K  + J - 1) \frac{\ln n}{n} + o \left( \frac{\ln n}{n} \right) 
\end{align*}
for all $p \in \mathcal{P}(A)$ and $\rho \in \mathcal{D}$, where the term $o(\ln n / n)$, when divided by $\ln n / n$, tends to zero at a rate determined only by alphabet sizes $J$ and $K$.  
\label{lemmagolden2}
\end{lemma}
For the proof of Lemma \ref{lemmagolden}, see \cite[Lemma 5]{mahmood2021lossy}. The proof of Lemma \ref{lemmagolden2} is similar to Lemma \ref{lemmagolden} and is briefly outlined in Appendix \ref{hiuhiuhiu}.  

Lemmas $\ref{lemmagolden}$ and $\ref{lemmagolden2}$ in conjunction with Theorems $\ref{upper_bound}$$-$$\ref{upper_bound4}$ imply that for an optimal $d$-semifaithful code, the difference between its expected rate and $\mathbb{E}_p \left [ R(T, d, \rho) \right ] $ tends to zero uniformly. Therefore, a necessary and sufficient condition for minimax convergence of the expected rate of a $d$-semifaithful code to the rate-distortion function $R(p,d,\rho)$ is uniform convergence of $\mathbb{E}_p[R(T,d,\rho)]$ to $R(p,d,\rho)$, over all $p$ and $\rho$. This condition is indeed satisfied by virtue of the uniform continuity of the rate-distortion function with respect to $(p, \rho)$ (Lemma \ref{lemmaunifcont}). The following lemma synthesizes Lemmas \ref{lemmatypes} and \ref{lemmaunifcont} to establish uniform convergence of $\mathbb{E}_p[R(T,d,\rho)]$ to $R(p,d,\rho)$ over all $p$ and $\rho$.  
\begin{lemma}
For any fixed $d > 0$, we have 
\begin{align*}
    \lim_{n \to \infty}\,\, \sup_{p \in \mathcal{P}(A), \rho \in \mathcal{D}}\,\, \big | \mathbb{E}_p \left [ R(T,d,\rho)  \right ]  - R(p,d,\rho) \big | = 0.
\end{align*}
\label{modcont}
\end{lemma}
\textit{Proof:} The proof is given in Appendix \ref{modcontproof}

While the rate-distortion function is a strict lower bound for the expected rate of a $d$-semifaithful code with a prefix encoder, Lemmas \ref{lemmagolden2} and \ref{modcont} imply that the rate-distortion function is an asymptotic lower bound for the expected rate of a $d$-semifaithful code with an arbitrary encoder.  Let $\mathcal{C}_{d,pr}$ be the set of all deterministic $d$-semifaithful codes with a prefix encoder and $\mathcal{C}_{d,npr}$ be the set of all deterministic $d$-semifaithful codes with an arbitrary encoder.   The following two corollaries directly follow from the results of Theorems $\ref{upper_bound}$$-$$\ref{upper_bound4}$ and Lemmas $\ref{lemmagolden}$$-$$\ref{modcont}$. 

\begin{corollary}[Existence of minimax prefix codes]
For any fixed $d > 0$, we have 
\begin{align*}
    &\lim_{n \to \infty}\,\, \inf_{(\phi_n,f_n, g_n) \in \mathcal{C}_{d, pr}} \,\, \sup_{p \in \mathcal{P}(A), \rho \in \mathcal{D}} \\
    &\left [  \frac{\ln(2)}{n} \mathbb{E}_p \left [ l (f_n(\phi_n(X^n, \rho)))  \right ] - R(p,d,\rho) \right] = 0.
\end{align*}
    \label{corr:prefix}
\end{corollary}

\begin{corollary}[Existence of minimax arbitrary codes]
For any fixed $d > 0$, we have 
\begin{align*}
    &\lim_{n \to \infty}\,\, \inf_{(\phi_n,f_n, g_n) \in \mathcal{C}_{d, npr}} \,\, \sup_{p \in \mathcal{P}(A), \rho \in \mathcal{D}} \\
    &\left [  \frac{\ln(2)}{n} \mathbb{E}_p \left [ l (f_n(\phi_n(X^n, \rho)))  \right ] - R(p,d,\rho) \right]  = 0.
\end{align*}
    \label{corr:arb}
\end{corollary}

These corollaries do not have explicit bounds on the rate of minimax convergence to the rate-distortion function owing to the absence of explicit bounds for the convergence in Lemma \ref{modcont}. However, Lemma \ref{lemmataylor}, which uses uniform continuity bounds from \cite[Lemma 2]{hari_unifcont}, resolves this shortcoming. While Lemma \ref{lemmataylor} is stronger than Lemma \ref{modcont} because it provides an explicit rate of convergence, it is weaker because it is not uniform over distortion measures.

\begin{lemma}
Fix $d > 0$ and a distortion measure $\rho$. Then for sufficiently large $n$, we have for all $a \geq \sqrt{2 J  + 2}$, 
\begin{align*}
    &\sup_{p \in \mathcal{P}(A)} \big | \mathbb{E}_p \left [ R(T,d,\rho)  \right ]  - R(p,d,\rho) \big |  \\
    &\leq \frac{7 \rho_{\max}}{\rho_{\min}} \left(a \sqrt{J} \sqrt{\frac{\ln n}{n}} \right) \ln \left( \frac{J\,K \sqrt{n}}{a \sqrt{J \ln n}} \right) +  \ln(K) \frac{e^{J-1}}{n^2}, 
\end{align*}
where 
\begin{align*}
    \rho_{\min}  = \min_{(j, k) : \rho(j, k) > 0} \rho(j, k).
\end{align*}
\label{lemmataylor}
\end{lemma}

\begin{remark}
For any function $f(n) \in O \left(\sqrt{\frac{\ln n}{n}} \ln \left( \sqrt{\frac{n}{\ln n}} \right) \right )$, we have $f(n) \in O\left( \frac{\ln^{3/2}(n)}{\sqrt{n}} \right)$.
\label{ordersimpl}
\end{remark}
\begin{IEEEproof}[Proof of Lemma \ref{lemmataylor}]
Fix $d > 0$ and a distortion measure $\rho$. For some $a \geq \sqrt{2 J + 2}$, we start by writing 
\begin{align}
&\mathbb{E}_p [R(T,d,\rho)] \notag\\
    &= \sum_{t \in \mathcal{P}_n(A)} p^n(T^n_A(t)) R(t,d,\rho) \notag\\
    &= \sum_{t:||t - p||_2 \leq a \sqrt{\ln n / n}}  p^n(T^n_A(t)) R(t,d,\rho) \notag\\
    & \,\,\,\,\,\,\,\,\,\,\,\,+ \sum_{t:||t - p||_2 > a \sqrt{\ln n / n}}  p^n(T^n_A(t)) R(t,d,\rho) \notag \\
    &\leq \sum_{t:||t - p||_2 \leq a \sqrt{\ln n / n}}  p^n(T^n_A(t)) R(t,d,\rho) +   \ln(K) \frac{e^{J-1}}{n^2}, \label{geenida}
\end{align}
where the last inequality follows from Lemma \ref{lemmatypes} and the fact that $R(t,d,\rho) \leq \ln(K)$ from the assumption in $(\ref{dist_assump})$.

We now invoke \cite[Lemma 2]{hari_unifcont} which states that for any $p, q \in \mathcal{P}(A)$ satisfying $||p - q||_1 \leq \frac{\rho_{\min}}{4 \rho_{\max}}$ and for any $d > 0$, 
\begin{align}
    |R(p,d,\rho) - R(q, d, \rho) | \leq \frac{7 \rho_{\max}}{\rho_{\min}} ||p - q||_1 \ln \left( \frac{J\,K}{||p-q||_1} \right). \label{hariskalemm}
\end{align}
For sufficiently large $n$, we can ensure 
\begin{align}
    ||t - p||_1 &\leq \sqrt{J} ||t  -p||_2 \notag \\
    &\leq a\sqrt{J} \sqrt{\frac{\ln n}{n} } \label{joooooooo}\\
    &\leq \frac{\rho_{\min}}{4 \rho_{\max}}. \notag 
\end{align}
Therefore, using $(\ref{hariskalemm})$ in $(\ref{geenida})$, we obtain 
\begin{align*}
    &\mathbb{E}_p \left [ R(T, d, \rho) \right ]\\
    &\leq \sum_{t:||t - p||_2 \leq a \sqrt{\ln n / n}}  p^n(T^n_A(t)) \left( R(p,d,\rho) + \frac{7 \rho_{\max}}{\rho_{\min}} \times \right. \\
    & \left . ||p - t||_1 \ln \left( \frac{J\,K}{||p-t||_1} \right) \right) +   \ln(K) \frac{e^{J-1}}{n^2}\\
    &\leq R(p,d,\rho) + \frac{7 \rho_{\max}}{\rho_{\min}} \left(a \sqrt{J} \sqrt{\frac{\ln n}{n}} \right) \ln \left( \frac{J\,K \sqrt{n}}{a \sqrt{J \ln n}} \right) \\
    & \,\,\,\,\,\,\,\,\,\,\,\,\,\,\,\,\,\,\,\,\,\,+  \ln(K) \frac{e^{J-1}}{n^2}.
\end{align*}
In the last inequality above, we use the fact that $x \ln (JK/ x)$ is an increasing function in $x$ for all $x \leq JK / e$ and it is easy to ensure $||t - p||_1 \leq JK / e$ for sufficiently large $n$ using the upper bound in $(\ref{joooooooo})$.

For the lower bound, we can write  
\begin{align}
    &\mathbb{E}_p \left [ R(T, d, \rho) \right ] \notag \\
    &\geq \sum_{t:||t - p||_2 \leq a \sqrt{\ln n / n}}  p^n(T^n_A(t)) R(t,d,\rho). \label{opopli}
\end{align}
Then, for sufficiently large $n$, we can again apply the result in $(\ref{hariskalemm})$ and, using a similar argument as before, obtain
\begin{align*}
    &\mathbb{E}_p \left [ R(T, d, \rho) \right ] \notag \\
    &\geq R(p,d,\rho) - \frac{7 \rho_{\max}}{\rho_{\min}} \left(a \sqrt{J} \sqrt{\frac{\ln n}{n}} \right) \ln \left( \frac{J\,K \sqrt{n}}{a \sqrt{J \ln n}}  \right)\\
    & \,\,\,\,\,\,\,\,\,\,\,\,\,\,\,\,\,\,\,\,\,\, - \ln(K)\frac{e^{J-1}}{n^2}.
\end{align*}
\end{IEEEproof}

Using the simplification from Remark \ref{ordersimpl}, the following two corollaries follow from the results of Theorems $\ref{upper_bound}$$-$$\ref{upper_bound4}$ and Lemmas $\ref{lemmagolden}$, $\ref{lemmagolden2}$ and $\ref{lemmataylor}$.  

\begin{corollary}[Minimax redundancy with prefix codes]
Fix $d > 0$ and some distortion measure $\rho$. Then 
\begin{align*}
     &\inf_{(\phi_n,f_n, g_n) \in \mathcal{C}_{d, pr}} \,\, \sup_{p \in \mathcal{P}(A)} \\
     &\left [  \frac{\ln(2)}{n} \mathbb{E}_p \left [ l (f_n(\phi_n(X^n)))  \right ] - R(p,d,\rho) \right] \\
     &= O \left( \frac{\ln^{3/2} (n)}{\sqrt{n}}\right).
\end{align*}
    \label{corr:prefix2}
\end{corollary}

\begin{corollary}[Minimax redundancy with arbitrary codes]
Fix $d > 0$ and some distortion measure $\rho$. Then 
\begin{align*}
    & \inf_{(\phi_n,f_n, g_n) \in \mathcal{C}_{d, npr}} \,\, \sup_{p \in \mathcal{P}(A)} \\
    &\left [  \frac{\ln(2)}{n} \mathbb{E}_p \left [ l (f_n(\phi_n(X^n)))  \right ] - R(p,d,\rho) \right]\\
    &= O \left( \frac{\ln^{3/2} (n)}{\sqrt{n}}\right).
\end{align*}
    \label{corr:arb2}
\end{corollary}
\noindent
One can obtain explicit bounds in Corollaries \ref{corr:prefix2} and \ref{corr:arb2} from the statements of Theorems $\ref{upper_bound}$$-$$\ref{upper_bound4}$ and 
Lemmas $\ref{lemmagolden}$, $\ref{lemmagolden2}$ and $\ref{lemmataylor}$.

We turn to impossibility results. We first show in Lemma \ref{lemma:expectedlower} that the upper bound in Lemma \ref{lemmataylor} cannot be improved more than logarithmically, i.e., 
\begin{align*}
    \sup_{p \in \mathcal{P}(A)} \left | \mathbb{E}_p [R(T,d,\rho)] - R(p,d,\rho) \right | = \Omega\left( \frac{1}{\sqrt{n}}\right).
\end{align*}

\begin{lemma}
Consider alphabets $A = B = \{ 0,1 \}$ with distortion measure $\rho(0,0) = \rho(1,1) = 0$ and $\rho(0, 1) = \rho(1, 0) = \rho_{\max} > 0$. Then for any
distortion level $d \in (0, \rho_{\max}/2)$, if the source distribution
is Bernoulli($\bar{d}$), where $\bar{d} = d/\rho_{\max}$, we have\footnote{We use the notation $x^+ = \max(x, 0)$. Also note that the two terms on the right-hand side of $(\ref{chandlermonica})$ are greater than zero for sufficiently large $n$. }
\begin{align}
    \mathbb{E}_p \left [ R(T,d,\rho) \right ] -
    R(p,d,\rho) \ge \left[ \sqrt{\frac{\bar{d}(1-\bar{d})}{n}}
    \ln \left(\frac{1-\bar{d}}{\bar{d}}\right) - \frac{1}{2n} \right]^+
       \cdot \left[\Phi(2) - \Phi(1) - \frac{(1 - \bar{d})^2
           + {\bar{d}}^2}{\sqrt{n \bar{d} (1-\bar{d})}}
              \right]^+
              \label{chandlermonica}
\end{align}
    for any $n$ satisfying $2\bar{d} + 3\sqrt{\frac{\bar{d}(1-\bar{d})}{n}} < 1$. 
    \label{lemma:expectedlower}
\end{lemma}
\begin{IEEEproof}
Denoting the binary entropy function by $H_b(\cdot)$,
we have 
\begin{align}
    \mathbb{E}_p \left [ R(T,d,\rho) \right ]
    &= \sum_{t \in \mathcal{P}_n(A)} p^n(T^n_A(t)) R(t,d,\rho)\notag \\
    &\geq \sum_{\bar{d} + \sqrt{\frac{\bar{d}(1-\bar{d})}{n}} < t(1) \leq \bar{d} + 2\sqrt{\frac{\bar{d}(1-\bar{d})}{n}}} p^n(T^n_A(t)) \left [ H_b(t(1)) - H_b(\bar{d}) \right ] \notag \\
    &\geq \left( H_b\left(\bar{d} + \sqrt{\frac{\bar{d}(1-\bar{d})}{n}}  \right) - H_b\left(\bar{d}\right) \right) \mathbb{P}\left( \bar{d} + \sqrt{\frac{\bar{d}(1-\bar{d})}{n}} < \frac{1}{n} \sum_{i=1}^n X_i \leq \bar{d} + 2\sqrt{\frac{\bar{d}(1-\bar{d})}{n}} \right), \label{exp_low_bndmp}
\end{align}
where the second inequality uses the assumption that $2\bar{d} + 3\sqrt{\frac{\bar{d}(1-\bar{d})}{n}} < 1$.
By a simple Taylor series expansion,
\begin{align}
    H_b\left (\bar{d} + \sqrt{\frac{\bar{d}(1-\bar{d})}{n}}\right ) - H_b\left(\bar{d}\right) & \ge \sqrt{\frac{\bar{d}(1-\bar{d})}{n}} \ln \left( \frac{1-\bar{d}}{\bar{d}} \right) - \frac{1}{2n}.
    \label{entropa_tayl}
\end{align}
    A standard application of the Berry-Esseen theorem (with constant $1/2$~\cite{Korolev:Shevtsova:2010,Tyurin2010}) yields 
\begin{align}
    \mathbb{P}\left( \bar{d} + \sqrt{\frac{\bar{d}(1-\bar{d})}{n}} < \frac{1}{n} \sum_{i=1}^n X_i \leq \bar{d} + 2\sqrt{\frac{\bar{d}(1-\bar{d})}{n}}  \right)
    \ge \left[ \Phi(2) - \Phi(1) - \frac{(1-\bar{d})^2 + {\bar{d}}^2}
    {\sqrt{n \bar{d} (1-\bar{d})}} \right].
      \label{berry_conv_prop} 
\end{align}
Substituting $(\ref{entropa_tayl})$ and $(\ref{berry_conv_prop})$ into $(\ref{exp_low_bndmp})$ completes the proof. 
\end{IEEEproof}

By combining Lemma \ref{lemma:expectedlower} with Lemmas~\ref{lemmagolden} and~\ref{lemmagolden2}, we obtain
in particular that, up to logarithmic factors, the redundancy bounds
in Corollaries~\ref{corr:prefix2} and~\ref{corr:arb2} cannot be improved.

\begin{corollary}
    Under the choice of $\rho$ and $d$ assumed in Lemma~\ref{lemma:expectedlower},
    \begin{align}
        \liminf_{n \rightarrow \infty} \sup_{p \in \mathcal{P}(A)} \inf_{(\phi_n,f_n, g_n) \in \mathcal{C}_{d, pr}} \,\,  \left [  \frac{\ln(2)}{n} \mathbb{E}_p \left [ l (f_n(\phi_n(X^n)))  \right ] - R(p,d,\rho) \right] \sqrt{n} & > 0  \\
        \liminf_{n \rightarrow \infty} \sup_{p \in \mathcal{P}(A)} \inf_{(\phi_n,f_n, g_n) \in \mathcal{C}_{d, npr}} \,\,  \left [  \frac{\ln(2)}{n} \mathbb{E}_p \left [ l (f_n(\phi_n(X^n)))  \right ] - R(p,d,\rho) \right] \sqrt{n} & > 0.  
    \end{align}
    \label{corollary:redundancylower}
\end{corollary}

Note that the lower bounds in Corollary~\ref{corollary:redundancylower}
apply to the max-min redundancy, i.e., the non-universal setup. 
Corollaries \ref{corr:prefix2} and \ref{corr:arb2},
on the other hand, provide achievable results for the min-max
redundancy, i.e., the universal setup. It follows that 
the discrepancy between the max-min
and min-max redundancies, which is related to the price
of universality, is no more than logarithmic. Also note
that the choice of $\rho$ and $d$ in crucial in 
Lemma~\ref{lemma:expectedlower}
and Corollary~\ref{corollary:redundancylower}. 
If $d$ is zero then the problem
reduces to the lossless case, for which (for prefix codes) the
max-min redundancy is $O(1/n)$~\cite[Thm.~5.4.1]{thomas_cov} and the min-max redundancy
is $O(\log n/n)$~\cite{rissanen1}. If $\rho(j,k) = 0$ for all $j$ and $k$,
then all forms of the redundancy are obviously zero.

\section{Lagrange Formulation of Rate-Distortion Problem \label{lagrangeformulation}}

The proofs of the main theorems rely on a Lagrangian characterization of the rate-distortion function. 
%The minimization problem in $(\ref{rdfunc}) - (\ref{rdfunc3})$ defining the rate-distortion function can be analyzed using the Lagrange multiplier method as shown in \cite{thomas_cov} and \cite{berger1}. The results are provided below for convenience. 
For a given $t \in \mathcal{P}(A)$, $d > 0$ and $\rho \in \mathcal{D}$, an optimal solution $Q^*_{B|A}$ to the rate-distortion problem satisfies the following system of equations: 
\begin{align}
    &Q^*_{B|A}(k|j) = \frac{Q^{t,d,\rho}(k) \exp \left(- \lambda^* \rho(j,k)  \right)}{\sum \limits_{k' \in B} Q^{t,d,\rho}(k') \exp \left( -\lambda^* \rho(j,k') \right) } \label{compslack444}\\
    &\sum_{j \in A} \frac{t(j)}{\sum \limits_{k' \in B} Q^{t,d,\rho}(k') \exp \left( -\lambda^* \rho(j,k') \right) } e^{-\lambda^* \rho(j,k)} \notag \\
    &\,\,\,\,\,\,\,\,\,\,\,\,\,\,\,\,\,\,\,\,\,\,\,\,\,\,\,\,\,\,\begin{cases}
    =  1 & \text{ if } Q^{t,d,\rho}(k) > 0\\
     \leq 1 & \text{ if } Q^{t,d,\rho}(k) = 0
    \end{cases}\\
    &\lambda^* \left( \sum_{j \in A, k \in B} t(j) Q^*_{B|A}(k|j) \rho(j, k) - d \right) = 0 \label{jiiiuy}\\
    &R(t,d,\rho) = -\lambda^* d - \sum_{j \in A} t(j) \ln \left( \sum_{k' \in B} Q^{t,d,\rho}(k') e^{-\lambda^* \rho(j, k')} \right), \label{liop2}
\end{align}
where
\begin{align}
   - \lambda^* \in  \frac{\partial R(t,d,\rho)}{\partial d} \Bigg. \label{slopeeeguy}
\end{align}
and the right-hand side refers to the subdifferential of $R(t,d,\rho)$ with respect to $d$.
Note that $R(t, d,\rho)$ is differentiable in $d$ except possibly at the distortion
associated with zero rate:
\begin{equation}
    \min_{k \in B} \sum_{j \in A} p(j) \rho(j,k),
\end{equation}
as noted earlier. The existence of the Lagrange multiplier $\lambda^*$ satisfying~(\ref{slopeeeguy})
follows from, e.g.,~\cite[Thm.~29.1]{rockafellar}.  Then 
\cite[Thm.~28.4]{rockafellar} guarantees 
that $Q^*_{B|A}$ minimizes the Lagrangian,
in which case complementary slackness (\ref{jiiiuy}) must hold.
Then \cite[Thm.~8.7]{korner1} establishes the remaining assertions.  

\section{Random $d$-ball lemmas \label{mimplemmas}}

Fix $d > 0$. For any given type $t \in \mathcal{P}_n(A)$ and distortion measure $\rho \in \mathcal{D}$, let $(Q^*_{B|A}, \lambda^*)$ be a solution to the Lagrange formulation of the rate-distortion problem in $(\ref{compslack444})$$-$$(\ref{liop2})$. Let $Q^{t,d,\rho}$ be the corresponding optimal reconstruction distribution on $B$ defined in $(\ref{vqye})$. The proofs of the main theorems of this paper use a lower bound\footnote{This lower bound holds uniformly over all types $t \in \mathcal{P}_n(A)$, distortion measures $\rho \in \mathcal{D}$ and all sequences $x^n \in T^n_A(t)$.} on 
\begin{align}
    \mathbb{P}\left (\rho_n(x^n, Y^n) \leq d + \epsilon\right),
\end{align}
where $x^n \in T^n_A(t)$, $Y^n$ is an i.i.d.\ sequence generated according to $Q^{t,d,\rho}$ and $\epsilon$ is a real parameter. We derive this lower bound through several successive lemmas.

For any given $(t,d,\rho)$, define a sequence of independent random variables $U_1, U_2, \ldots, U_n $ as  
\begin{align}
    U_i &\triangleq \rho(x_i, \tilde{Y}_i) - \sum_{k \in B} Q^*_{B|A}(k|x_i) \rho(x_i, k), \label{loopaUi's}
\end{align}
where $\tilde{Y}_i \sim Q^*_{B|A}(\cdot | x_i)$.
\begin{lemma}[Refined Lucky-Strike Lemma]
Fix $d > 0$. For any real number parameters $\epsilon$ and $C$, we have 
\begin{align}
    &\mathbb{P}\left (\rho_n(x^n, Y^n) \leq d + \epsilon\right)\notag\\ 
    &\geq \exp \left(-n R(t,d,\rho) - C \lambda^* \right)   \mathbb{P} \left( - C \leq \sum_{i=1}^n U_i \leq \epsilon\,n \right) \label{further14}
\end{align}
for all integers $n$, for all $t \in \mathcal{P}_n(A)$, $\rho \in \mathcal{D}$ and $x^n \in T^n_A(t)$, where $Y^n$ is distributed according to $(Q^{t,d,\rho})^n$.
\label{rootnode}
\end{lemma}
The proof of Lemma \ref{rootnode} is given in Appendix \ref{pandakakuta}. 

By making appropriate choices of parameters $\epsilon$ and $C$ in Lemma \ref{rootnode}, we can further lower bound $(\ref{further14})$ using concentration results and the Berry-Esseen theorem.

\begin{lemma}
Fix $d > 0$. For any nonnegative numbers $C_1, C_2$ and $\alpha$,
we have 
\begin{align*}
    &\mathbb{P} \left( \rho_n(x^n, Y^n) \leq d  + \frac{C_1 }{ n^{\alpha}}  \right) \\
    &\geq \exp \left(-n R(t,d,\rho) - C_1 \lambda^* n^{1-\alpha}  \right)  \xi \left( C_1, C_2, \alpha \right)
\end{align*}
for 
\begin{align*}
    n &\geq \left( \frac{(C_1)^2}{3 (\rho_{\max})^2} \right)^{\frac{1}{2 \alpha - 1}}, 
\end{align*}
for all $t \in \mathcal{P}_n(A)$, $\rho \in \mathcal{D}$ and $x^n \in T^n_A(t)$, where $Y^n$ is distributed according to $(Q^{t,d,\rho})^n$,  
\begin{align*}
    &\xi \left( C_1, C_2, \alpha \right) \\
    &= \min \left(\left(1 - \frac{C_2 }{C_1^2 } \right),  \left(\frac{C_1}{\sqrt{2 \pi} n^{\alpha - 1/2}\rho_{\max}}  - \frac{2C_0(\rho_{\max})^3}{(C_2)^{3/2} n^{2-3\alpha}}  \right) \right), 
\end{align*}
and $C_0$ is the absolute constant from Berry-Esseen theorem \cite{Shevtsova}. 
\label{lemlowerbnd}
\end{lemma}
The proof of Lemma \ref{lemlowerbnd} is given in Appendix \ref{pandakakuta2}. 

The final lemma in this sequence, which will be directly used in proving the main theorems, follows as a direct corollary of Lemma \ref{lemlowerbnd}. Specifically, using the upper bound $C_0 \leq 0.56$ \cite{Shevtsova} and choosing $\alpha = 5/8$, $C_1 = 2 \rho_{\max}$ and $C_2 = 2.5 (\rho_{\max})^2$ in Lemma \ref{lemlowerbnd}, we obtain the following.

\begin{lemma}
Fix $d > 0$. Then for $n \geq 10$, we have 
\begin{align}
    \nonumber
    &\mathbb{P} \left( \rho_n(x^n, Y^n) \leq d  + \frac{2 \rho_{\max} }{ n^{5/8}}  \right)\\
    &\geq \exp \left(-n R(t,d,\rho) - 2 \rho_{\max} \lambda^* n^{3/8} - \frac{1}{8} \ln n + \ln \left (\frac{1}{2} \right)  \right),
    \label{eq:lemmalowerbnd}
\end{align}
for all $t \in \mathcal{P}_n(A)$, $\rho \in \mathcal{D}$, $x^n \in T^n_A(t)$ and $Y^n$ distributed according to $(Q^{t,d,\rho})^n$.
\label{lemlowerbnd55}
\end{lemma}
Since $\lambda^*$ satisfies~(\ref{slopeeeguy}) and $R(t,d,\rho)$ is convex in $d$
and satisfies $R(t,d,\rho) \le \min(\log J,\log K)$, we 
have\footnote{This observation was credited by Yu and Speed~\cite{yu1} to T.~Linder.}
\begin{equation}
    \lambda^* \le \frac{\min(\log J, \log K)}{d}.
\end{equation}
Substituting this into (\ref{eq:lemmalowerbnd}) gives the bound
\begin{align}
    \frac{\mathbb{P} \left( \rho_n(x^n, Y^n) \leq d  + \frac{2 \rho_{\max} }{ n^{5/8}}  \right)}{e^{-n R(t,d,\rho)}}
    \geq \exp \left(- \frac{2 \rho_{\max} \min(\log J, \log K) n^{3/8}}{d} - \frac{1}{8} \ln n + \ln \left (\frac{1}{2} \right)  \right),
    \label{CODMsucks}
\end{align}
which has the crucial property that the right-hand side decays to zero subexponentially 
independently of $t$ and $\rho$.
This uniformity relies on the leeway afforded by allowing the code to violate the
distortion constraint by $2 \rho_{\max}/n^{5/8}$.
The following proposition shows that without such freedom, it is not possible to
have a nonvanishing lower bound, even for a fixed $n$, that holds uniformly over 
source sequences and distortion measures.
\begin{proposition}
    Fix $\rho_{\max} = 3$, $d = 1$, and alphabets
    $A = \{0,1\}$ and $B = \{0,1,2\}$.
    Then for all even $n$ 
    and $\epsilon > 0$, there exists $x^n$ and $\rho$ such that
    for any optimal output distribution $Q^{t,d,\rho}$, we have
    \begin{equation*}
        \frac{\mathbb{P}(\rho_n(x^n,Y^n) \le d)}{e^{-nR(t,d,\rho)}} \le \epsilon,
    \end{equation*}
    where $Y^n$ is i.i.d. $Q^{t,d,\rho}$ and $t$ is the type
      of $x^n$.
      \label{prop:counterexample}
\end{proposition}
\begin{IEEEproof} 
Fix an even integer $n$ and some $\epsilon > 0$. First consider the rate-distortion problem with distortion measure
\begin{equation*}
    \rho' = \left[ \begin{array}{ccc}
            0 & 3 & 1 \\
            3 & 0 & 1 
    \end{array} \right],
\end{equation*}
and a uniform source distribution $p$ over $A$. 
Evidently $R(p,1,\rho') = 0$ and since the rate-distortion function is
    continuous in the distortion level~\cite[Lemma~7.2]{korner1},
\begin{equation}
    \label{eq:firstzero}
    \lim_{\delta \rightarrow 0} R(p,1-\delta,\rho') = 0.
\end{equation}
For some $\delta > 0$, consider the perturbed distortion measure\footnote{We suppress the dependence of $\rho$ on $\delta$.}
\begin{equation*}
    \rho = \left[ \begin{array}{ccc}
            0 & 3 & 1 + \delta \\
        3 & 0 & 1 + \delta \\
    \end{array} \right],
\end{equation*}
with distortion constraint $d = 1$. The rate-distortion function
for this problem, $R(p,1,\rho)$, is clearly upper bounded by
that of the problem with distortion measure 
\begin{equation*}
    \left[ \begin{array}{ccc}
            \delta & 3+\delta & 1 + \delta \\
        3+\delta & \delta & 1 + \delta \\
    \end{array} \right],
\end{equation*}
with distortion constraint $d = 1$, for which the rate-distortion
function is $R(p,1-\delta,\rho')$. Thus, from (\ref{eq:firstzero}), we have
\begin{equation}
    \lim_{\delta \rightarrow 0} R(p,d,\rho) = 0.
\end{equation}
Given any $\epsilon_1 > 0$, choose $\delta > 0$ such that $R(p,d,\rho) < \epsilon_1$.
Let $x^n$ be any sequence with half zeros and
half ones, and let $t$ denote its type. For the given $(t,d,\rho)$, let
$(Q^*_{B|A}, \lambda^*)$ be a solution to the Lagrange formulation of the rate-distortion problem as described in Section \ref{lagrangeformulation} and let $Q^{t,d,\rho}$ be the corresponding optimal output distribution on $\{0, 1, 2 \}$ defined via $(\ref{vqye})$. Let $Y^n$ be i.i.d.  $Q^{t,d,\rho}$. From~(\ref{cruxinequal}), we have (by choosing $\epsilon = 0$)
\begin{align}
    \frac{\mathbb{P}(\rho_n(x^n,Y^n) \le d)}{e^{-n R(t,d,\rho)}} \leq
        \mathbb{P}\left(\sum_{i = 1}^n U_i \le 0\right),
\end{align}
where $U_1,\ldots,U_n$ are as defined in $(\ref{loopaUi's})$. Now since 
\begin{align}
    &\,D(t(\cdot) Q^*_{B|A}(\cdot|\cdot)||t(\cdot) Q^{t,d,\rho}(\cdot)) \\
    &= \frac{1}{2} D(Q^*_{B|A}(\cdot | 0) || Q^{t,d,\rho}(\cdot)) + \frac{1}{2} D(Q^*_{B|A}(\cdot | 1) || Q^{t,d,\rho}(\cdot))\\ 
    &= R(t,d,\rho) < \epsilon_1,
\end{align}
we have $D(Q^*_{B|A}(\cdot | j) || Q^{t,d,\rho}(\cdot)) < 2 \epsilon_1$ for all $j \in \{0,1 \}$. By Pinsker's inequality~\cite[Prob.~3.18]{korner1}, we have 
\begin{align}
    |Q^*_{B|A}(k|j) - Q^{t,d,\rho}(k)| \le \sqrt{4 \epsilon_1}, \label{kqzpv}
\end{align}
for all $k$ and $j$.  Since $Q^*_{B|A}(\cdot | \cdot)$ satisfies
the distortion constraint,
\begin{align}
    1 & \ge \frac{1}{2}\left[ 3 Q^*_{B|A}(1|0) + (1+ \delta)
         Q^*_{B|A}(2|0) + 3 Q^*_{B|A}(0|1) + (1+ \delta)
         Q_{B|A}(2|1) \right] \notag \\
    &\stackrel{(a)}{\geq} \frac{1}{2} \left [ 3 Q^{t,d,\rho}(1) + 3 Q^{t,d,\rho}(0) + 2(1+\delta) Q^{t,d,\rho}(2) - \sqrt{4 \epsilon_1} (8 + 2 \delta) \right ] \notag \\
    &\geq \frac{1}{2} \left [ 3 - Q^{t,d,\rho}(2) - \sqrt{4 \epsilon_1}(8 + 2 \delta) \right ], \notag 
\end{align}
which implies that
\begin{align*}
    Q^{t,d,\rho}(2) &\geq 1 - \sqrt{4 \epsilon_1}(8 + 2 \delta).
\end{align*}
Inequality $(a)$ follows from $(\ref{kqzpv})$. Then by the union bound
\begin{align}
    \mathbb{P}\left (\sum_{i = 1}^n U_i \le 0\right ) &\le \sum_{i = 1}^n \mathbb{P}(U_i \le 0) \notag \\
    &= \frac{n}{2} \left [ Q^*_{B|A}(0 | 0) + Q^*_{B|A}(1 | 1) \right ] \notag\\
    &\stackrel{(a)}{\leq} \frac{n}{2} \left [ Q^{t,d,\rho}(0) + Q^{t,d,\rho}(1) + 2 \sqrt{4 \epsilon_1}  \right ] \notag\\
    &= \frac{n}{2} \left [ 2 \sqrt{4 \epsilon_1} + 1 - Q^{t,d,\rho}(2) \right ] \notag\\
    &\leq \frac{n}{2} \left [ 2 \sqrt{4 \epsilon_1} + \sqrt{4\epsilon_1} (8 + 2 \delta) \right ], \label{vildayanida}
\end{align}
where inequality $(a)$ above uses $(\ref{kqzpv})$. Finally, we can choose $\epsilon_1$ and $\delta$ small enough so that $(\ref{vildayanida})$ is less than $\epsilon$. 

\end{IEEEproof}

\section{Proof of Theorems \ref{upper_bound} and \ref{upper_bound2} \label{thms12}}

Fix $d > 0$. Let $x^n$ be the input source sequence and $\rho$ be the input distortion measure. Let 
\begin{align}
d' \triangleq d + \frac{2 \rho_{\max}}{n^{5/8}}.
\label{d'kidef}
\end{align}
We first encode the sequence $x^n$ using a random $d'$-semifaithful code $\tilde{C}_n = (\phi_n, f_n, g_n)$ and then use a (deterministic) post-correction scheme to reduce distortion from $d'$ to $d$. Let $Q^{\text{NML}} \in \mathcal{P}(B^n)$ be defined as
\begin{align}
    Q^{\text{NML}}(y^n) &= \frac{\sup \limits_{q \in \mathcal{P}(B)} q^n(y^n) }{S_n}, 
    \label{tutu}
\end{align}
where 
\begin{align}
    S_n = \sum \limits_{z^n \in B^n} \sup \limits_{p \in \mathcal{P}(B)} p^n(z^n). 
    \label{tutu2}
\end{align}
Let $Z_1^n, Z_2^n, Z_3^n, \ldots $ be i.i.d. random vectors each distributed according to $Q^{\text{NML}}$. The random codebook $B_{\phi_n} \subset B^n$,
\begin{align*}
    B_{\phi_n} &= \{Z_1^n, Z_2^n, Z_3^n, \ldots \},
\end{align*}
is available to both the encoder and decoder. 
\\
\\
Let $t = t(x^n)$ be the type of $x^n$, where $t \in \mathcal{P}_n(A)$. For the given $(t,d,\rho)$, let $(Q^*_{B|A}, \lambda^*)$ be a solution to the Lagrange formulation of the rate-distortion problem in $(\ref{compslack444})$$-$$(\ref{liop2})$. Let $Q^{t, d, \rho }$ be the corresponding optimal reconstruction distribution on $B$. From the sequence $\{Z_i^n \}_{i=1}^\infty$, the encoder uses acceptance-rejection method to derive a subsequence $\{Z_{i_j}^n \}_{j=1}^\infty$, where ${Z}_{i_1}^n, Z_{i_2}^n, Z_{i_3}^n, \ldots $ are i.i.d.\ random vectors each distributed according to $(Q^{t,d,\rho})^n$. It is easy to see that
\begin{align*}
    \max_{y^n \in B^n} \frac{\prod_{i=1}^n Q^{t,d,\rho}(y_i)  }{Q^{\text{NML}}(y^n)} \leq S_n.
\end{align*}
The acceptance-rejection algorithm to construct the subsequence $\{Z_{i_j}^n\}_{j=1}^\infty$ is described below. 
\begin{enumerate}
    \item Set $i = 1$; $j = 1$. 
    \item Generate $U \sim \text{Unif} \left([0,1] \right)$.
    \item If $$U < \frac{ (Q^{t,d,\rho})^n(Z_i^n)}{S_n Q^{\text{NML}}(Z_{i}^n)}, \,\,\,\,\,\,\,\,\,\, (\text{success if true} )$$ then set $ i_j = i$. Set $i := i + 1$ ; $j := j + 1$. Go back to step $2$. 
    \item Else set $i := i + 1$. Go back to step $2$.
\end{enumerate}
In each iteration of the above algorithm, Step $3$ has success probability of $1/S_n$ independent of other iterations. 

Let $J(x^n)$
be the smallest integer such that $Z_{i_{J(x^n)}}^n$ satisfies 
\begin{align*}
\rho_n(x^n, Z_{i_{J(x^n)}}^n) \leq d + \frac{2 \rho_{\max}}{ n^{5/8}}   = d'.
\end{align*}
We set 
\begin{align}
    \phi_n(x^n, \rho) &=  Z_{i_{J(x^n)}}^n. \label{doosrawalapan}
\end{align}
We can now either use a non-prefix fixed-to-variable encoder $(\ref{codeen111})$ or a prefix Elias encoder $(\ref{codeen1})$ to encode the index $i_{{J(x^n)}}$ of the codeword. Therefore, the length of the binary encoding satisfies 
\begin{align}
    &l(f_n(\phi_n(x^n,\rho))) \leq  1 + \log i_{{J(x^n)}} \label{ftvgg}
\end{align}
if $f_n$ is a fixed-to-variable encoder or 
\begin{align}
    &l(f_n(\phi_n(x^n,\rho))) \leq \lfloor \log(i_{{J(x^n)}}) \rfloor + 2 \lfloor \log \left( \lfloor \log(i_{{J(x^n)}}) \rfloor + 1 \right) \rfloor + 1 \label{ftvgg09}
\end{align}
if $f_n$ is an Elias encoder. 
The decoder $g_n$ then outputs $Z_{i_{{J(x^n)}}}^n$ as the reconstruction sequence. 

We now evaluate the expected rate of this $d'$-semifaithful coding scheme. For every sequence $x^n$ with type $t = t(x^n)$, it is easy to see that ${J(x^n)}$ is a geometric random variable with success parameter 
$$ \mathbb{P}\left (\rho_n(x^n, Y^n) \leq d + \frac{2 \rho_{\max}}{ n^{5/8}}  \right ),$$ 
where $Y^n$ is an i.i.d. sequence with distribution $(Q^{t,d,\rho})^n$ and $\mathbb{P}$ is the probability law associated with $Y^n$. It also follows that 
\begin{align}
    \mathbb{E} \left [ i_{{J(x^n)}} \,| \, {J(x^n)} = 1  \right ] &= S_n \notag \\
    \mathbb{E} \left [ i_{{J(x^n)}} \,| \, {J(x^n)} = 2  \right ] &= 2 S_n \notag \\
    &\,\,\vdots \notag \\
    \mathbb{E} \left [ i_{{J(x^n)}} \,| \, {J(x^n)}  \right ] &= {J(x^n)}  \, S_n \notag \\
    \implies \mathbb{E} \left [i_{{J(x^n)}} \right ] &= S_n \mathbb{E} \left [ {J(x^n)} \right ] = \frac{S_n}{ \mathbb{P}\left (\rho_n(x^n, Y^n) \leq d + \frac{2 \rho_{\max}}{ n^{5/8}}  \right )}. \label{Expec_i} 
\end{align}
Alternatively, we can see that $i_{{J(x^n)}}$ is a geometric random variable with success parameter 
\begin{align}
    \frac{ \mathbb{P}\left (\rho_n(x^n, Y^n) \leq d + \frac{2 \rho_{\max}}{ n^{5/8}}  \right )}{S_n}, \label{Expec_i2}     
\end{align}
where in both $(\ref{Expec_i})$ and $(\ref{Expec_i2})$, $Y^n$ is an i.i.d. sequence with distribution $(Q^{t,d,\rho})^n$.
From $(\ref{S_n bound })$, we obtain
\begin{align}
    \mathbb{E} [i_{{J(x^n)}}] = \frac{\exp \left( \frac{K - 1}{2} \ln n + \ln \left(  \frac{\Gamma\left( \frac{1}{2}\right)^K}{(2\pi)^{\frac{K-1}{2}} \, \Gamma \left(\frac{K}{2} \right)} \right) +  o_K(1)\, \ln (2)\right)}{ \mathbb{P}\left (\rho_n(x^n, Y^n) \leq d + \frac{2 \rho_{\max}}{ n^{5/8}}  \right )},
    \label{iopxz}
\end{align}
where $o_K(1) \to 0$ as $n \to \infty$ at the rate determined only by $K$. For $n \geq 10$, we can use the lower bound from Lemma \ref{lemlowerbnd55} in $(\ref{iopxz})$ to obtain 
\begin{align}
     &\mathbb{E} [i_{{J(x^n)}}] \notag \\
     &\leq \exp \left( \frac{K - 1}{2} \ln n + \ln \left(  \frac{\Gamma\left( \frac{1}{2}\right)^K}{(2\pi)^{\frac{K-1}{2}} \, \Gamma \left(\frac{K}{2} \right)} \right) +  o_K(1)\, \ln (2) + n R(t,d,\rho) + \mbox{} \right . \notag \\ 
     &\,\,\,\,\,\,\,\,\,\,\,\,\,\,\,\,\,\,\,\,\,\,\,\,\,\,\,\,\left. 2 \lambda^* \rho_{\max} n^{3/8} + \frac{1}{8}\ln n + \ln(2)  \right) \label{hopopnida} \\
     &\leq \exp \left( \frac{K - 1}{2} \ln n + \ln \left(  \frac{\Gamma\left( \frac{1}{2}\right)^K}{(2\pi)^{\frac{K-1}{2}} \, \Gamma \left(\frac{K}{2} \right)} \right) +  o_K(1)\, \ln (2) + n R(t,d,\rho) + \mbox{} \right. \notag \\
     &\,\,\,\,\,\,\,\,\,\,\,\,\,\,\,\,\,\,\,\,\,\,\,\,\,\,\,\,\left . 2\frac{\min \left(\ln(K), \ln(J) \right)}{d} \rho_{\max} n^{3/8} + \frac{1}{8} \ln n +  \ln \left( 2 \right) \right) \label{kaisha1}\\
     &\leq \exp \left( \frac{K - 3/4}{2} \ln n + \ln \left(  \frac{\Gamma\left( \frac{1}{2}\right)^K}{(2\pi)^{\frac{K-1}{2}} \, \Gamma \left(\frac{K}{2} \right)} \right)  + n R(t,d,\rho) + 2\frac{\min \left(\ln(K), \ln(J) \right)}{d} \rho_{\max} n^{3/8} + 2\ln \left( 2 \right) \right) \label{jioqw}\\
     &= \exp \left(n R(t,d,\rho) + \frac{K - 3/4}{2} \ln n  + 2\frac{\min \left(\ln(K), \ln(J) \right)}{d} \rho_{\max} n^{3/8} + V_1 \right), \label{aleenanida}
\end{align}
where we have defined the constant $V_1$ for convenience: 
\begin{align}
    V_1 &\triangleq   \ln \left(  \frac{\Gamma\left( \frac{1}{2}\right)^K}{(2\pi)^{\frac{K-1}{2}} \, \Gamma \left(\frac{K}{2} \right)} \right) + 2\ln(2).
    \label{u1u2consts}
\end{align}
In inequality $(\ref{kaisha1})$, we use the assumption that $(\ref{dist_assump})$ holds which implies that $R(t,d,\rho) \leq \min \left( \ln(K), \ln(J) \right)$. Then it is easy to see by convexity of the rate-distortion function in $d$ that any subderivative $\lambda
^*$ is upper bounded by $\min\left(\ln(K), \ln(J) \right) / d$ which explains inequality $(\ref{kaisha1})$. Inequality $(\ref{jioqw})$ follows by taking $n$ sufficiently large such that $o_K(1) \leq 1$. Since $o_K(1)$ tends to zero at a rate \textit{independent} of $t$ and $\rho$, the bound in $(\ref{jioqw})$ and, therefore, in $(\ref{aleenanida})$  hold uniformly over all $t \in \mathcal{P}_n(A)$, all $\rho \in \mathcal{D}$ and all sequences $x^n \in T^n_A(t)$.  

Hence, when $X^n$ is i.i.d. according to $p$, the total expected rate in nats of the random $d'$-semifaithful code $\tilde{C}_n$ with a fixed-to-variable encoder is 
\begin{align}
    &\frac{\ln(2)}{n}\mathbb{E}_p \left [ \mathbb{E}_c \left [l (f_n(\phi_n(X^n, \rho)))  \right] \right ] \notag \\
    &\leq \frac{\ln(2)}{n} \mathbb{E}_p \left [ \mathbb{E}_c\left [1 + \log i_{J(X^n)} \right] \right ] \label{randathm3} \\
    &\leq \frac{1}{n} \mathbb{E}_p \left [\ln(2) + \ln \left( \mathbb{E}_c\left [i_{J(x^n)} \right] \right) \right ] \label{jensenaa}\\
    &\leq \frac{\ln(2)}{n} + \frac{1}{n} \mathbb{E}_p \left [n R(t(X^n),d,\rho) + \frac{K - 3/4}{2} \ln n  + 2\frac{\min\left(\ln(K), \ln(J) \right)}{d} \rho_{\max} n^{3/8} + V_1 \right ] \notag\\
    &= \mathbb{E}_p \left [R(T, d,\rho) \right ]  +    \frac{K - 3/4}{2} \frac{\ln n}{n}  + \frac{2 \min\left(\ln(K), \ln(J) \right)}{d \, n^{5/8}} \rho_{\max}  + \frac{V_1 + \ln(2)}{n}. \label{subskarisko}
\end{align}
In inequality $(\ref{randathm3})$, we used $(\ref{ftvgg})$. In inequality $(\ref{jensenaa})$, we used Jensen's inequality. 

Similarly, the total expected rate in nats of the random $d'$-semifaithful code $\tilde{C}_n$ with an Elias encoder is
\begin{align}
    &\frac{\ln(2)}{n}\mathbb{E}_p \left [ \mathbb{E}_c \left [l (f_n(\phi_n(X^n, \rho)))  \right] \right ] \notag \\
    &\leq \frac{\ln(2)}{n} \mathbb{E}_p \left [ \mathbb{E}_c \left [\lfloor \log(i_{{J(X^n)}}) \rfloor + 2 \lfloor \log \left( \lfloor \log(i_{{J(X^n)}}) \rfloor + 1 \right) \rfloor + 1  \right] \right ]  \label{eveach}\\
    &\leq \frac{\ln(2)}{n} + \frac{\ln(2)}{n} \mathbb{E}_p \left [ \mathbb{E}_c \left [ \log(i_{{J(X^n)}})  + 2  \log \left(  \log(i_{{J(X^n)}})  + 1 \right)    \right] \right ] \notag\\
    &\leq \frac{\ln(2)}{n} + \frac{1}{n} \mathbb{E}_p \left [   \ln(\mathbb{E}_c \left [ i_{{J(X^n)}} \right] )  + 2  \ln \left(  \log( \mathbb{E}_c \left [ i_{{J(X^n)}} \right])  + 1 \right)     \right ] \label{evll}
\end{align}
In equality $(\ref{eveach})$, we used $(\ref{ftvgg09})$. In inequality $(\ref{evll})$, we used Jensen's inequality.  

For convenience, we evaluate the last two terms in $(\ref{evll})$ separately and then add them together later. Using the same definition of the constant $V_1$ in $(\ref{u1u2consts})$, we have from $(\ref{aleenanida})$ that 
\begin{align}
    \frac{1}{n}\ln  \mathbb{E}_c[i_{J(X^n)}] &\leq  R(t(X^n),d,\rho) + \frac{K - 3/4}{2}\frac{ \ln n}{n}  + \frac{2\min\left(\ln(K), \ln(J) \right)}{d\, n^{5/8}} \rho_{\max}  + \frac{V_1}{n} \label{nidah201}
\end{align}
and
\begin{align}
     &\frac{2}{n}\ln \left ( \log \mathbb{E}_c[i_{J(X^n)}] + 1 \right) \notag\\ 
     &\leq \frac{2}{n} \ln \left(\frac{1}{\ln 2} \left(n R(t(X^n),d,\rho) + \frac{K - 3/4}{2} \ln n  + 2\frac{\min \left(\ln(K), \ln(J)\right)}{d} \rho_{\max} n^{3/8} + V_1 + \ln(2) \right) \right) \notag\\
    &= \frac{2}{n} \ln \left(n R(t(X^n),d,\rho) + \frac{K - 3/4}{2} \ln n  + 2\frac{\min \left(\ln(K), \ln(J)\right)}{d} \rho_{\max} n^{3/8} + V_1 + \ln(2)\right) - \frac{2\ln \ln(2)}{n}.  \label{nidah6}
\end{align}
We now use the bounds in $(\ref{nidah201})$ and $(\ref{nidah6})$ in $(\ref{evll})$. Since $R(t,d,\rho) \leq \ln(K)$, it is easy to see that there exist an integer $\mathcal{Z}$ and a constant $\mathcal{G}$ such that for $n \geq \mathcal{Z}$, we have 
\begin{align}
     &\frac{\ln(2)}{n}\mathbb{E}_p \left [ \mathbb{E}_c \left [l (f_n(\phi_n(X^n, \rho)))  \right] \right ] \notag \\
     &\leq  \mathbb{E}_p \left [  R(T,d,\rho) \right]  + \frac{K + 13/4}{2}\frac{ \ln n}{n}  + \frac{2 \min\left(\ln(K), \ln(J)\right)}{d\, n^{5/8}} \rho_{\max}  + \mathcal{G} \frac{\ln \ln n}{n}  . \label{balahi}
\end{align}
Note that $\mathcal{Z}$ and $\mathcal{G}$ are independent of $t$ and $\rho$.

So far, we have constructed random non-prefix and prefix $d'$-semifaithful codes with expected rates upper bounded by $(\ref{subskarisko})$ and $(\ref{balahi})$, respectively. We now use post-correction to obtain $d$-semifaithful codes. Let $y^n$ be the reconstruction sequence corresponding to $x^n$ such that
\begin{align*}
 \rho_n(x^n, y^n) &= \frac{1}{n} \sum_{i=1}^n \rho(x_i, y_i) \leq d'.
\end{align*}
For any integer $M < n$, let $\{l_m \}_{m=1}^M$ be a sequence indexing the $M$ highest distortion letter pairs $(x_i, y_i)$, i.e., 
\begin{align*}
    \rho(x_{l_1}, y_{l_1}) \geq \rho(x_{l_2}, y_{l_2}) \geq  \cdots \geq \rho(x_{l_M}, y_{l_M}) \geq \rho(x_i, y_i) \,\,\,\, \forall i \notin \{l_m \}_{m=1}^M.
\end{align*}
We replace $$M = \left \lceil \frac{2 \rho_{\max}}{d} n^{3/8} \right \rceil $$
symbols, namely $y_{l_1}, y_{l_2}, \ldots , y_{l_M}$, in $y^n$ to obtain another sequence $\hat{y}^n$ which satisfies $\rho_n(x^n, \hat{y}^n) \leq d$. From the assumption in $(\ref{dist_assump})$, we can choose the replacement symbols $\hat{y}_{l_1}, \hat{y}_{l_2}, \ldots , \hat{y}_{l_M}$ such that $\rho(x_{l_m}, \hat{y}_{l_m} ) = 0$ for all $1 \leq m \leq M$. The post-corrected sequence $\hat{y}^n$ reproduces $x^n$ within distortion $d$ since 
\begin{align*}
    &\rho_n(x^n, \hat{y}^n) \\
    &= \frac{1}{n} \sum_{i \in \{l_m \}_{m=1}^M } \rho(x_i, y_i) + \frac{1}{n} \sum_{i \notin \{l_m \}_{m=1}^M } \rho(x_i, y_i)\\
    &= \frac{1}{n} \sum_{i \notin \{l_m \}_{m=1}^M } \rho(x_i, y_i)\\
    &= \frac{n - M}{n} \frac{1}{n - M} \sum_{i \notin \{l_m \}_{m=1}^M } \rho(x_i, y_i)\\
    &\leq \frac{n - M}{n} \left(d + \frac{2\rho_{\max}}{n^{5/8}} \right)\\
    &\leq d.
\end{align*}
The encoder will need at most
\begin{align}
     M \left( \log(n) + \min\left(\log(K), \log(J) \right) + 2 \right)  \label{opwxz}
\end{align}
bits using fixed-length encoding to convey this "post-correction" information. Note that for the non-prefix $d'$-semifaithful coding scheme from before, this "post-correction" information needs to be sent before the variable length encoding of the reconstruction sequence. The rate increment from the post-correction bits in $(\ref{opwxz})$ is upper bounded by 
\begin{align}
    &\,\frac{\ln(2)}{n} \left (  M \left( \log(n) + \min\left(\log(K), \log(J) \right) + 2 \right) \right ) \notag \\
    &\leq  \frac{\ln(n)}{n} + \ln(n) \frac{2 \rho_{\max}}{d \, n^{5/8}}  + \frac{\min\left(\ln(K), \ln(J) \right)}{n} + \min\left(\ln(K), \ln(J) \right) \frac{2 \rho_{\max}}{d \, n^{5/8}} + \frac{\ln(4)}{n} + \frac{4 \rho_{\max}}{d \, n^{5/8}}  \ln(2). \label{rate_incr}
\end{align}
Adding $(\ref{rate_incr})$ to $(\ref{subskarisko})$ and $(\ref{balahi})$ establishes the results of Theorems \ref{upper_bound} and \ref{upper_bound2}, respectively.

\section{Proof of Theorem \ref{upper_bound3}  \label{pfthms3}}

Fix $d > 0$. In the proof of Theorem \ref{upper_bound} (see $(\ref{subskarisko})$), we showed that for sufficiently large $n$, there exists a universal random, non-prefix $d'$-semifaithful code $\tilde{C}_n = (\phi_n, f_n, g_n)$ satisfying  
\begin{align*}
    &\sup_{p \in \mathcal{P}(A), \rho \in \mathcal{D}} \left [ \frac{1}{n} \mathbb{E}_p \left [ \ln(2) \mathbb{E}_c\left [l (f_n(\phi_n(X^n, \rho))) \right ] \right ] - \mathbb{E}_p \left [R(T,d,\rho) \right ] \right ]   \\
    & \,\,\,\,\,\,\,\,\,\, \leq  \frac{K - 3/4}{2} \frac{\ln n}{n}  + \frac{2 \min\left(\ln(K), \ln(J) \right)}{d \, n^{5/8}} \rho_{\max}  + \frac{V_1 + \ln(2)}{n} ,
\end{align*}
where $V_1$ is given in $(\ref{u1u2consts})$ and $d' = d + 2 \rho_{\max} / n^{5/8}$. 

We will now use uppercase $\Phi_n$ to distinguish the random $d'$-quantizer from a deterministic one for which we will use lowercase $\phi_n$.  

For any sequence $x^n$, we have 
\begin{align*}
    &l(f_n(\Phi_n(x^n,\rho))) \leq  1 + \log i_{J(x^n)}
\end{align*}
from $(\ref{ftvgg})$. If $x^n \in T^n_A(t)$, then we know (see $(\ref{Expec_i2})$ and $(\ref{aleenanida})$) that $i_{J(x^n)}$ is a geometric random variable with parameter 
\begin{align}
    g_{t,\rho} &\geq \exp \left(-n R(t,d,\rho) - \frac{K - 3/4}{2} \ln n  - 2\frac{\min \left(\ln(K), \ln(J) \right)}{d} \rho_{\max} n^{3/8} - V_1 \right).
    \label{lbboundgeo}
\end{align}  
Let 
\begin{align}
    \gamma_n \triangleq 1 + \frac{\ln(J^2 K^2 + J - 1)}{ \ln \ln n}.
    \label{gammandef}
\end{align}
Denoting the probability law associated with the random code $\tilde{C}_n$ by $\mathbb{P}_c(\cdot)$, we have  
\begin{align}
    &\,\,\,\,\mathbb{P}_c \left( l(f_n(\Phi_n(x^n,\rho))) > \frac{1}{\ln(2)} \left( n R(t, d, \rho) + \ln(2) + \frac{K - 3/4}{2} \ln n + \frac{2 n^{3/8} \min \left(\ln(K), \ln(J)\right)}{d} \rho_{\max} + V_1 + \gamma_n \ln \ln n  \right)   \right) \notag \\
    &\leq \mathbb{P}_c \left( 1 + \log i_{J(x^n)} > \frac{1}{\ln(2)} \left( n R(t, d, \rho) + \ln(2) + \frac{K - 3/4}{2} \ln n + \frac{2 n^{3/8} \min\left(\ln(K), \ln(J)\right)}{d} \rho_{\max} + V_1 + \gamma_n \ln \ln n  \right)   \right) \notag \\
    &= \mathbb{P}_c \left(  i_{J(x^n)} >  \exp \left(  n R(t, d, \rho)  + \frac{K - 3/4}{2} \ln n + \frac{2 n^{3/8}\min\left( \ln(K), \ln(J)\right)}{d} \rho_{\max} + V_1 + \gamma_n \ln \ln n \right) \right) \label{thm4maine}   \\
     &\leq (1 - g_{t,\rho})^{\exp \left( n\,R(t,d,\rho) + \frac{2 \rho_{\max} \min\left(\ln(K), \ln(J)\right)}{d} n^{3/8} + \frac{K - 3/4}{2} \ln n + V_1 + \gamma_n \ln \ln n \right) - 1} \notag \\
     &\leq \left (1 - \exp \left(-n R(t,d,\rho) - \frac{K - 3/4}{2} \ln n - V_1  - \mbox{} \right . \right. \notag \\
     &\,\,\,\,\,\,\,\,\,\,\,\,\,\,\,\,\,\,\,\,\,\,\,\,\,\,\,\,\,\, \left. \left . 2\frac{\min\left(\ln(K), \ln(J) \right)}{d} \rho_{\max} n^{3/8}  \right) \right )^{\exp \left( n\,R(t,d,\rho) + \frac{2 \rho_{\max} \min\left(\ln(K), \ln(J)\right)}{d}n^{3/8} + \frac{K - 3/4}{2} \ln n + V_1 + \gamma_n \ln \ln n \right) - 1} \notag \\
     &\stackrel{(a)}{\leq} e^{-e^{  \gamma_n \ln \ln n } + \exp \left(-n R(t,d,\rho) - \frac{K - 3/4}{2} \ln n - V_1  - 2\frac{\min\left(\ln(K), \ln(J) \right)}{d} \rho_{\max} n^{3/8} \right)} \notag \\
     &\leq \frac{e}{n^{J^2 K^2 + J - 1}}, \label{kalaadeel}
\end{align}
where the inequality $(a)$ above uses the inequality $(1- x)^y \leq e^{-xy}$. Now if we let $X_t^n  \sim \text{Unif}(T^n_A(t))$ be a random sequence uniformly distributed over the type class $T^n_A(t)$, then  
\begin{align}
      &\,\,\,\,\mathbb{P}_{t,c} \left( l(f_n(\Phi_n(X_t^n,\rho))) > \frac{1}{\ln(2)} \left( n R(t, d, \rho) + \ln(2) + \frac{K - 3/4}{2} \ln n + \mbox{}  \right. \right. \notag \\
      & \,\,\,\,\,\,\,\,\,\,\,\,\,\,\,\,\,\,\,\,\,\,\,\,\,\,\,\,\,\,\,\,\,\,\,\,\,\,\,\,\,\, \left. \left . \frac{2 n^{3/8} \min \left(\ln(K), \ln(J)\right)}{d} \rho_{\max} + V_1 + \gamma_n \ln \ln n  \right)   \right) \leq \frac{e}{n^{J^2 K^2 + J - 1}}, \label{thm4koot}
\end{align}
where the last inequality above follows from $(\ref{kalaadeel})$. 

We used $\mathbb{P}_t$ to denote the probability law associated with the random sequence $X_t^n \sim \text{Unif}(T^n_A(t))$. We next use $\mathbb{P}_T$ to denote the probability law associated with the collection of random sequences $\{X_t^n : t \in \mathcal{P}_n(A) \}$.
Taking a union bound over all types gives
\begin{align}
      &\,\,\,\,\mathbb{P}_{T,c} \left(\bigcup_{t \in \mathcal{P}_n(A)} \left \{ l(f_n(\Phi_n(X_t^n,\rho))) > \frac{1}{\ln(2)} \left( n R(t, d, \rho) + \ln(2) + \frac{K - 3/4}{2} \ln n + \mbox{} \right . \right. \right. \notag \\
      & \left . \left . \left . \,\,\,\,\,\,\,\,\,\,\,\,\,\,\,\,\,\,\,\,\,\,\,\,\,\,\,\,\,\,\,\,\,\, \frac{2 n^{3/8}\min\left( \ln(K), \ln(J)\right)}{d} \rho_{\max} + V_1 + \gamma_n \ln \ln n  \right) \right \}  \right) \leq (n+1)^{J-1} \frac{e}{n^{J^2 K^2 + J - 1}} \to 0 \text{ as } n \to \infty. \label{naiveunibhai}
\end{align}
The above result proves that for large enough $n$, we have with high probability that the length of binary encoding used by the random $d'$-semifaithful code $(\Phi_n, f_n, g_n)$ to encode a randomly chosen sequence $X_t^n$ from any type class $t$ does not exceed
\begin{align}
    \frac{1}{\ln(2)} \left( n R(t, d, \rho) + \ln(2) + \frac{K - 3/4}{2} \ln n + \frac{2 n^{3/8} \min\left(\ln(K), \ln(J)\right)}{d} \rho_{\max} + V_1 + \gamma_n \ln \ln n  \right).
\end{align}
As we will show later, this result implies the existence of a deterministic $d'$-semifaithful code which has uniformly good performance over all types. However, our goal is to prove the existence of a deterministic $d'$-semifaithful code in the universal distortion framework, i.e., one which has uniformly good performance over all types as well as all distortion measures. Since the set of distortion measures $\mathcal{D}$ is uncountably infinite, a naive union bound similar to $(\ref{naiveunibhai})$ fails. Instead, we invoke the fact that the space of distortion measures $\mathcal{D}$ can be partitioned into a polynomial number of equivalence classes. For full explanation, we refer the reader to \cite[Theorem 1]{mahmood2021lossy}. Here, we merely state and use the following proposition which is a straightforward corollary of  \cite[Proposition 1]{mahmood2021lossy}:  
\begin{proposition} 
For a given blocklength $n$ and distortion level $d$, there are $ \xi \leq (n+1)^{J^2 K^2 - 1} + 1$ equivalence classes of $\mathcal{D}$, denoted by $[\mathcal{D}]_{\rho_1}$, $[\mathcal{D}]_{\rho_2}$, \ldots, $[\mathcal{D}]_{\rho_{\xi}}$, where $\rho_1, \rho_2, \ldots, \rho_\xi$ are arbitrarily chosen representative distortion measures. A $d$-semifaithful code $\tilde{C}_n$ with respect to a distortion measure $\rho$ is also $d$-semifaithful with respect to all distortion measures $\rho' \in [\mathcal{D}]_{\rho}$ in the same equivalence class.  
\label{propo2}
\end{proposition}
We will make the choice of representative distortion measures $\rho_1, \rho_2, \ldots, \rho_{\xi} $ be a function of the type $t$. For every type $t$ and every equivalence class $[\mathcal{D}]_{\rho_i}$, we choose the representative distortion measure $\rho_i^t \in [\mathcal{D}]_{\rho_i}$ which satisfies 
\begin{align}
    R(t,d,\rho_i^t) \leq R(t,d,\rho) \label{phoit}
\end{align}
for all $\rho \in [\mathcal{D}]_{ \rho_i  }$. Henceforth, the representative distortion measures, chosen differently for each type, will be $\rho_1^t, \rho_2^t, \ldots, \rho^t_{\xi}$.

Now applying the union bound over the types and distortion measures gives 
\begin{align}
      &\,\,\,\,\mathbb{P}_{T,c} \left( \bigcup_{i=1}^\xi \bigcup_{t \in \mathcal{P}_n(A)} \left \{ l(f_n(\Phi_n(X_t^n,\rho_i^t))) > \frac{1}{\ln(2)} \left( n R(t, d, \rho_i^t) + \ln(2) + \frac{K - 3/4}{2} \ln n + V_1 + \mbox{} \right . \right. \right. \notag \\
      & \left . \left . \left . \,\,\,\,\,\,\,\,\,\,\,\,\,\,\,\,\,\,\,\,\,\, \frac{2 n^{3/8} \min\left(\ln(K), \ln(J)\right)}{d} \rho_{\max} + \gamma_n \ln \ln n  \right) \right \}  \right) \leq \left( (n+1)^{J^2K^2 - 1} + 1 \right)(n+1)^{J-1} \frac{e}{n^{J^2 K^2 +J - 1}} \to 0 \text{ as } n \to \infty.
      \label{nidahamaya}
\end{align}
Also note that 
\begin{align}
    &\,\,\,\,\,\,\mathbb{P}_{T,c} \left( \bigcup_{i=1}^\xi \bigcup_{t \in \mathcal{P}_n(A)} \left \{ l(f_n(\Phi_n(X_t^n,\rho_i^t))) > \frac{1}{\ln(2)} \left( n R(t, d, \rho_i^t) + \ln(2) + \frac{K - 3/4}{2} \ln n + \mbox{} \right . \right. \right. \notag \\
      & \left . \left . \left . \,\,\,\,\,\,\,\,\,\,\,\,\,\,\,\,\,\,\,\,\,\, \frac{2 n^{3/8} \min\left(\ln(K), \ln(J)\right)}{d} \rho_{\max} + V_1 + \gamma_n \ln \ln n  \right) \right \} \right)  \label{repeatanalysis}\\
    &= \mathbb{E}_{T,c} \left [ \mathds{1}\left( \bigcup_{i=1}^\xi \bigcup_{t \in \mathcal{P}_n(A)} \left \{ l(f_n(\Phi_n(X_t^n,\rho_i^t))) > \frac{1}{\ln(2)} \left( n R(t, d, \rho_i^t) + \ln(2) + \frac{K - 3/4}{2} \ln n + \mbox{} \right . \right. \right. \right. \notag \\
      & \left . \left . \left . \left . \,\,\,\,\,\,\,\,\,\,\,\,\,\,\,\,\,\,\,\,\,\, \frac{2 n^{3/8} \min\left( \ln(K),\ln(J) \right)}{d} \rho_{\max} + V_1 + \gamma_n \ln \ln n  \right) \right \} \right) \right ] \notag \\
    &= \mathbb{E}_{c} \left [ \mathbb{E}_T  \left [ \mathds{1}\left( \bigcup_{i=1}^\xi \bigcup_{t \in \mathcal{P}_n(A)} \left \{ l(f_n(\Phi_n(X_t^n,\rho_i^t))) > \frac{1}{\ln(2)} \left( n R(t, d, \rho_i^t) + \ln(2) + \frac{K - 3/4}{2} \ln n + \mbox{} \right . \right. \right. \right. \right.  \notag \\
      & \left . \left . \left .  \left . \left . \,\,\,\,\,\,\,\,\,\,\,\,\,\,\,\,\,\,\,\,\,\, \frac{2 n^{3/8} \min\left(\ln(K),\ln(J)\right)}{d} \rho_{\max} + V_1 + \gamma_n \ln \ln n  \right) \right \} \right)  \Bigg | \Phi_n \right] \right ]  \notag \\
    &\leq \left( (n+1)^{J^2K^2 - 1} + 1 \right) \left( (n+1)^{J-1} \right) \frac{e}{n^{J^2 K^2 + J - 1}}. \label{qwjewq}
\end{align}
The above inequality implies that that there exists a deterministic $d'$-quantizer $\phi_n$ such that 
\begin{align}
     & \mathbb{E}_T  \left [ \mathds{1}\left( \bigcup_{i=1}^\xi \bigcup_{t \in \mathcal{P}_n(A)} \left \{ l(f_n(\Phi_n(X_t^n,\rho_i^t))) > \frac{1}{\ln(2)} \left( n R(t, d, \rho_i^t) + \ln(2) + \frac{K - 3/4}{2} \ln n + \mbox{} \right . \right. \right. \right.   \notag \\
      &  \left . \left .  \left . \left . \,\,\,\,\,\,\,\,\,\,\,\,\,\,\,\,\,\,\,\,\,\, \frac{2 n^{3/8} \min\left(\ln(K), \ln(J)\right)}{d} \rho_{\max} + V_1 + \gamma_n \ln \ln n  \right) \right \} \right)  \Bigg | \Phi_n = \phi_n \right ] \notag \\
      &\,\,\,\,\,\,\,\,\,\,\,\,\,\,\,\,\,\,\,\,\,\,\leq  \left( (n+1)^{J^2K^2-1} + 1 \right) \left( (n+1)^{J-1} \right) \frac{e}{n^{J^2 K^2 + J - 1}}. \label{nxmnvm}
\end{align}
This in turn implies that 

\begin{align}
    &\,\,\,\,\,\,\,\mathbb{P}_T\left( \bigcup_{i=1}^\xi \bigcup_{t \in \mathcal{P}_n(A)} \left \{ l(f_n(\Phi_n(X_t^n,\rho_i^t))) > \frac{1}{\ln(2)} \left( n R(t, d, \rho_i^t) + \ln(2) + \frac{K - 3/4}{2} \ln n + \mbox{} \right . \right. \right.    \notag \\
      &   \left .  \left . \left . \,\,\,\,\,\,\,\,\,\,\,\,\,\,\,\,\,\,\,\,\,\, \frac{2 n^{3/8} \min\left(\ln(K), \ln(J)\right)}{d} \rho_{\max} + V_1 + \gamma_n \ln \ln n  \right) \right \} \Big | \Phi_n = \phi_n \right) \notag  \\
    &\stackrel{(a)}{=}  
    \mathbb{P}_T\left( \bigcup_{i=1}^\xi \bigcup_{t \in \mathcal{P}_n(A)} \left \{ l(f_n(\phi_n(X_t^N,\rho_i^t))) > \frac{1}{\ln(2)} \left( n R(t, d, \rho_i^t) + \ln(2) + \frac{K - 3/4}{2} \ln n + \mbox{} \right . \right. \right.    \notag \\
      &   \left .  \left . \left . \,\,\,\,\,\,\,\,\,\,\,\,\,\,\,\,\,\,\,\,\,\, \frac{2 n^{3/8} \min \left(\ln(K) , \ln(J)\right)}{d} \rho_{\max} + V_1 + \gamma_n \ln \ln n  \right) \right \}   \right)  \notag  \\
    &\leq \left( (n+1)^{J^2K^2-1} + 1 \right) \left( (n+1)^{J-1} \right) \frac{e}{n^{J^2 K^2 + J - 1}} . \label{nsfka}
\end{align}
Equality $(a)$ above follows from the independence of the random code $\Phi_n$ and the random source sequence $X_t^n$. 

Now we have a deterministic $d'$-semifaithful code $(\phi_n, f_n, g_n)$ which, with high probability, has uniformly good performance in encoding a random sequence $X_t^n \sim \text{Unif}(T^n_A(t))$ for any type $t$ and any of the chosen representative distortion measures $\rho_1^t, \rho_2^t, \ldots, \rho_\xi^t$. 

But we are interested in encoding an i.i.d. sequence $X^n \sim p^n$ with respect to an arbitrary distortion measure. To accomplish this, we can use the code $(\phi_n, f_n, g_n)$ to construct another code $(\tilde{\phi}_n, \tilde{f}_n, \tilde{g}_n)$ as described next. For any given sequence $x^n$ and input distortion measure $\rho$, let $t = t(x^n)$ be the type and let $\rho \in [\mathcal{D}]_{\rho_i^t}$ for some $1 \leq i \leq \xi$. The new code $(\tilde{\phi}_n, \tilde{f}_n, \tilde{g}_n)$ uses $(\phi_n, f_n, g_n)$ for encoding if 
\begin{align*}
    l(f_n(\phi_n(x^n,\rho_i^t))) &\leq \frac{1}{\ln(2)} \left( n R(t, d, \rho_i^t) + \ln(2) + \frac{K - 3/4}{2} \ln n + \right . \\
    &\,\,\,\,\,\,\,\,\left . \frac{2 n^{3/8} \min\left(\ln(K), \ln(J)\right)}{d} \rho_{\max} + V_1 + \gamma_n \ln \ln n  \right) \,\,\,\,\,\,\,\,\,\,\,\,\,\, \left( \text{ Case } 1 \,\right)
\end{align*}
and otherwise (in Case $2$), searches the entire $B^n$ space to send the index of a $y^n$ satisfying $(\rho_i^t)_n(x^n, y^n) \leq d'$. In both cases, the distortion measure $\rho_i^t$ is used because of the equivalence $\rho_i^t \sim \rho$ from Proposition \ref{propo2}. The two cases can be indicated to the decoder using a flag bit $F$, where $F = 1$ in Case $1$ and $F = 0$ in Case $2$. 

Hence, for any source distribution $p \in \mathcal{P}(A)$ and for any $\rho \in \mathcal{D}$, the expected rate in nats of $(\tilde{\phi}_n, \tilde{f}_n, \tilde{g}_n)$ is  
\begin{align}
    &\frac{\ln(2)}{n}\mathbb{E}_p \left [l(\tilde{f}_n(\tilde{\phi}_n(X^n,\rho))) \right ] \notag\\
    &\stackrel{(a)}{=} \frac{\ln(2)}{n} \sum_{t \in \mathcal{P}_n(A)} p^n(T^n_A(t)) \mathbb{E}_t \left [l(\tilde{f}_n(\tilde{\phi}_n(X_t^n,\rho))) \right ] \notag \\
    &\leq \sum_{t \in \mathcal{P}_n(A)} p^n(T^n_A(t)) \left [  R(t, d, \rho_i^t)  + \frac{K - 3/4}{2} \frac{\ln n}{n} + \frac{2  \min\left(\ln(K), \ln(J)\right)}{d \, n^{5/8}} \rho_{\max} + \frac{V_1 + \ln(2)}{n} + \gamma_n \frac{\ln \ln n}{n}   + \mbox{}  \right. \notag \\
    &\,\,\,\,\,\,\,\,\,\,\,\,\,\,\,\,\,\,\left . \left( (n+1)^{J^2K^2-1} + 1 \right) \left( (n+1)^{J-1} \right) \frac{e}{n^{J^2 K^2 + J - 1}} \left( \ln K + \frac{\ln(2)}{n} \right) + \frac{\ln(2)}{n}  \right ]  \notag \\
    &=   \mathbb{E}_p \left [ R(T, d, \rho_i^T) \right ]  + \frac{K - 3/4}{2} \frac{\ln n}{n} + \frac{2 \min\left( \ln(K),\ln(J)\right)}{d \, n^{5/8}} \rho_{\max} + \frac{V_1 + \ln(2)}{n} + \gamma_n \frac{\ln \ln n }{n}  + \notag \\
    & \,\,\,\,\,\,\,\,\,\,\,\,\,\,\,\,\,\,\,\,\,\,\,\,\,\,\,\,\,\,\,\,\,\,\,\,\,\,\left( (n+1)^{J^2K^2-1} + 1 \right) \left( (n+1)^{J-1} \right) \frac{e}{n^{J^2 K^2 + J - 1}} \left( \ln K + \frac{\ln(2)}{n} \right) + \frac{\ln(2)}{n} \notag \\
    &\stackrel{(b)}{\leq}   \mathbb{E}_p \left [ R(T, d, \rho) \right ]  + \frac{K - 3/4}{2} \frac{\ln n}{n} + \frac{2  \min \left( \ln(K), \ln(J) \right)}{d \, n^{5/8}} \rho_{\max} + \frac{V_1 + \ln(2)}{n} + \gamma_n \frac{\ln \ln n}{n}   + \notag \\
    & \,\,\,\,\,\,\,\,\,\,\,\,\,\,\,\,\,\,\,\,\,\,\,\,\,\,\,\,\,\,\,\,\,\,\,\,\,\,\left( (n+1)^{J^2K^2-1} + 1 \right) \left( (n+1)^{J-1} \right) \frac{e}{n^{J^2 K^2 + J - 1}} \left( \ln K + \frac{\ln(2)}{n} \right) + \frac{\ln(2)}{n}. \label{joopapee}
\end{align}
In equality $(a)$, we use the fact that conditioned on the type, $X^n$ is distributed uniformly over the type class $t$, which we denote by writing $X_t^n$. Equality $(b)$ follows from $(\ref{phoit})$.

Finally, we use post-correction to make the code $d$-semifaithful. This post-correction was described in the proof of Theorem 1 (Section \ref{thms12}), specifically $(\ref{opwxz})$ and $(\ref{rate_incr})$. By adding the rate increment from post-correction in $(\ref{rate_incr})$ to the expression in $(\ref{joopapee})$, the expected rate of the overall code is upper bounded by 
\begin{align}
    &\, \mathbb{E}_p \left [ R(T, d, \rho) \right ] + \frac{2 \rho_{\max} \ln(n)}{d \, n^{5/8}}   + \frac{4 \rho_{\max}\left(  \min \left( \ln(K), \ln(J) \right)  + \ln(2)\right)}{d \, n^{5/8}} +  \frac{K + 5/4}{2} \frac{\ln n}{n}   + \gamma_n \frac{\ln \ln n}{n}   + \notag \\
    & \,\,\,\,\,\,\,\,\,\,\,\,\,\,\,\,\,\,\,\,\,\,\,\,\,\,\,\,\,\,\,\,\,\,\,\,\,\,\left( (n+1)^{J^2K^2-1} + 1 \right) \left( (n+1)^{J-1} \right) \frac{e}{n^{J^2 K^2 + J - 1}} \left( \ln K + \frac{\ln(2)}{n} \right)      +  \frac{V_1 + \ln(16) + \min\left(\ln(K), \ln(J) \right)}{n}. \label{namayaw}
\end{align}
The above bounds holds uniformly over 
$\mathcal{P}(A) \times \mathcal{D}$ for sufficiently large $n$. This finishes the proof of Theorem \ref{upper_bound3}.

\section{Proof of Theorem \ref{upper_bound4} \label{thm4proof}}

The proof of Theorem \ref{upper_bound4} is similar to the proof of Theorem \ref{upper_bound3}. 
Fix $d > 0$. In the proof of Theorem \ref{upper_bound2} (see $(\ref{balahi})$), we showed that for sufficiently large $n$, there exists a universal random, prefix $d'$-semifaithful code $\tilde{C}_n = (\phi_n, f_n, g_n)$ satisfying  
\begin{align*}
    &\sup_{p \in \mathcal{P}(A), \rho \in \mathcal{D}} \left [ \frac{1}{n} \mathbb{E}_p \left [ \ln(2) \mathbb{E}_c\left [l (f_n(\phi_n(X^n, \rho))) \right ] \right ] - \mathbb{E}_p \left [R(T,d,\rho) \right ] \right ]   \\
    & \,\,\,\,\,\,\,\,\,\, \leq \frac{K + 13/4}{2}\frac{ \ln n}{n}  + \frac{2 \min\left(\ln(K), \ln(J)\right)}{d\, n^{5/8}} \rho_{\max}  + \mathcal{G} \frac{\ln \ln n}{n},
\end{align*}
where $\mathcal{G}$ is a constant depending only on $J, K, \rho_{\max}$ and $d$, and $d' = d + 2 \rho_{\max} / n^{5/8}$. 

We will now use uppercase $\Phi_N$ to distinguish the random $d'$-quantizer from a deterministic one for which we will use lowercase $\phi_N$.  

For any sequence $x^n$, we have 
\begin{align}
    &l(f_n(\Phi_n(x^n,\rho))) \leq \lfloor \log(i_{{J(x^n)}}) \rfloor + 2 \lfloor \log \left( \lfloor \log(i_{{J(x^n)}}) \rfloor + 1 \right) \rfloor + 1 \label{collll}
\end{align}
from $(\ref{ftvgg09})$. If $x^n \in T^n_A(t)$, then we know (see $(\ref{Expec_i2})$ and $(\ref{aleenanida})$) that $i_{J(x^n)}$ is a geometric random variable with parameter 
\begin{align}
    g_{t,\rho} &\geq \exp \left(-n R(t,d,\rho) - \frac{K - 3/4}{2} \ln n  - 2\frac{\min\left(\ln(K), \ln(J)\right)}{d} \rho_{\max} n^{3/8} - V_1 \right).
    \label{lbboundgeo5}
\end{align}  
Let $\gamma_n$ be as defined in $(\ref{gammandef})$. Then as shown in $(\ref{thm4maine})$$-$$(\ref{kalaadeel})$, we have for any $x^n \in T^n_A(t)$, 
\begin{align}
    &= \mathbb{P}_c \left(  i_{J(x^n)} >  \exp \left(  n R(t, d, \rho)  + \frac{K - 3/4}{2} \ln n + \frac{2 n^{3/8} \min\left(\ln(K), \ln(J) \right)}{d} \rho_{\max} + V_1 + \gamma_n \ln \ln n \right) \right) \notag    \\
     &\leq \frac{e}{n^{J^2 K^2 + J - 1}}. \label{kalaadeel2}
\end{align}
In view of $(\ref{collll})$ and $(\ref{kalaadeel2})$, we have that with probability at least $1 - e/n^{J^2 K^2 + J - 1}$, 
\begin{align*}
    &l(f_n(\Phi_n(x^n,\rho))) \\
    &\leq \log \left [ \exp \left(  n R(t, d, \rho)  + \frac{K - 3/4}{2} \ln n + \frac{2 n^{3/8} \min\left( \ln(K), \ln(J) \right)}{d} \rho_{\max} + V_1 + \gamma_n \ln \ln n \right) \right ]  \\
    & + 2\log\left( \log \left [ \exp \left(  n R(t, d, \rho)  + \frac{K - 3/4}{2} \ln n + \frac{2 n^{3/8} \min\left(\ln(K),\ln(J) \right)}{d} \rho_{\max} + V_1 + \gamma_n \ln \ln n \right) \right]  + 1 \right) + 1.
\end{align*}
Since $R(t,d,\rho) \leq \ln(K)$ and $V_1$ and $\gamma$ are independent of $t = t(x^n)$ and $\rho$, it is easy to see that there exist an integer $\mathcal{Z}$ and a constant $\mathcal{G}$ such that for $n \geq \mathcal{Z}$, we have 
\begin{align*}
    &l(f_n(\Phi_n(x^n,\rho)))\\
    &\leq \frac{1}{\ln(2)} \left( n R(t,d,\rho) + \frac{K + 13/4}{2} \ln n + \frac{2 n^{3/8} \min\left( \ln(K), \ln(J) \right)}{d} \rho_{\max}  + \mathcal{G} \ln \ln n \right)
\end{align*}
with probability at least $1 - e/n^{J^2 K^2 + J - 1}$. Note that $\mathcal{Z}$ and $\mathcal{G}$ depend on $J$, $K$ and $\rho_{\max}$ but do not depend on $t$ and $\rho$. Hence, for sufficiently large $n$, we have for any $x^n \in T^n_A(t)$, 
\begin{align}
    &\mathbb{P}_c \left ( l(f_n(\Phi_n(x^n,\rho))) > \frac{1}{\ln(2)} \left( n R(t,d,\rho) + \frac{K + 13/4}{2} \ln n + \mbox{} \right . \right. \notag \\ 
    &\,\,\,\,\,\,\,\,\,\,\,\,\,\,\,\,\,\,\,\,\,\,\,\,\,\,\,\,\,\,\left . \left . \frac{2 n^{3/8} \min\left( \ln(K), \ln(J) \right)}{d} \rho_{\max} + \mathcal{G} \ln \ln n \right) \right) \leq \frac{e}{n^{J^2 K^2 + J - 1}}. \label{hjdhfsk}
\end{align}
Now if we let $X_t^n  \sim \text{Unif}(T^n_A(t))$ be a random sequence uniformly distributed over the type class $T^n_A(t)$, then 
\begin{align}
      &\,\,\,\,\mathbb{P}_{t,c} \left( l(f_n(\Phi_n(X^n_t,\rho))) > \frac{1}{\ln(2)} \left( n R(t,d,\rho) + \frac{K + 13/4}{2} \ln n + \mbox{} \right. \right. \notag \\
      &\,\,\,\,\,\,\,\,\,\,\,\,\,\,\,\,\,\,\,\,\,\,\,\,\,\,\,\,\,\,\,\,\,\,\,\,\,\,\,\,\,\,\,\,\,\, \left. \left.  \frac{2 n^{3/8} \min \left( \ln(K), \ln(J) \right)}{d} \rho_{\max} + \mathcal{G} \ln \ln n \right)   \right) \leq \frac{e}{n^{J^2 K^2 + J - 1}},
\end{align}
where the inequality above follows from $(\ref{hjdhfsk})$. Then similar to the proof of Theorem \ref{upper_bound3} (see Proposition \ref{propo2}), applying a union bound over the types and the specially chosen representative distortion measures from their respective equivalence classes (see $(\ref{phoit})$) gives 
\begin{align}
      &\,\,\,\,\mathbb{P}_{T,c} \left( \bigcup_{i=1}^\xi \bigcup_{t \in \mathcal{P}_n(A)} \left \{ l(f_n(\Phi_n(X_t^n,\rho_i^t))) > \frac{1}{\ln(2)} \left( n R(t, d, \rho_i^t) +  \frac{K + 13/4}{2} \ln n + \mbox{} \right . \right. \right. \notag \\
      & \left . \left . \left . \,\,\,\,\,\,\,\,\,\,\,\,\,\,\,\, \frac{2 n^{3/8} \min \left(\ln(K),\ln(J)\right)}{d} \rho_{\max} + \mathcal{G} \ln \ln n  \right) \right \}  \right) \leq \left( (n+1)^{J^2K^2 - 1} + 1 \right)(n+1)^{J-1} \frac{e}{n^{J^2 K^2 + J - 1}} \to 0 \text{ as } n \to \infty.
      \label{nidahamaya88}
\end{align}
Then following the same line of argument as in $(\ref{nidahamaya}), (\ref{repeatanalysis}), (\ref{qwjewq}), (\ref{nxmnvm})$ and $(\ref{nsfka})$, we have that for sufficiently large $n$, there exists a deterministic $d'$-quantizer $\phi_n$ satisfying 
\begin{align*}
    & \mathbb{P}_T\left( \bigcup_{i=1}^\xi \bigcup_{t \in \mathcal{P}_n(A)} \left \{ l(f_n(\phi_n(X_t^n,\rho_i^t))) > \frac{1}{\ln(2)} \left( n R(t, d, \rho_i^t) + \frac{K + 13/4}{2} \ln n + \mbox{} \right . \right. \right.    \notag \\
      &   \left .  \left . \left . \,\,\,\,\,\,\,\,\,\,\,\,\,\,\,\,\,\,\,\,\,\, \frac{2 n^{3/8}  \min\left(\ln(K),\ln(J)\right)}{d} \rho_{\max} + \mathcal{G} \ln \ln n  \right) \right \}   \right)  \notag  \\
    &\leq \left( (n+1)^{J^2K^2-1} + 1 \right) \left( (n+1)^{J-1} \right) \frac{e}{n^{J^2 K^2 + J - 1}} .
\end{align*}
Now we have a deterministic $d'$-semifaithful code $(\phi_n, f_n, g_n)$ which, with high probability, has uniformly good performance in encoding a random sequence $X_t^n \sim \text{Unif}(T^n_A(t))$ for any type $t$ and any of the chosen representative distortion measures $\rho_1^t, \rho_2^t, \ldots, \rho_\xi^t$. 

As in the proof of Theorem \ref{upper_bound3}, using this code $(\phi_n, f_n, g_n)$, we can construct another $d'$-semifaithful code $(\tilde{\phi}_n, \tilde{f}_n, \tilde{g}_n)$ which can encode an i.i.d. sequence $X^n \sim p^n$ with respect to an arbitrary distortion measure. For any given sequence $x^n$ and input distortion measure $\rho$, let $t = t(x^n)$ be the type and let $\rho \in [\mathcal{D}]_{\rho_i^t}$ for some $1 \leq i \leq \xi$. The new code $(\tilde{\phi}_n, \tilde{f}_n, \tilde{g}_n)$ uses $(\phi_n, f_n, g_n)$ for encoding if 
\begin{align*}
    l(f_n(\phi_n(x^n,\rho_i^t))) &\leq \frac{1}{\ln(2)} \left( n R(t, d, \rho_i^t) + \frac{K + 13/4}{2} \ln n + \right. \\
    &\,\,\,\,\,\,\,\,\,\,\,\,\,\,\left . \frac{2 n^{3/8} \min \left( \ln(K), \ln(J)\right)}{d} \rho_{\max} +  \mathcal{G} \ln \ln n  \right) \,\,\,\,\,\,\,\,\,\,\,\,\,\, \left( \text{ Case } 1 \right)
\end{align*}
and otherwise (in Case $2$), searches the entire $B^n$ space to send the index of a $y^n$ satisfying $(\rho_i^t)_n(x^n, y^n) \leq d'$. In both cases, the distortion measure $\rho_i^t$ is used because of the equivalence $\rho_i^t \sim \rho$ from Proposition \ref{propo2}. The two cases can be indicated to the decoder using a flag bit $F$, where $F = 1$ in Case $1$ and $F = 0$ in Case $2$. 

Hence, for any source distribution $p \in \mathcal{P}(A)$ and for any $\rho \in \mathcal{D}$, the expected rate in nats of  $(\tilde{\phi}_n, \tilde{f}_n, \tilde{g}_n)$ is  
\begin{align}
    &\frac{\ln(2)}{n}\mathbb{E}_p \left [l(\tilde{f}_n(\tilde{\phi}_n(X^n,\rho))) \right ] \notag\\
    &\stackrel{(a)}{=} \frac{\ln(2)}{n} \sum_{t \in \mathcal{P}_n(A)} p^n(T^n_A(t)) \mathbb{E}_t \left [l(\tilde{f}_n(\tilde{\phi}_n(X_t^n,\rho))) \right ] \notag \\
    &\leq \sum_{t \in \mathcal{P}_n(A)} p^n(T^n_A(t)) \left [   R(t, d, \rho_i^t)  + \frac{K + 13/4}{2}\frac{ \ln n}{n} + \frac{2 \min\left( \ln(K), \ln(J)\right)}{d \, n^{5/8}} \rho_{\max} +  \mathcal{G} \frac{\ln \ln n}{n}  + \mbox{}  \right. \notag \\
    &\,\,\,\,\,\,\,\,\,\,\,\,\,\,\,\,\,\,\left . \left( (n+1)^{J^2K^2-1} + 1 \right) \left( (n+1)^{J-1} \right) \frac{e}{n^{J^2 K^2 + J - 1}} \left( \ln K + \frac{\ln(2)}{n} \right) + \frac{\ln(2)}{n}  \right ]  \notag \\
    &=   \mathbb{E}_p \left [ R(T, d, \rho_i^T) \right ]  + \frac{K + 13/4}{2} \frac{\ln n}{n} + \frac{2 \min \left( \ln(K), \ln(J) \right)}{d \, n^{5/8}} \rho_{\max} +  \mathcal{G} \frac{\ln \ln n}{n}   + \notag \\
    & \,\,\,\,\,\,\,\,\,\,\,\,\,\,\,\,\,\,\,\,\,\,\,\,\,\,\,\,\,\,\,\,\,\,\,\,\,\,\left( (n+1)^{J^2K^2-1} + 1 \right) \left( (n+1)^{J-1} \right) \frac{e}{n^{J^2 K^2 + J - 1}} \left( \ln K + \frac{\ln(2)}{n} \right) + \frac{\ln(2)}{n} \notag \\
    &\stackrel{(b)}{\leq}  \mathbb{E}_p \left [ R(T, d, \rho) \right ]  + \frac{K + 13/4}{2} \frac{\ln n}{n} + \frac{2 \min \left( \ln(K), \ln(J) \right)}{d \, n^{5/8}} \rho_{\max} +  \mathcal{G} \frac{\ln \ln n}{n}   + \notag \\
    & \,\,\,\,\,\,\,\,\,\,\,\,\,\,\,\,\,\,\,\,\,\,\,\,\,\,\,\,\,\,\,\,\,\,\,\,\,\,\left( (n+1)^{J^2K^2-1} + 1 \right) \left( (n+1)^{J-1} \right) \frac{e}{n^{J^2 K^2 + J - 1}} \left( \ln K + \frac{\ln(2)}{n} \right) + \frac{\ln(2)}{n}. \label{joopapee00j}
\end{align}
In equality $(a)$, we use the fact that conditioned on the type, $X^n$ is distributed uniformly over the type class $t$, which we denote by writing $X_t^n$. Equality $(b)$ follows from $(\ref{phoit})$.

Finally, we use post-correction to make the code $d$-semifaithful. This post-correction was described in the proof of Theorem 1 (Section \ref{thms12}), specifically $(\ref{opwxz})$ and $(\ref{rate_incr})$. By adding the rate increment from post-correction in $(\ref{rate_incr})$ to the expression in $(\ref{joopapee00j})$, the expected rate of the overall code is upper bounded by 
\begin{align*}
    &\leq \mathbb{E}_p \left [ R(T, d, \rho) \right ] + \frac{2 \rho_{\max} \ln(n)}{d \, n^{5/8}}  +  \frac{4 \rho_{\max}\left(  \min \left( \ln(K), \ln(J) \right) + \ln(2) \right)}{d \, n^{5/8}} +  \frac{K + 21/4}{2} \frac{\ln n}{n} +    \mathcal{G} \frac{\ln \ln n}{n}   + \notag \\
    & \,\,\,\,\,\,\,\,\,\,\,\,\,\,\,\,\,\,\,\,\,\,\,\,\,\left( (n+1)^{J^2K^2-1} + 1 \right) \left( (n+1)^{J-1} \right) \frac{e}{n^{J^2 K^2 + J - 1}} \left( \ln K + \frac{\ln(2)}{n} \right) + \frac{\ln(16) + \min\left(\ln(K), \ln(J) \right) }{n}. 
\end{align*}
This finishes the proof of Theorem \ref{upper_bound4}.

\appendices

\section{Proof of Lemma \ref{lemmaunifcont} \label{facebookdafarhai}}

Since $\mathcal{P}(A) \times \mathcal{D}$ is a compact set, it suffices to show that $R(p,d,\rho)$ is a continuous function of the pair $(p, \rho)$. With some abuse of notation, we define  
\begin{align*}
    \rho(p, W) \triangleq \sum_{j \in A, k \in B} p(j) W(k|j) \rho(j, k)
\end{align*}
for any $p \in \mathcal{P}(A)$, $W \in \mathcal{P}(B|A)$ and $\rho \in \mathcal{D}$.

Fix any $(p^*, \rho^*) \in \mathcal{P}(A) \times \mathcal{D}$ and let $(p^{(m)}, \rho^{(m)}) \to (p^*, \rho^*)$ as $m \to \infty$ with respect to the metric defined in $(\ref{newmetric})$. Since $R(p,d,\rho)$ is continuous in $d$, it is possible to choose, for every $\epsilon > 0$, a $Q \in \mathcal{P}(B|A)$ satisfying $\rho^*(p^*, Q) < d$ and $I(p^*, Q) < R(p^*, d, \rho^*) + \epsilon$. By continuity of $I(p, W)$ in $p$ and $\rho(p, W)$ in both $p$ and $\rho$, it follows that for sufficiently large $m$, we have $\rho^{(m)}(p^{(m)}, Q) < d$ and $I(p^{(m)}, Q) < R(p^*, d, \rho^*) +  \epsilon$. Since $R(p^{(m)}, d, \rho^{(m)}) \leq I(p^{(m)}, Q)$ eventually, we obtain 
\begin{align}
    \limsup_{m \to \infty} R(p^{(m)}, d, \rho^{(m)}) \leq R(p^*, d, \rho^*). \label{batman1}
\end{align}
On the other hand, let $Q^{(m)} \in \mathcal{P}(B|A)$ achieve the minimum in the definition of $R(p^{(m)}, d, \rho^{(m)})$. Let $\{m_l\}$ be a subsequence such that $Q^{(m_l)} \to Q$ for some $Q$ and 
\begin{align*}
    \lim_{l \to \infty} R(p^{(m_l)}, d, \rho^{(m_l)}) = \liminf_{m \to \infty} R(p^{(m)}, d, \rho^{(m)}).
\end{align*}
If $d \geq \min_{k \in B} \sum_{j \in A} p^*(j) \rho^*(j,k)$, then 
\begin{align}
    0 = R(p^*, d, \rho^*) \leq \liminf_{m \to \infty} R(p^{(m)}, d, \rho^{(m)}). \label{bat22}
\end{align}
If $d < \min_{k \in B} \sum_{j \in A} p^*(j) \rho^*(j,k)$, then for sufficiently large $m$, we have $d < \min_{k \in B} \sum_{j \in A} p^{(m)}(j) \rho^{(m)}(j, k)$ and therefore, $\rho^{(m)}(p^{(m)}, Q^{(m)}) = d$. The last assertion follows from the fact that $R(p,d,\rho)$ is strictly decreasing in $d$ for $d \in (0, \min_{k \in B} \sum_{j \in A} p(j) \rho(j,k) )$. Now since 
\begin{align*}
    \lim_{l \to \infty} \rho^{(m_l)}(p^{(m_l)}, Q^{(m_l)}) = \rho^*(p^*, Q) = d,
\end{align*}
we obtain 
\begin{align}
    R(p^*, d, \rho^*) &\leq I(p^*, Q) \notag \\
    &= \lim_{l \to \infty} I(p^{(m_l)}, Q^{(m_l)}) \notag \\
    &= \liminf_{m \to \infty} R(p^{(m)}, d, \rho^{(m)}). \label{bat30}
\end{align}
The result of Lemma \ref{lemmaunifcont} follows from $(\ref{batman1})$, $(\ref{bat22})$ and $(\ref{bat30})$.

\section{Proof of Lemma \ref{lemmagolden2} \label{hiuhiuhiu}}

Fix $d > 0$ and let $\rho$ be a fixed distortion measure. Let $X^n$ be an i.i.d. source sequence distributed according to some distribution $p \in \mathcal{P}(A)$. For any $d$-semifaithful code $(\phi_n, f_n, g_n)$, let $Y^n = g_n(f_n(\phi_n(X^n)))$. It was shown in \cite[Appendix E]{mahmood2021lossy} that 
\begin{align}
    \frac{1}{n} H(Y^n) &\geq  \mathbb{E}_p \left [ R(T,d,\rho) \right] - (JK + J - 2)\frac{\ln n }{n} - \frac{JK + J - 2}{n}. \label{hoopeenurban}
\end{align}
To prove Lemma \ref{lemmagolden2}, we only use the fact that the optimal expected length $L^*_n$ for a non-prefix code \cite[Theorem 1]{verdu2} satisfies $L^*_n \leq H(Y^n)$ and 
\begin{align}
    H(Y^n) \leq L_n^* + (L^*_n+1) \ln (L^*_n + 1) - L^*_n \ln (L^*_n). \label{kameenaadeel}
\end{align}
Since $(x+1) \ln(x+1)  - x \ln(x)$ is non-decreasing in $x$, we can use $L^*_n \leq H(Y^n) \leq n \ln K$ to write $(\ref{kameenaadeel})$ as
\begin{align*}
    H(Y^n) \leq L_n^* + (n \ln K + 1) \ln (n \ln K + 1) - (n \ln K ) \ln (n \ln K ).
\end{align*}
Hence, we have 
\begin{align*}
   \frac{1}{n} \mathbb{E}_p \left [ l(f_n(\phi_n(X^n, \rho))) \right ] &\geq \frac{1}{n}L_n^*\\
   &\geq \frac{1}{n}H(Y^n) -  \left( \ln K + \frac{1}{n} \right)  \ln (n \ln K + 1) + (\ln K) \ln(n \ln K)\\
   &\stackrel{(1)}{\geq} \mathbb{E}_p \left [ R(T,d,\rho) \right] - (JK + J - 2)\frac{\ln n }{n} - \frac{JK + J - 2}{n} - \\
   & \,\,\,\,\,\,\,\,\,\,\,\,\,\,\,\,\,\,\,\,\,\,\,\,\,\,\,\left( \ln K + \frac{1}{n} \right)  \ln (n \ln K + 1) + (\ln K) \ln(n \ln K)\\
   &\stackrel{(2)}{\geq} \mathbb{E}_p \left [ R(T,d,\rho) \right] - (JK + J - 2)\frac{\ln n }{n} - \frac{JK + J - 2}{n} - \\
   & \,\,\,\,\,\,\,\,\,\,\,\,\,\,\,\,\,\,\,\,\,\,\,\,\,\,\, (\ln K) \ln \left(1 + \frac{1}{n \ln K} \right) - \frac{\ln n}{n} - \frac{\ln (2 \ln K)}{n} \\
   &\stackrel{(3)}{=} \mathbb{E}_p \left [ R(T,d,\rho) \right ] - (JK + J - 1) \frac{\ln n}{n} + o \left(\frac{\ln n}{n} \right).
\end{align*}
Inequality $(1)$ above follows from $(\ref{hoopeenurban})$. Inequality $(2)$ above holds for $n > \frac{1}{\ln K}$. Equality $(3)$ above holds because as $n \to \infty$, $\ln(1+1/n)$ approaches zero faster than $\ln n / n$. It is easy to see that the $o(\ln n / n)$ term, when divided by $\ln n / n$, tends to zero at a rate determined only by alphabet sizes $J$ and $K$.

\section{Proof of Lemma \ref{modcont} \label{modcontproof} }

For any $a \geq \sqrt{2J+2}$, we have 
\begin{align}
&\mathbb{E}_p [R(T,d,\rho)] \notag\\
    &= \sum_{t \in \mathcal{P}_n(A)} p^n(T^n_A(t)) R(t,d,\rho) \notag\\
    &= \sum_{t:||t - p||_2 \leq a \sqrt{\ln n / n}}  p^n(T^n_A(t)) R(t,d,\rho) \notag\\
    & \,\,\,\,\,\,\,\,\,\,\,\,+ \sum_{t:||t - p||_2 > a \sqrt{\ln n / n}}  p^n(T^n_A(t)) R(t,d,\rho) \notag \\
    &\leq \sum_{t:||t - p||_2 \leq a \sqrt{\ln n / n}}  p^n(T^n_A(t)) R(t,d,\rho) +   \ln(K) \frac{e^{J-1}}{n^2}, \label{jobbuu}
\end{align}
where the last inequality follows from Lemma \ref{lemmatypes} and the fact that $R(t,d,\rho) \leq \ln(K)$ from the assumption in $(\ref{dist_assump})$. Now since $R(p,d,\rho)$ is uniformly continuous on $\mathcal{P}(A) \times \mathcal{D}$ by Lemma \ref{lemmaunifcont}, it admits a modulus of continuity $\omega(\cdot)$ satisfying $\lim_{t \to 0} \omega(t) = \omega(0) = 0$ and 
\begin{align}
    |R(p_1,d,\rho_1) - R(p_2, d, \rho_2)| \leq \omega \left(||(p_1, \rho_1) - (p_2, \rho_2)|| \right). \label{kyeaz}
\end{align}
Therefore, we can use $(\ref{kyeaz})$ in $(\ref{jobbuu})$ to obtain 
\begin{align*}
    &\mathbb{E}_p \left [ R(T, d, \rho) \right ]\\
    &\leq \sum_{t:||t - p||_2 \leq a \sqrt{\ln n / n}}  p^n(T^n_A(t)) \left( R(p,d,\rho) + \omega\left( a \sqrt{\frac{\ln n}{n}} \right) \right) \\
    & \,\,\,\,\,\,\,\,\,\,\,\,\,\,\,\,\,\,\,\,\,\,\,\,+   \ln(K) \frac{e^{J-1}}{n^2}\\
    &\leq R(p,d,\rho) + \omega\left( a \sqrt{\frac{\ln n}{n}} \right) +  \ln(K) \frac{e^{J-1}}{n^2}.
\end{align*}
Similarly, we have 
\begin{align*}
    &\mathbb{E}_p \left [ R(T, d, \rho) \right ] \notag \\
    &\geq \sum_{t:||t - p||_2 \leq a \sqrt{\ln n / n}}  p^n(T^n_A(t)) R(t,d,\rho)\\
    &\geq R(p,d,\rho) - \omega \left( a \sqrt{\frac{\ln n}{n}} \right) - \ln(K) \frac{e^{J-1}}{n^2}. 
\end{align*}

\section{Strongly Universal Codes over a restricted set of source distributions   \label{newresult447}}

The $O(\ln n/n)$ convergence rate for weakly universal 
$d$-semifaithful codes in prior works holds under certain
regularity conditions on the source distribution and the
distortion measure. Corollary \ref{corollary:redundancylower} shows that eliminating
these conditions slows convergence rate to $1/\sqrt{n}$,
even in the non-universal context. Here we show that with
the regularity conditions of \cite{yang3} in place, upgrading
to strong universality also slows the convergence rate to
$O(1/\sqrt{n})$.

For a given $p \in \mathcal{P}(A)$, $d > 0$ and $\rho \in \mathcal{D}$, let $(Q^*_{B|A}, \lambda^*)$ be a solution to the Lagrange formulation of the rate-distortion problem $R(p,d,\rho)$ as in $(\ref{compslack444})$$-$$(\ref{slopeeeguy})$, and let $Q^{p, d, \rho }$ be the corresponding optimal reconstruction distribution on $B$. The assumed regularity conditions in \cite{yang3} are, in
our notation (cf.~\cite{yang2}),
\begin{enumerate}
    \item The matrix $\mathcal{E}(\lambda^*)$, defined by
     $[\mathcal{E}(\lambda^*)]_{j,k} = e^{-\lambda^* \rho(j,k)}$
     is full column rank.
 \item $p$ and $Q^{p,d,\rho}$ are both full support.
 \item $0 < \lambda^* < \infty$.
 \item The determinant of the Jacobian, 
     \begin{equation}
         \frac{\partial F(p,\lambda^*)}{\partial p_{j_1} \partial p_{j_2}
         \cdots p_{j_K} \partial \lambda^*},
     \end{equation}
     is nonzero for some $1 \le j_1 < j_2 \cdots < j_K \leq J$,
     where $F$ is the vector-valued function
     \begin{equation}
        F(p,\lambda^*) = \left[ \begin{array}{c}
              Q^{p,d,\rho}(1) \\
              Q^{p,d,\rho}(2) \\
              \vdots \\
              Q^{p,d,\rho}(k) \\
              d 
      \end{array} \right],
     \end{equation}
     where we have used the implicit one-to-one mapping between
     $\lambda^*$ and $d$ for a given $p$ and $\rho$.
\end{enumerate}

In fact, it is impossible to satisfy the fourth condition
because the first $K$ components of $F(p,\lambda^*)$ sum to
one; thus, their derivative with respect to any input
must sum to zero. This could potentially be rectified by
redefining $F$ as 
     \begin{equation}
        F(p,\lambda^*) = \left[ \begin{array}{c}
              Q^{p,d,\rho}(1) \\
              Q^{p,d,\rho}(2) \\
              \vdots \\
              Q^{p,d,\rho}(k-1) \\
              d 
      \end{array} \right],
     \end{equation}
     and modifying the proofs accordingly. In any event,
     assumption 4) is assumed in both the converse and
     achievability results in~\cite{yang3} (and similarly
     in~\cite{yang2}), but is only used in the proof of
     the converse result (and similarly in~\cite{yang2}).
     As such, we will only consider the first three
     assumptions.

\begin{lemma}
Consider alphabets $A = B = \{ 0,1 \}$, fix the distortion measure 
\begin{align*}
    \rho = \begin{bmatrix} 
    0 & \rho_{\max}\\
    \rho_{\max} & 0
    \end{bmatrix}
\end{align*}
and distortion level $d \in (0, \rho_{\max}/2)$. Let $\mathcal{P}_{\rho,d} \subset \mathcal{P}(A)$ be the set of source distributions satisfying conditions 1)-3) above for this choice of $\rho$. Then
\begin{align*}
    \liminf_{n \rightarrow \infty}  \inf_{(\phi_n,f_n, g_n) } \,\, \sup_{p \in \mathcal{P}_{\rho,d}} \,\,  \left [  \frac{\ln(2)}{n} \mathbb{E}_p \left [ l (f_n(\phi_n(X^n)))  \right ] - R(p,d,\rho) \right] \sqrt{n} & > 0,
\end{align*}
where the infimum is over all (prefix or non-prefix) $d$-semifaithful codes. 
\label{baskardoyar}
\end{lemma}

\begin{IEEEproof}
    The proof is similar to that of Lemma~\ref{lemma:expectedlower} and Corollary \ref{corollary:redundancylower}.
  Let $p_n \in \mathcal{P}(A)$ be a sequence of source distributions given by $p_n(1) = \bar{d} + 1/n$, where $\bar{d} \triangleq d/\rho_{\max}$, which is well-defined for large $n$. Since $\bar{d} \in (0, 1/2)$, we have, for sufficiently large $n$, $p_n(1) \in (\bar{d}, 1/2)$. 
Denoting the binary entropy function by $H_b(\cdot)$,
we have 
\begin{align}
    &\,\mathbb{E}_{p_n} \left [ R(T,d,\rho) \right ] \notag \\
    &= \sum_{t \in \mathcal{P}_n(A)} p_n^n(T^n_A(t)) R(t,d,\rho)\notag \\
    &\geq \sum_{p_n(1) +  \sqrt{\frac{p_n(1)(1-p_n(1))}{n}} < t(1) \leq p_n(1) +  2\sqrt{\frac{p_n(1)(1-p_n(1))}{n}}} p_n^n(T^n_A(t)) \left [ H_b(t(1)) - H_b(\bar{d}) \right ] \notag \\
    &\geq \left( H_b\left(p_n(1) +  \sqrt{\frac{p_n(1)(1-p_n(1))}{n}}  \right) - H_b\left(\bar{d}\right) \right) \cdot \notag \\
    & \,\,\,\,\,\,\,\,\,\,\,\,\mathbb{P}\left( p_n(1) +  \sqrt{\frac{p_n(1)(1-p_n(1))}{n}} < \frac{1}{n} \sum_{i=1}^n X_i \leq p_n(1) +  2\sqrt{\frac{p_n(1)(1-p_n(1))}{n}} \right), \label{exp_low_bndmp22}
\end{align}
where the second inequality above assumes sufficiently large $n$.
By a simple Taylor series expansion,
\begin{align}
    &\, H_b\left (p_n(1) +  \sqrt{\frac{p_n(1)(1-p_n(1))}{n}}\right ) - H_b\left(\bar{d}\right) \notag\\
%    &\geq \left ( \frac{1}{n} + \sqrt{\frac{p_n(1)(1-p_n(1))}{n}} \right) \ln \left( \frac{1-\bar{d}}{\bar{d}} \right) - \frac{1}{2 \tilde{d}(1-\tilde{d})} \left ( \frac{1}{n} + \sqrt{\frac{p_n(1)(1-p_n(1))}{n}} \right)^2 \notag \\
    &\geq \left ( \frac{1}{n} + \sqrt{\frac{p_n(1)(1-p_n(1))}{n}} \right) \ln \left( \frac{1-\bar{d}}{\bar{d}} \right) - \frac{1}{2 \bar{d}(1-\bar{d})} \left ( \frac{1}{n} + \sqrt{\frac{p_n(1)(1-p_n(1))}{n}} \right)^2 \notag \\
    &\geq \left ( \frac{1}{n} + \sqrt{\frac{\bar{d}(1-\bar{d})}{n}} \right) \ln \left( \frac{1-\bar{d}}{\bar{d}} \right) - \frac{1}{2 \bar{d}(1-\bar{d})} \left ( \frac{1}{n} + \sqrt{\frac{p_n(1)(1-p_n(1))}{n}} \right)^2. \label{elephantrabbit}
\end{align}
    A standard application of the Berry-Esseen theorem (with constant $1/2$~\cite{Korolev:Shevtsova:2010,Tyurin2010}) yields 
\begin{align}
    &\,\mathbb{P}\left( p_n(1) +  \sqrt{\frac{p_n(1)(1-p_n(1))}{n}} < \frac{1}{n} \sum_{i=1}^n X_i \leq p_n(1) +  2\sqrt{\frac{p_n(1)(1-p_n(1))}{n}}  \right) \notag \\
    &\ge \left[ \Phi(2) - \Phi(1) - \frac{(1-p_n(1))^2 + {p_n(1)}^2}
    {\sqrt{n p_n(1) (1-p_n(1))}} \right] \notag \\
   % &\geq \left[ \Phi(2) - \Phi(1) - \frac{(1-\bar{d})^2 + {\bar{d}}^2}
   % {\sqrt{n \bar{d} (1-\bar{d})}} \right] \notag \\
    &\geq \frac{1}{10},
      \label{berry_conv_prop22} 
\end{align}
for sufficiently large $n$.
Substituting $(\ref{elephantrabbit})$ and $(\ref{berry_conv_prop22})$ into $(\ref{exp_low_bndmp22})$, we have 
\begin{align}
    &\,\mathbb{E}_{p_n}[R(T,d,\rho)] - R(p_n,d,\rho) \notag \\
    &\geq \frac{1}{10} \left ( \frac{1}{n} + \sqrt{\frac{\bar{d}(1-\bar{d})}{n}} \right) \ln \left( \frac{1-\bar{d}}{\bar{d}} \right) - \frac{1}{20\, \bar{d}(1-\bar{d})} \left ( \frac{1}{n} + \sqrt{\frac{p_n(1)(1-p_n(1))}{n}} \right)^2 - R(p_n,d,\rho) \notag \\
    &= \Omega\left(\frac{1}{\sqrt{n}} \right), \label{houseofdead4}
\end{align}
where the last equality above follows from the upper bound 
\begin{align*}
    &\,R(p_n,d,\rho)\\
    & \le H_b(p_n(1)) \\
    &\leq \frac{1}{n} \ln \left(\frac{1 - \bar{d}}{\bar{d}} \right).
\end{align*}

Now consider the subset $\mathcal{P}_{\rho,d}^* \subset \mathcal{P}(A)$ defined as 
\begin{align}
    \mathcal{P}_{\rho,d}^* \triangleq \{p \in \mathcal{P}(A): \bar{d} <  p(1) < 1/2 \}. \label{Pdstar}
\end{align}
We first check that the set $\mathcal{P}_{\rho,d}^*$ satisfies the assumptions 1)-3), i.e., $\mathcal{P}_{\rho,d}^* \subset \mathcal{P}_{\rho,d}$. 
Fix any $p \in \mathcal{P}_{\rho,d}^*$ and let $(Q^*_{B|A}, \lambda^*)$ be a solution to the Lagrange formulation of the rate-distortion problem in $(\ref{compslack444})$$-$$(\ref{slopeeeguy})$. Obviously $p$ is full-support, and we have 
\begin{equation}
    \label{eq:ratenotzero}
    0 < d < \min_{k \in B}\,\sum_{j \in A} p(j) \rho(j, k).
\end{equation}
The matrix $\mathcal{E}(\lambda^*)$ associated with $\lambda^*$ and $\rho$, 
\begin{align*}
    \begin{bmatrix}
    1 & e^{- \lambda^* \rho_{\max}}\\
    e^{-\lambda^* \rho_{\max}} & 1
    \end{bmatrix},
\end{align*}
is of full rank; hence, from \cite[Lemma 7]{yang2}, the optimal output distribution $Q^{p,d,\rho}$ is unique. From \cite[Theorem 10.3.1]{thomas_cov}, one can infer that $Q^{p,d,\rho}$ is given by 
\begin{align}
\left(Q^{p,d,\rho}(0), Q^{p,d,\rho}(1)\right ) &= \left(\frac{1 - p(1) - \bar{d}}{1 - 2\bar{d}}, \frac{p(1) - \bar{d}}{1 - 2\bar{d}}
    \right), \label{iopafaraz} 
\end{align} 
and it is easy to check that $Q^{p,d,\rho}$ is full-support for the specified $p,d$ and $\rho$. Furthermore, we have 
\begin{align*}
    \lambda^* &= -\frac{\partial}{\partial d} R(p,d,\rho)\\
    &= -\frac{\partial}{\partial d} \left [ H_b(p(1)) - H_b(\bar{d}) \right ]\\
    &= \frac{\partial}{\partial d} H_b(d/\rho_{\max}), \\
    & = \frac{1}{\rho_{\max}} \ln \left(\frac{\rho_{\max}}{d} - 1\right).
\end{align*}
Thus $0 < d < \rho_{\max}/2$ implies that $0 < \lambda^* < \infty$.
We conclude that $\mathcal{P}_{\rho,d}^* \subset \mathcal{P}_{\rho,d}$. 
Now the expected rate of a strongly universal code (prefix or non-prefix) with uniform convergence over $\mathcal{P}_{\rho,d}^*$ satisfies  
\begin{align*}
    &\,\sup_{p \in \mathcal{P}_{\rho,d}^*}\, \left [ \frac{\ln(2)}{n} \mathbb{E}_p\left [ l(f_n(\phi_n(X^n))) \right] - R(p,d,\rho)\right]\\
    &\stackrel{(a)}{\geq} \sup_{p \in \mathcal{P}_{\rho,d}^*} \left [ \mathbb{E}_p \left [ R(T,d,\rho) \right]- R(p,d,\rho) - (JK + J)\frac{\ln n}{n}  \right ]\\
    &\stackrel{(b)}{\geq}  \mathbb{E}_{p_n} \left [ R(T,d,\rho) \right]- R(p_n,d,\rho) - (JK + J) \frac{\ln n}{n} \\
    &\stackrel{(c)}{=} \Omega \left(\frac{1}{\sqrt{n}} \right). 
\end{align*}
Inequality $(a)$ holds for sufficiently large $n$ where we used Lemmas $\ref{lemmagolden}$ and $\ref{lemmagolden2}$ for prefix and non-prefix codes, respectively. In inequality $(b)$, we used the fact that the sequence of $p_n$ satisfies $p_n \in \mathcal{P}_{\rho,d}^*$ for every $n$. In equality $(c)$, we used $(\ref{houseofdead4})$. 
\end{IEEEproof}

\section{Proof of Lemma \ref{rootnode} \label{pandakakuta}}

Fix $d > 0$. Let $x^n$ be a source sequence with type $t = t(x^n)$ and $\rho$ be a distortion measure. Let $(Q^*_{B|A}, \lambda^*)$ be a solution to the Lagrange formulation of the rate-distortion problem as in $(\ref{compslack444})$$-$$(\ref{liop2})$ and $Q^{t, d, \rho }$ be the corresponding optimal reconstruction distribution on $B$. Define $Z_i = \rho(x_i, Y_i)$ where $Y_i \sim Q^{t,d,\rho}$. Letting $\epsilon$ be any real number, we can write   
\begin{align*}
      \mathbb{P} \left( \rho_n(x^n, Y^n) \leq d + \epsilon  \right) &= \mathbb{P} \left( \frac{1}{n} \sum_{i=1}^n Z_i \leq d + \epsilon \right). 
\end{align*}
Let $f_i$ be the probability mass function of $Z_i$. The cumulant generating function of $Z_i$ is defined as 
\begin{align*}
    \Lambda_i(\lambda) &\triangleq \ln \left(\mathbb{E}\left[e^{\lambda Z_i}\right ] \right) = \ln \left( \sum_{k \in B} Q^{t,d,\rho}(k) e^{\lambda \rho(x_i, k)} \right) . 
\end{align*}
The distribution of $Z_i$ depends on $x_i$ only through its value, not the index. Hence, for each $j \in A$, define  
\begin{align*}
    \Lambda^{(j)}(\lambda) &\triangleq \ln \left( \sum_{k \in B} Q^{t,d,\rho}(k) e^{\lambda \rho(j, k)} \right),
\end{align*}
which is the cumulant generating function of $Z_i$ if $x_i = j$. We apply the exponential tilting technique to form the distribution $r_i$ given by 
\begin{align*}
    \frac{r_i(z)}{f_i(z)} &= e^{\lambda z - \Lambda_i(\lambda)},
\end{align*}
where $\lambda$ is a parameter which will be chosen later. Further define for each $1 \leq i \leq n$
\begin{align*}
    d_i &\triangleq \frac{\sum \limits_{k \in B} Q^{t,d,\rho}(k)\, \rho(x_i, k)\, e^{\lambda \rho(x_i, k) }}{\sum \limits_{k' \in B} Q^{t,d,\rho}(k') e^{\lambda \rho(x_i, k')}}
\end{align*}
and for each $j \in A$
\begin{align*}
    d^{(j)} &\triangleq \frac{\sum \limits_{k \in B} Q^{t,d,\rho}(k)\, \rho(j, k)\, e^{\lambda \rho(j, k) }}{\sum \limits_{k' \in B} Q^{t,d,\rho}(k') e^{\lambda \rho(j, k')}}. 
\end{align*}
Then we have 
\begin{align}
    &\,\,\mathbb{P} \left( \rho_n(x^n, Y^n) \leq d + \epsilon \right) \notag\\
    &= \sum_{z^n : \frac{1}{n}  \sum \limits_{i=1}^n z_i \leq d + \epsilon} f_1(z_1) f_2(z_2) \cdots f_n(z_n) \notag \\
    &= \exp \left(\sum \limits_{i=1}^n \Lambda_i(\lambda)\right)  \sum_{z^n : \frac{1}{n} \sum \limits_{i=1}^n z_i \leq d + \epsilon} \exp \left( -\sum_{i=1}^n \lambda z_i \right)  r_1(z_1) r_2(z_2) \cdots r_n(z_n) \notag\\
    &= \exp \left(\sum \limits_{i=1}^n \Lambda_i(\lambda)\right)  \sum_{z^n : \frac{1}{n} \sum \limits_{i=1}^n z_i \leq d + \epsilon} \exp \left( - \lambda \sum_{i=1}^n  (z_i -d_i +d_i) \right)  r_1(z_1) r_2(z_2) \cdots r_n(z_n) \notag \\
    &= \exp \left(- \sum_{i=1}^n \left [\, \lambda d_i - \Lambda_i(\lambda) \,\right ] \right)    \sum_{z^n : \frac{1}{n} \sum \limits_{i=1}^n z_i \leq d + \epsilon} \exp \left( - \lambda \sum_{i=1}^n  (z_i -d_i) \right)  r_1(z_1) r_2(z_2) \cdots r_n(z_n) \notag \\
    &= \exp \left(- n \left [ \lambda \sum_{j \in A} t(j) d^{(j)} - \sum_{j \in A} t(j) \Lambda^{(j)}(\lambda) \right ]  \right)    \sum_{z^n : \frac{1}{n} \sum \limits_{i=1}^n z_i \leq d + \epsilon} \exp \left( - \lambda \sum_{i=1}^n  (z_i -d_i) \right)  r_1(z_1) r_2(z_2) \cdots r_n(z_n). \label{kiopw}
\end{align}
Now we fix $\lambda = -\lambda^*$ throughout. Then, from $(\ref{compslack444})$, we have the following simplified expressions for $d_i$ and $d^{(j)}$:
\begin{align}
    d_i &= \sum_{k \in B} Q^*_{B|A}(k|x_i) \rho(x_i, k) \label{diane} \text{ and }\\
    d^{(j)} &= \sum_{k \in B} Q^*_{B|A}(k|j) \rho(j, k). \label{djjane} 
\end{align}
From $(\ref{jiiiuy})$, we have 
\begin{align*}
    -\lambda^*\sum_{j \in A} t(j) d^{(j)} &= -\lambda^*\sum_{j \in A} t(j) \sum_{k \in B} Q^*_{B|A}(k|j) \rho(j, k)  = -\lambda^* d, 
\end{align*}
Hence, we have from $(\ref{liop2})$ that
\begin{align*}
    \exp \left(- n \left [ -\lambda^* \sum_{j \in A} t(j) d^{(j)} - \sum_{j \in A} t(j) \Lambda^{(j)}(-\lambda^*) \right ]  \right) &= \exp \left(- n \left [ -\lambda^* d - \sum_{j \in A} t(j) \Lambda^{(j)}(-\lambda^*) \right ]  \right) \\
    &= \exp \left( -nR(t,d,\rho) \right).
\end{align*}
Hence, with $\lambda = -\lambda^*$ in $(\ref{kiopw})$, we have  
\begin{align}
     &\,\,\mathbb{P} \left( \rho_n(x^n, Y^n) \leq d + \epsilon \right) \notag \\
     &= e^{-n R(t,d,\rho)} \, \sum_{z^n : \frac{1}{n} \sum \limits_{i=1}^n z_i \leq d + \epsilon} \exp \left( \lambda^* \sum_{i=1}^n  (z_i -d_i) \right)  r_1(z_1) r_2(z_2) \cdots r_n(z_n). \label{changeof}
\end{align}
In $(\ref{changeof})$, performing a change of variable $u_i = z_i - d_i$ and defining $\tilde{r}_i(u) = r_i(u+d_i)$ for each $1 \leq i \leq n$, we obtain 
\begin{align}
    \mathbb{P}\left (\rho_n(x^n, Y^n) \leq d + \epsilon\right) &= e^{-n R(t,d,\rho)} \sum_{u^n : \sum_{i=1}^n u_i  \leq \epsilon\,n  } \exp \left( \lambda^* \sum_{i=1}^n u_i \right) \tilde{r}_1(u_1) \tilde{r}_2(u_2) \cdots \tilde{r}_n(z_n) \notag \\
    &= e^{-n R(t,d,\rho)} \, \mathbb{E} \left   [ \exp\left( \lambda^* \sum_{i=1}^n U_i \right) \mathds{1} \left( \sum_{i=1}^n U_i \leq \epsilon\,n  \right) \right ], \label{cruxinequal}
\end{align}
where $U_1, U_2, ..., U_n$ are independent random variables and $U_i$ is distributed according to $\tilde{r}_i(\cdot)$. We next need to show that the distribution of $U_i$ can be written as 
\begin{align}
    U_i &= \rho(x_i, \tilde{Y}_i) - \sum_{k \in B} Q^*_{B|A}(k|x_i) \rho(x_i, k), \label{distequal}
\end{align}
where the random variable $\tilde{Y}_i \sim Q^*_{B|A}(\cdot | x_i)$. We have 
\begin{align*}
    r_i(z) &= f_i(z) e^{-\lambda^* z - \Lambda_i(-\lambda^*)}\\
    &= \sum_{k \in B} Q^{t,d,\rho} (k) \mathds{1} \left( \rho(x_i, k) = z \right) e^{-\lambda^* z - \Lambda_i(-\lambda^*)} \\
    &= \frac{\sum_{k \in B} Q^{t,d,\rho}(k) \mathds{1}(\rho(x_i, k) = z) e^{-\lambda^* \rho(x_i, k)}}{\sum_{k' \in B} Q^{t,d,\rho}(k') e^{-\lambda^* \rho(x_i, k')}}\\
    &= \sum_{k \in B} Q^*_{B|A}(k|x_i) \mathds{1}(\rho(x_i, k) = z),
\end{align*}
where the last equality follows from $(\ref{compslack444})$. This shows that $\rho(x_i, \tilde{Y}_i)$ has the same distribution as $r_i(\cdot)$. Hence, the assertion in $(\ref{distequal})$ follows from the fact that $\tilde{r}_i(u) = r_i(u + d_i)$.

For any real number $C$, we can lower bound $(\ref{cruxinequal})$ as 
\begin{align}
    &\mathbb{P}\left (\rho_n(x^n, Y^n) \leq d + \epsilon\right) \notag\\
    &\geq e^{-n R(t,d,\rho)} \, \mathbb{E} \left   [ \exp\left( \lambda^* \sum_{i=1}^n U_i \right) \mathds{1} \left( - C  \leq \sum_{i=1}^n U_i \leq \epsilon\,n \right) \right ] \notag \\
    &\geq e^{-n R(t,d,\rho) - C \lambda^* } \mathbb{P} \left( - C \leq \sum_{i=1}^n U_i \leq \epsilon\,n \right). \notag
\end{align}
This finishes the proof of Lemma \ref{rootnode}.

\section{Proof of Lemma \ref{lemlowerbnd} \label{pandakakuta2}}

We start with the result of Lemma \ref{rootnode} and reparametrize $\epsilon$ and $C$ in terms of nonnegative numbers $C_1$ and $\alpha$ as follows: 
\begin{align*}
    \epsilon &= \frac{C_1}{n^\alpha}\\
    C &= C_1 n^{1-\alpha}
\end{align*}
Then we obtain 
\begin{align}
    &\mathbb{P}\left (\rho_n(x^n, Y^n) \leq d + \frac{C_1}{n^\alpha}\right) \notag\\
      &\geq e^{-n R(t,d,\rho) - C_1 \lambda^* n^{1-\alpha}} \mathbb{P} \left( - C_1 n^{1-\alpha} \leq \sum_{i=1}^n U_i \leq C_1 n^{1-\alpha} \right). \label{salamrooh}
\end{align}

To proceed further, we consider two cases parametrized by a nonnegative number $C_2$: 
\begin{enumerate}
    \item $\text{var}\left( \sum_{i=1}^n U_i \right) < C_2 \,n^{2-2\alpha}$
    \item $\text{var}\left( \sum_{i=1}^n U_i \right) \geq C_2 \,n^{2-2\alpha}$
\end{enumerate}
where $\text{var}(\cdot)$ denotes the variance. In the first case above, a simple application of Chebyshev's inequality to $(\ref{salamrooh})$ yields 
\begin{align}
     &\mathbb{P}\left (\rho_n(x^n, Y^n) \leq d + \frac{C_1}{n^{\alpha}}\right) \notag\\
     &\geq e^{-n R(t,d,\rho) - C_1 \lambda^* n^{1-\alpha} } \left(1 - \frac{C_2 }{C_1^2 } \right). \label{chebyshevwala}
\end{align}
For the second case, we use the Berry-Esseen theorem.
Each $U_i$ has support set 
\begin{align*}
    \text{supp}(U_i) =  \left \{ \rho(x_i, k) - d_i : k \in B \right \},
\end{align*}
where $d_i$ is as defined in $(\ref{diane})$.
Since we are only considering the space of uniformly bounded distortion measures, it is easy to see from the definition of $d_i$ that 
\begin{align*}
    \text{supp}(U_i) \subset [-\rho_{\max}, \rho_{\max}].
\end{align*}
Clearly, each $U_i$ has finite second- and third-order moments which we denote by $\mathbb{E}[U_i^2] = \sigma_i^2 $ and $\mathbb{E}[|U_i|^3] = \eta_i$. Hence, we can apply the Berry-Esseen theorem for non-identically distributed summands \cite{esseen11}:
\begin{align}
     &\mathbb{P}\left (\rho_n(x^n, Y^n) \leq d + \frac{C_1}{n^\alpha}\right) \notag\\
      &\geq e^{-n R(t,d,\rho) - C_1 \lambda^* n^{1-\alpha}} \mathbb{P} \left( - C_1 n^{1-\alpha} \leq \sum_{i=1}^n U_i \leq C_1 n^{1-\alpha} \right) \notag \\
      &= e^{-n R(t,d,\rho) - C_1 \lambda^* n^{1-\alpha}} \mathbb{P} \left( - \frac{C_1 n^{1-\alpha}}{\sqrt{\sum_{i=1}^n \sigma_{i}^2}} \leq \frac{\sum_{i=1}^n U_i}{\sqrt{\sum_{i=1}^n \sigma_{i}^2}} \leq \frac{C_1 n^{1-\alpha}}{\sqrt{\sum_{i=1}^n \sigma_{i}^2}} \right) \notag \\
      &\geq e^{-n R(t,d,\rho) - C_1 \lambda^* n^{1-\alpha}} \left( F_n \left(\frac{C_1 n^{1-\alpha}}{\sqrt{\sum_{i=1}^n \sigma_{i}^2}} \right) - F_n \left(- \frac{C_1 n^{1-\alpha}}{\sqrt{\sum_{i=1}^n \sigma_{i}^2}} \right) \right ), \label{jioz}
\end{align}
where $F_n$ denotes the cumulative distribution function of $\frac{\sum_{i=1}^n U_i}{\sqrt{\sum_{i=1}^n \sigma_{i}^2}}$. Now by Berry-Esseen theorem, we have that for all $n$ there exists an absolute constant $C_0$ such that 
\begin{align*}
    \sup_{s \in \mathbb{R}} |F_n(s) - \Phi(s)| \leq C_0 \left( \sum_{i=1}^n \sigma_i^2 \right)^{-3/2} \sum_{i=1}^n \eta_i.
\end{align*}
Since we have 
$$\sum_{i=1}^n \sigma_i^2 \geq C_2 n^{2-2\alpha}$$
and $\eta_i \leq \rho_{\max}^3$ for all $1 \leq i \leq n$, we can write 
\begin{align*}
    \sup_{s \in \mathbb{R}} |F_n(s) - \Phi(s)| \leq C_0 \left(C_2 n^{2 - 2\alpha}  \right)^{-3/2} n \rho_{\max}^3 \leq \frac{C_0 (\rho_{\max})^3}{(C_2)^{3/2}n^{2- 3\alpha}}.
\end{align*}
Using the above bound in $(\ref{jioz})$, we obtain
\begin{align}
      &\mathbb{P}\left (\rho_n(x^n, Y^n) \leq d + \frac{C_1}{n^{\alpha}}\right) \notag\\ 
      &\geq e^{-n R(t,d,\rho) - C_1 \lambda^*  n^{1-\alpha}} \left(\Phi\left( \frac{C_1 n^{1-\alpha}}{\sqrt{\sum_{i'=1}^n \sigma_{i}^2}}\right) - \Phi\left(- \frac{C_1 n^{1-\alpha}}{\sqrt{\sum_{i'=1}^n \sigma_{i}^2}}\right) -  \frac{2C_0 (\rho_{\max})^3}{(C_2)^{3/2}n^{2- 3\alpha}} \right) \notag \\
      &\geq  e^{-n R(t,d,\rho) - C_1 \lambda^*  n^{1-\alpha}} \left( \Phi\left( \frac{C_1  }{n^{\alpha - 1/2}\rho_{\max}}\right) - \Phi\left(- \frac{C_1 }{n^{\alpha - 1/2}\rho_{\max}}\right) -  \frac{2C_0 (\rho_{\max})^3}{(C_2)^{3/2}n^{2-3\alpha}} \right), 
      \label{kiwu}
\end{align}
where we used the upper bound $$\sum_{i=1}^n \sigma_i^2 \leq n \rho_{\max}^2$$ in the last inequality above. We now evaluate the expression in $(\ref{kiwu})$ as follows: 
\begin{align}
    &\Phi\left( \frac{C_1  }{n^{\alpha-1/2}\rho_{\max}}\right) - \Phi\left(- \frac{C_1  }{n^{\alpha-1/2}\rho_{\max}}\right)\notag\\
    &= \frac{1}{\sqrt{2 \pi}} \int_{- \frac{C_1  }{n^{\alpha-1/2}\rho_{\max}}}^{\frac{C_1  }{n^{\alpha-1/2}\rho_{\max}}} e^{-x^2/2} dx \notag\\
    &\geq \frac{1}{\sqrt{2 \pi}} \int_{- \frac{C_1  }{n^{\alpha-1/2}\rho_{\max}}}^{\frac{C_1  }{n^{\alpha-1/2}\rho_{\max}}} \left(1 - \frac{x^2}{2} \right) dx \notag \\
    &= \frac{2C_1}{\sqrt{2 \pi} n^{\alpha - 1/2}\rho_{\max}} - \frac{C_1^3}{3 \sqrt{2 \pi} n^{3 \alpha - 3/2}(\rho_{\max})^3}. \label{truncsum}
\end{align}
We can use $(\ref{truncsum})$ in $(\ref{kiwu})$ to obtain 
\begin{align}
     &\mathbb{P}\left (\rho_n(x^n, Y^n) \leq d + \frac{C_1 }{n^{\alpha}}\right) \notag\\ &\geq e^{-n R(t,d,\rho) - C_1 \lambda^*  n^{1-\alpha}} \left(\frac{2C_1}{\sqrt{2 \pi} n^{\alpha - 1/2}\rho_{\max}} - \frac{C_1^3}{3 \sqrt{2 \pi} n^{3 \alpha - 3/2}(\rho_{\max})^3} - \frac{2C_0 (\rho_{\max})^3}{(C_2)^{3/2} n^{2-3\alpha}} \right) \notag \\
     &\geq e^{-n R(t,d,\rho) - C_1 \lambda^* n^{1-\alpha}} \left(\frac{C_1}{\sqrt{2 \pi} n^{\alpha - 1/2}\rho_{\max}}  - \frac{2C_0(\rho_{\max})^3}{(C_2)^{3/2} n^{2-3\alpha}}  \right), \label{nlargeenough}
\end{align}
where inequality $(\ref{nlargeenough})$ follows by assuming 
\begin{align*}
    n &\geq \left( \frac{(C_1)^2}{3 (\rho_{\max})^2} \right)^{\frac{1}{2 \alpha - 1}}.
\end{align*}
Note that the lower bounds in $(\ref{chebyshevwala})$ and $(\ref{nlargeenough})$ hold uniformly for all $t \in \mathcal{P}_n(A)$ and $\rho \in \mathcal{D}$. Taking the minimum of $(\ref{chebyshevwala})$ and $(\ref{nlargeenough})$ gives the result of Lemma \ref{lemlowerbnd}.  


\section*{Acknowledgment}

The authors wish to thank the reviewers and
the associate editor, whose suggestions have
improved the paper. This
research was supported by the
US National Science Foundation under grants
 CCF-2008266 and CCF-1934985, by the US Army
 Research Office under grant W911NF-18-1-0426,
and by a gift from Google.


\ifCLASSOPTIONcaptionsoff
  \newpage
\fi



\bibliographystyle{IEEEtran}
\end{document}